\documentclass[sigconf]{acmart}
\usepackage{algorithmic}
\usepackage[ruled, vlined, commentsnumbered, linesnumbered]{algorithm2e}
\usepackage{mathdots}
\usepackage{amsfonts}
\usepackage{graphicx}
\usepackage{CJKutf8}
\usepackage[normalsize]{subfigure}
\usepackage{wrapfig}
\usepackage{mathdots}
\usepackage{appendix}
\usepackage{multirow}

\newcommand{\DEL}[1]{\iffalse #1 \fi}
\newcommand{\rd}{\color{red}}
\newcommand{\bl}{\color{blue}}
\newcommand{\squishlist}{
\begin{list}{$\bullet$}
  { \setlength{\itemsep}{0pt}
     \setlength{\parsep}{0pt}
     \setlength{\topsep}{0pt}
     \setlength{\partopsep}{0pt}
     \setlength{\leftmargin}{0em}
     \setlength{\labelwidth}{0em}
     \setlength{\labelsep}{0.2em} } }

\newcommand{\squishlisttwo}{
\begin{list}{$\bullet$}
  { \setlength{\itemsep}{0pt}
     \setlength{\parsep}{0pt}
    \setlength{\topsep}{0pt}
    \setlength{\partopsep}{0pt}
    \setlength{\leftmargin}{2em}
    \setlength{\labelwidth}{1.5em}
    \setlength{\labelsep}{0.5em} } }

\newcommand{\squishend}{
  \end{list}  }

\newtheorem{definition}{\bf{Definition}}[section]
\AtBeginDocument{%
  }

\copyrightyear{2024}
\acmYear{2024}
\setcopyright{acmlicensed}\acmConference[SIGSPATIAL '24]{The 32nd ACM
International Conference on Advances in Geographic Information
Systems}{October 29-November 1, 2024}{Atlanta, GA, USA}
\acmBooktitle{The 32nd ACM International Conference on Advances in
Geographic Information Systems (SIGSPATIAL '24), October 29-November 1,
2024, Atlanta, GA, USA}
\acmDOI{10.1145/3678717.3691211}
\acmISBN{979-8-4007-1107-7/24/10}

\begin{document}

\title{Protecting Vehicle Location Privacy with Contextually-Driven Synthetic Location Generation
\vspace{-0.00in}
}
\DEL{
\author{Sourabh Yadav}
\affiliation{%
  \institution{Department of Computer Science and Engineering, University of North Texas}
  \city{Denton}
  \state{Texas}
  \country{USA}}
\email{sourabhyadav@my.unt.edu}

\author{Chenyang Yu}
\affiliation{%
  \institution{Department of Computer Science and Engineering, University of North Texas}
  \city{Denton}
  \state{Texas}
  \country{USA}}
\email{chenyangyu@my.unt.edu}

\author{Xinpeng Xie}
\affiliation{%
  \institution{Department of Computer Science and Engineering, University of North Texas}
  \city{Denton}
  \state{Texas}
  \country{USA}}
\email{xinpengxie@my.unt.edu}

\author{Yan Huang}
\affiliation{%
  \institution{Department of Computer Science and Engineering, University of North Texas}
  \city{Denton}
  \state{Texas}
  \country{USA}}
\email{yan.huang@unt.edu}

\author{Chenxi Qiu}
\affiliation{%
  \institution{Department of Computer Science and Engineering, University of North Texas}
  \city{Denton}
  \state{Texas}
  \country{USA}}
\email{chenxi.qiu@unt.edu}}
\author{Sourabh Yadav, Chenyang Yu, Xinpeng Xie, Yan Huang, Chenxi Qiu}
\affiliation{%
  \institution{Department of Computer Science and Engineering, University of North Texas, USA }
\city{}
  \country{}}

\begin{abstract}

Geo-obfuscation is a \emph{Location Privacy Protection Mechanism} used in location-based services that allows users to report obfuscated locations instead of exact ones. A formal privacy criterion, \emph{geo-indistinguishability (Geo-Ind)}, requires real locations to be hard to distinguish from nearby locations (by attackers) based on their obfuscated representations. However, Geo-Ind often fails to consider context, such as road networks and vehicle traffic conditions, making it less effective in protecting the location privacy of vehicles, of which the mobility are heavily influenced by these factors.

In this paper, we introduce \emph{VehiTrack}, a new threat model to demonstrate the vulnerability of Geo-Ind in protecting vehicle location privacy from context-aware inference attacks. Our experiments demonstrate that VehiTrack can accurately determine exact vehicle locations from obfuscated data, reducing average inference errors by 61.20\% with Laplacian noise and 47.35\% with linear programming (LP) compared to traditional Bayesian attacks. By using contextual data like road networks and traffic flow, VehiTrack effectively eliminates a significant number of seemingly ``impossible'' locations during its search for the actual location of the vehicles. Based on these insights, we propose \emph{TransProtect}, a new geo-obfuscation approach that limits obfuscation to realistic vehicle movement patterns, complicating attackers' ability to differentiate obfuscated from actual locations. Our results show that TransProtect increases VehiTrack's inference error by 57.75\% with Laplacian noise and 27.21\% with LP, significantly enhancing protection against these attacks.

\end{abstract}

\ccsdesc[500]{Security and privacy~Formal security models}

\ccsdesc[300]{Security and privacy~Domain-specific security and privacy architectures}

\ccsdesc[300]{Security and privacy~Social network security and privacy}

\ccsdesc[500]{Security and privacy~Privacy-preserving protocols}

\ccsdesc[500]{Mathematics of computing~Bayesian computation}

\ccsdesc[300]{Mathematics of computing~Bayesian networks}

\ccsdesc[500]{Computing methodologies~Neural networks}

\ccsdesc[500]{Computing methodologies~Modeling methodologies}

\keywords{Geo-Indistinguishability, location privacy, location-based service}
\vspace{-0.05in}
\maketitle

\vspace{-0.05in}
\section{Introduction}
\vspace{-0.02in}
With the increasing availability of wireless connectivity and advances in positioning technologies, vehicles are heavily involved in various \emph{location-based services (LBS)} such as navigation \cite{Wang-WWW2017} and transportation systems \cite{Qiu-TMC2020}. These services often require vehicles to share their real-time locations with central servers, posing significant privacy risks, such as potential tracking and exposure of sensitive information like drivers' home addresses \cite{To-TMC2017}. Consequently, ensuring the location privacy of vehicles in LBS applications is crucial. \looseness = -1

The issue of location privacy in LBS has gained considerable attention over the past two decades. Many recent studies like \cite{Yu-NDSS2017, Andres-CCS2013, Bordenabe-CCS2014} have focused on geo-obfuscation, a \emph{location privacy protection mechanism (LPPM)} that allows users to report obfuscated locations instead of exact coordinates to servers. 
Notably, Andr{\'e}s et al. \cite{Andres-CCS2013} introduced a formal privacy criterion for geo-obfuscation, called \emph{geo-indistinguishability (Geo-Ind)}, which is extended from \emph{Differential Privacy (DP)}, requiring that nearby real locations remain indistinguishable based on their obfuscated representations.


Albeit effective in protecting sporadic location privacy, Geo-Ind is also known as a \emph{context-free} privacy criterion \cite{GAP2018}, without considering the impact of contextual information on mobile users' obfuscated locations. Such an assumption limits the applications of Geo-Ind in many practical scenarios, where mobile users' mobility is highly impacted by the surrounding environments. 
Recent endeavors \cite{LIAO2007311,Xu-WWW2017,Cao-ICDE2017,Emrich-ICDE2012,Li-Sigspatial2008,Arain-MTA2018, Cao-ICDE2019,Li-SigSpatial2017} have delved into exploring the vulnerabilities of Geo-Ind by taking into account the spatiotemporal correlation of users' reported locations. As a countermeasure, some other data privacy works \cite{Cao-ICDE2017, Cao-ICDE2019, Ghinita-Sigspatial2009, Xiao-SIGSAC2015} focus on devising new context-aware privacy criteria and solutions to protect users' location data. 

While elegant, those works mainly rely on explicit stochastic models such as Markov chains \cite{Qiu-SIGSPATIAL2022}, but they tend to overlook the implicit long-term correlation between locations that may be deeply embedded within the contextual data. In fact, the availability of vehicle traffic flow information is on the rise globally, especially within the urban and suburban contexts, sourced from road sensors and traffic cameras \cite{Kashinath-IEEEAccess2021, Ide-TITS2017}, mobile applications \cite{Liu-JNCA2019}, and official government or municipal websites \cite{Qiu-SIGSPATIAL2022}. This rich contextual data presents an opportunity for attackers to learn vehicles' implicit mobility patterns over long-term periods. Leveraging this knowledge, attackers can potentially refine the accuracy of location inference attacks, even when the vehciles' locations have been obfuscated. 

\vspace{0.05in}
\noindent \textbf{Our Contributions}. To fill the aforementioned research gap, in this paper, we aim to study new \textbf{context-aware threat model} to track vehicle locations and develop the \textbf{countermeasure}. Particularly, we focus on the scenario where the vehicle traffic flow information is available. By leveraging the recent fast advancement of deep neural networks, we aim to delve into the implicit relationships - both short-term and long-term - embedded in the vehicles' location data using the traffic flow information, without depending on explicit stochastic models. \looseness = -1


\vspace{-0.08in}
\subsubsection*{\textbf{Contribution 1: New context-aware threat model ``VehiTrack''}} 
To demonstrate the vulnerability of Geo-Ind when protecting vehicles' location privacy, we introduce a new threat model called \emph{VehiTrack}. VehiTrack seeks to recover the actual locations of a vehicle from its obfuscated data during a journey. It operates in two phases: In \emph{Phase 1}, VehiTrack applies Bayes' formula, considering the vehicle's mobility constraints over time on the road network, to estimate the real locations from obfuscated ones. In \emph{Phase 2}, it uses \emph{Long Short-Term Memory (LSTM)} neural networks, which are effective at recognizing both short-term and long-term correlation in sequence data, to refine these estimates using the vehicle traffic dataset. Our results show that VehiTrack significantly reduces the inference error by 61.51\% and 48.15\% compared to traditional Bayesian attacks when using Laplacian noise and linear programming (LP)-based geo-obfuscation methods, respectively.

\begin{figure}[t]
\centering
\begin{minipage}{0.48\textwidth}
\centering
  \subfigure{
\includegraphics[width=1.00\textwidth]{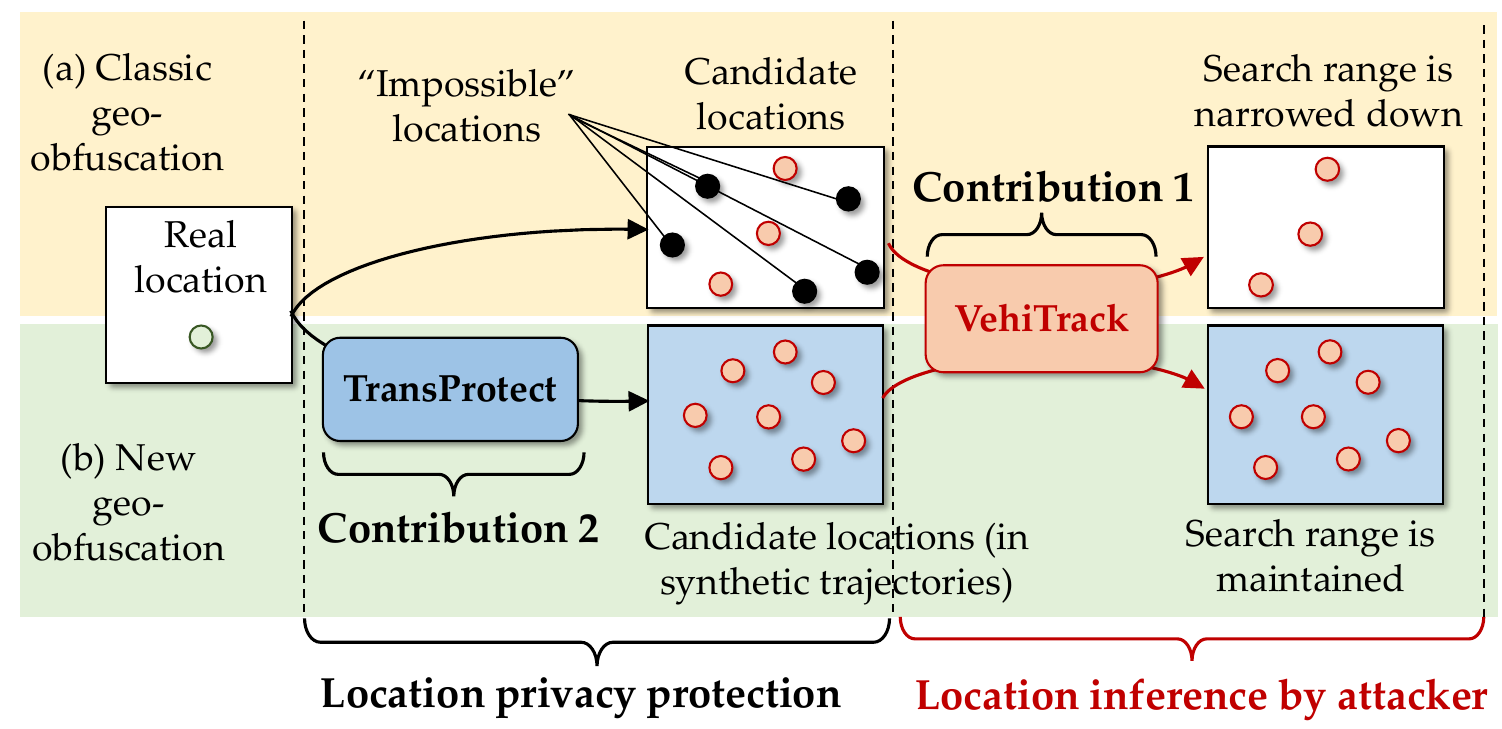}}
\vspace{-0.05in}
\caption{Introduction of VehiTrack and TransProtect. }
\label{fig:introduction}
\end{minipage}
\vspace{-0.08in}
\end{figure}

Our findings also reveal that the vulnerability of Geo-Ind to context-aware inference attacks largely stems from its failure to account for the constraints of road networks and traffic conditions on vehicle mobility. As depicted in Fig. \ref{fig:introduction}(a), many locations within the obfuscation range, though compliant with Geo-Ind, are deemed ``impossible'' when contextual data is considered. This allows the VehiTrack model to significantly narrow its search for actual locations-on average, 81.69\% of locations can be dismissed by incorporating road and traffic constraints. Motivated by this observation, we developed a new geo-obfuscation method called \emph{TransProtect}. This approach ensures that the chosen obfuscated locations conform to realistic vehicle movement patterns, making it challenging for VehiTrack to effectively reduce the search range using contextual information.

\vspace{-0.00in}

\subsubsection*{\textbf{Contribution 2: The countermeasure ``TransProtect''}} TransProtect is designed to create synthetic trajectories to closely emulate the vehicle's actual mobility. As Fig. \ref{fig:introduction}(b) shows, TransProtect can be integrated into the current geo-obfuscation framework to confine the obfuscation range to a set of locations within the generated synthetic trajectories, called ``candidate locations''. Consequently, the obfuscated location chosen from candidate locations has to adhere to the realistic vehicles' mobility patterns, which effectively prevents attackers from excluding any ``impossible locations'' using context-aware inference approaches like VehiTrack. \looseness = -1

To create such synthetic trajectories, TransProtect leverages a \emph{transformer} model, a deep neural network that learns context by handling long-range dependencies of the sequence data and has achieved great success in many artificial intelligence domains \cite{lin2021survey, Chen-ICDE2023}. To process the location sequence data of vehicles, we let TransProtect first conduct \emph{location embedding} to map the node (location) set of the road network to a lower dimensional vector space, preserving spatial features of road network locations. To achieve this objective, TransProtect applies \emph{Node2Vec} \cite{grover2016node2vec} to maximize the log probability of observing a network neighborhood for each node conditioned on its feature representation. It also utilizes a Graph Convolutional Network (GCN) \cite{kipf2017semisupervised} to integrate edge weights and neighborhood data into these embeddings. 

TransProtect uses a \emph{Transformer encoder} to evaluate each location's likelihood of being the vehicle's real position by capturing spatial patterns and adjusting scores based on data utility loss, then selects the top locations for obfuscation.



\vspace{-0.05in}
\subsubsection*{\textbf{Contribution 3: Empirical validation based on real-world dataset}} To evaluate the performance of both VehiTrack and \newline TransProtect, we conducted an extensive simulation using the Rome taxi trajectory datasets \cite{roma-taxi-20140717}, comparing both methods against state-of-the-art location inference and location privacy protection algorithms. The experimental results show that (1) When provided with vehicle locations obfuscated using Laplacian noise \cite{Andres-CCS2013} and LP, VehiTrack demonstrates remarkable accuracy in tracking vehicle locations. On average, its expected inference error (EIE), which reflects the location privacy level, is 66.58\% and 51.17\% lower for Laplacian noise and LP (averaged on all epsilon values), respectively, compared to classic Bayesian inference attacks \cite{Yu-NDSS2017}. (2) Our proposed countermeasure, TransProtect, can effectively protect the location privacy of vehicles against VehiTrack. On average, the synthetic location set generated by TransProtect increases EIE by 40.26\%
when using Laplacian noise and LP for obfuscation \cite{Qiu-TMC2020, Shokri-CCS2012}. \looseness = -1
\DEL{
Simply put, we summarize our contributions as follows:
\begin{itemize}
\item [1)] We develop a new threat model, called \textbf{VehiTrack}, which can accurately recover vehicles' real locations from obfuscated locations using the context information even if Geo-Ind has been satisfied.
\item [2)] As a countermeasure of VehiTrack, we develop a new LPPM, called \textbf{TransProtect}, to restrict the obfuscation range to a set of candidate locations that are hard to distinguish from real locations (by VehiTrack).
\item [3)] We carry out an extensive simulation based on a real-world vehicle trajectory dataset to test the performance of both VehiTrack and TransProtect. The experimental results demonstrate (1) the vulnerability of Geo-Ind to VehiTrack and (2) the effectiveness of TransProtect in protecting vehicles' location privacy against VehiTrack. 
\end{itemize}}

The rest of the paper is organized as follows: Section \ref{sec:background} gives the preliminaries of Geo-Ind. Section \ref{sec:threat} introduces the new threat model VehiTrack and Section \ref{sec:countermeasure} describes the countermeasure TransProtect. Section \ref{sec:exp} evaluates the performance of both VehiTrack and TransProtect. Section \ref{sec:related} presents the related work. Finally,  Section \ref{sec:conclude} makes a conclusion. 
\vspace{-0.00in}
\section{Preliminary}
\label{sec:background}
\vspace{-0.00in}

In this Section, we introduce the preliminary knowledge of geo-obfuscation (Section \ref{subsec:geoobfuscation}), its privacy criterion Geo-Ind (Section \ref{subsec:geoind}), 
and the limitation of Geo-Ind (Section \ref{subsec:limits}).  
\vspace{-0.10in}
\subsection{Geo-Obfuscation in LBS}
\label{subsec:geoobfuscation}
\vspace{-0.00in}
To ensure the quality of LBS, the server needs to collect the participating vehicles' location information in real-time. Like \cite{Wang-WWW2017}, we consider the scenario where \emph{the server is non-malicious but vulnerable to potential data breaches}. In such scenarios, unauthorized parties might gain access to the reported vehicle locations stored on the server. Accordingly, the precise locations of the vehicles should be kept hidden from the server. \looseness = -1

In geo-obfuscation \cite{Andres-CCS2013, Bordenabe-CCS2014}, privacy-conscious vehicles are allowed to use an \emph{obfuscation function} to perturb their actual locations before reporting the locations to the server. The obfuscation function takes the vehicle's precise location as the input and returns a probability distribution of the obfuscated location, based on which the vehicle can randomly select an obfuscated location to report. Beyond concealing the precise locations of vehicles, the obfuscated locations should be chosen in a manner that maintains the estimated travel cost to the destination reasonably close to the actual travel cost of the vehicle.  Assessing the distortion of travel costs resulting from obfuscated locations requires access to global information about the target region, including its real-time vehicle traffic conditions and the distribution of spatial tasks. However, managing this data on an individual basis by vehicles poses significant challenges. As such, previous studies \cite{Wang-WWW2017, Bordenabe-CCS2014, Qiu-TMC2020, Yu-NDSS2017} have mainly concentrated on the server-side computation of the obfuscation function. \looseness = -1

For the sake of computational efficiency, many geo-obfuscation methods \cite{Andres-CCS2013, Bordenabe-CCS2014} consider users' mobility on a set of discrete locations. When considering vehicle LBS, the existing works \cite{Qiu-TMC2020, Qiu-CIKM2020} discretize the road network into a set of road connections, denoted as $\mathcal{V} = \left\{v_1, ..., v_L\right\}$. Those connections include road intersections, forks, junctions where roads intersect with others, and points where the road changes direction. All the other locations within the road network are approximated to their nearest connections in $\mathcal{V}$. By discretizing the location field into the finite location set $\mathcal{V}$, the obfuscation function $Q$ can be described as an \emph{obfuscation matrix} $\mathbf{Z} = \{z_{i,k}\}_{(v_i,v_k) \in \mathcal{V}^2}$, where each $z_{i,k}$ denotes the probability of selecting $v_k$ as the obfuscated location given the actual location $v_i$ ($v_i, v_k \in \mathcal{V}$). \looseness = -1
\vspace{-0.05in}
\subsection{Geo-Indistinguishability}
\label{subsec:geoind}
\vspace{-0.00in}

Although the server takes charge of generating the obfuscation function, the vehicles' exact locations are still hidden from the server since the obfuscated locations are selected in a probabilistic manner \cite{Wang-WWW2017}. Specifically, the obfuscation function is designed to satisfy Geo-Ind \cite{Andres-CCS2013}, which requires that even if an attacker has obtained a vehicle's reported (obfuscated) location and the obfuscation function from the server, it is still hard for the attacker to distinguish the vehicle's real location $v_i$ from any nearby location $v_j$. \emph{Geo-Ind}  is formally defined in \emph{Definition \ref{def:GeoI}}: 
\vspace{-0.04in}
\begin{definition} 
\label{def:GeoI}
(Geo-Ind) An obfuscation matrix $\mathbf{Z}$ satisfies $\epsilon$-Geo-Ind if, for each pair of neighboring locations $v_i, v_j \in \mathcal{V}$ with $d_{i, j} \leq \gamma$, the following constraints are satisfied 
\begin{equation}
\label{eq:Geo-Ind-LP}
z_{i,k} - e^{\epsilon d_{i, j}}z_{j,k} \leq 0, ~\forall v_i, v_j, v_k \in \mathcal{V}~\mbox{with $d_{i, j} \leq \gamma$.}
\vspace{-0.00in}
\end{equation}
which means that the probability distributions of the obfuscated locations of $v_i$ and $v_j$ are sufficiently close. Here, $d_{i, j}$ denotes the \emph{Haversine distance} (the angular distance on the surface of a sphere) between $v_i$ and $v_j$, and $\gamma >0$ is a predetermined distance threshold. 
\end{definition}


\DEL{
\begin{figure}[t]
\centering
\begin{minipage}{0.42\textwidth}
\centering
  \subfigure{
\includegraphics[width=0.70\textwidth]{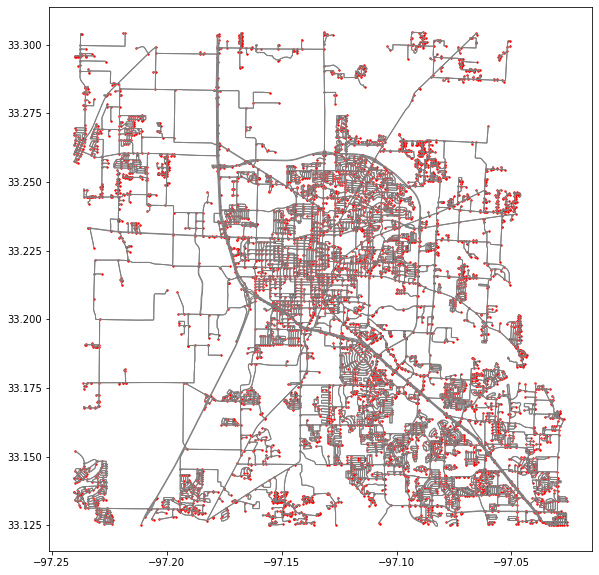}}
\vspace{-0.00in}
\caption{Road map of the Denton county, Texas (connections are marked by red dots). }
\label{fig:dentonmap}
\end{minipage}
\vspace{-0.00in}
\end{figure}}

\DEL{
\begin{wrapfigure}{r}{0.24\textwidth}
\vspace{-0.1in}
\begin{minipage}{0.24\textwidth}
\centering
    \subfigure{
\includegraphics[width=1.00\textwidth]{./fig/Denton_Nodal_Map(10000m diameter).png}}
\vspace{-0.15in}
\caption{Road map of Denton county, Texas (connections are marked by red dots).}
\vspace{-0.00in}
\label{fig:dentonmap}
\end{minipage}
\end{wrapfigure}}
\noindent \textbf{LP formulation}. Many recent works \cite{Wang-WWW2017, Qiu-TMC2020} address the quality issue caused by geo-obfuscation using \emph{linear programming (LP)}, of which the objective is to minimize the data quality loss while ensuring Geo-Ind is maintained. 
The LP is then formulated to optimize the values of $\mathbf{Z}$, which comprises $K^2$ decision variables (entries). 
Besides satisfying Geo-Ind in Equ. (\ref{eq:Geo-Ind-LP}), for each real location $v_i$, the sum probability of  the obfuscated locations should be 1 (probability unit measure), i.e., 
\vspace{-0.00in}
\begin{equation}
\label{eq:unitmeasure}
\sum_{k=1}^K z_{i,k} = 1,~\forall v_i \in \mathcal{V}.
\vspace{-0.00in}
\end{equation}
We let $\mathrm{UL}(\mathbf{Z})$ denote the \emph{utility loss} caused by the obfuscation matrix $\mathbf{Z}$, where $\mathrm{UL}(\mathbf{Z})$ is assumed to be a linear function of $\mathbf{Z}$ \cite{Wang-WWW2017, Qiu-TMC2020}. Finally, the LP is formulated to minimize $\mathrm{UL}\left(\mathbf{Z}\right)$ while satisfying the constraints of the probability unit measure (Equ. (\ref{eq:unitmeasure})) and Geo-Ind (Equ. (\ref{eq:Geo-Ind-LP})):
\vspace{-0.05in}
\begin{eqnarray}
\label{eq:OMGobj}
\min~ \mathrm{UL}\left(\mathbf{Z}\right)~ 
\mbox{s.t. } ~ \mbox{ Equ. (\ref{eq:Geo-Ind-LP})(\ref{eq:unitmeasure})  are  satisfied}. 
\vspace{-0.00in}
\end{eqnarray}

\vspace{-0.05in}

\DEL{
\vspace{0.00in}
\subsection{Geo-obfuscation Calculation} 
\label{subsec:calgeoobfuscation}

Many recent works \cite{Wang-WWW2017, Qiu-TMC2020} address the quality issue caused by geo-obfuscation using \emph{linear programming (LP)}, of which the objective is to minimize the data quality loss while ensuring Geo-Ind is maintained. For the sake of computational efficiency, LP-based geo-obfuscation methods typically consider users' mobility on a set of discrete locations. When considering vehicle LBS, the existing works \cite{Qiu-TMC2020, Qiu-CIKM2020} discretize the road network into a set of road connections, denoted as $\mathcal{V}$. Those connections include road intersections, forks, junctions where roads intersect with others, and points where the road changes direction. All the other locations within the road network are approximated to their nearest connections in $\mathcal{V}$. 

By discretizing the location field into the finite location set $\mathcal{V}$, the obfuscation function $Q$ can be described as an \emph{obfuscation matrix} $\mathbf{Z} = \{z_{i,k}\}_{(v_i,v_k) \in \mathcal{V}^2}$, where each $z_{i,k}$ denotes the probability of selecting $v_k$ as the obfuscated location given the actual location $v_i$ ($v_i, v_k \in \mathcal{V}$). The LP is then formulated to optimize the values of $\mathbf{Z}$, which comprises $K^2$ decision variables (entries). As such, the Geo-Ind constraints in Definition \ref{def:GeoI} can be rewritten as
\begin{equation}
\label{eq:Geo-Ind-LP}
z_{i,k} - e^{\epsilon d_{i, j}}z_{j,k} \leq 0, ~\forall v_i, v_j, v_k \in \mathcal{V}~\mbox{with $d_{i, j} \leq \gamma$.}
\vspace{-0.00in}
\end{equation}
Besides satisfying Geo-Ind, for each real location $v_i$, the sum probability of  the obfuscated locations should be 1 (probability unit measure), i.e., 
\vspace{-0.00in}
\begin{equation}
\label{eq:unitmeasure}
\textstyle    
\sum_{k=1}^K z_{i,k} = 1,~\forall v_i \in \mathcal{V}.
\vspace{-0.00in}
\end{equation}
We let $\mathrm{UL}(\mathbf{Z})$ denote the \emph{utility loss} caused by the obfuscation matrix $\mathbf{Z}$, where $\mathrm{UL}(\mathbf{Z})$ is assumed to be a linear function of $\mathbf{Z}$ \cite{Wang-WWW2017, Qiu-TMC2020}. Finally, the LP is formulated to minimize $\mathrm{UL}\left(\mathbf{Z}\right)$ while satisfying the constraints of the probability unit measure (Equ. (\ref{eq:unitmeasure})) and Geo-Ind (Equ. (\ref{eq:Geo-Ind-LP})):
\vspace{-0.0in}
\begin{eqnarray}
\label{eq:OMGobj}
\min~ \mathrm{UL}\left(\mathbf{Z}\right)~ 
\mbox{s.t. } ~ \mbox{ Equ. (\ref{eq:Geo-Ind-LP})(\ref{eq:unitmeasure})  are  satisfied}. 
\vspace{-0.00in}
\end{eqnarray}
}

\vspace{-0.10in}
\subsection{Limitations of Geo-Ind}
\label{subsec:limits}
\vspace{-0.02in}


\begin{figure}[t]
\centering
\begin{minipage}{0.38\textwidth}
\centering
  \subfigure{
\includegraphics[width=1.00\textwidth]{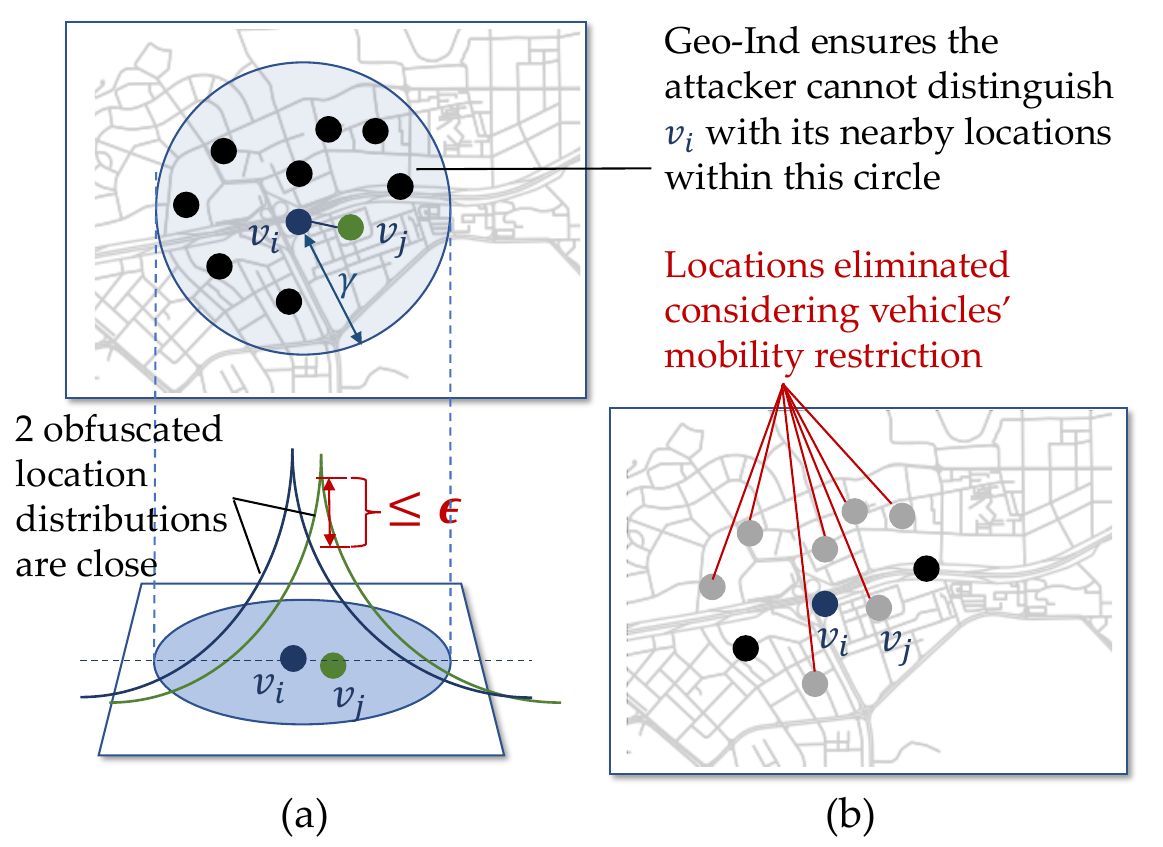}}
\label{}
\end{minipage}
\vspace{-0.15in}
\caption{Geo-indistinguishability and its limit. }
\label{fig:GeoIndLimit}
\vspace{-0.15in}
\end{figure}



\vspace{0.00in}
As Fig. \ref{fig:GeoIndLimit}(a) shows, Geo-Ind aims to guarantee that a vehicle's location $v_i$ remains indistinguishable from any other location $v_j$ within the circle centered at $v_i$ with a radius of $\gamma$ based on their obfuscated location distributions. However, Geo-Ind is a context-free privacy criterion without considering the context information that can be used in inference attacks. As Fig. \ref{fig:GeoIndLimit}(b) shows, an attacker can leverage context information, such as the vehicle's historical locations, speed limits, and surrounding traffic conditions, to eliminate ``impossible'' locations of the vehicle within the circle. This, in turn, narrows down the search range for the vehicle's actual location and increases the accuracy of its location tracking. 

In the next section, we will introduce a new threat model to demonstrate the vulnerability of Geo-Ind when protecting vehicles' location privacy. 

\vspace{-0.05in}
\section{VehiTrack: A Context-Aware Location Inference Algorithm}
\label{sec:threat}
\vspace{-0.00in}
In this section, we introduce a new location inference algorithm, called \emph{VehiTrack}, to accurately recover the real locations of a target vehicle from its obfuscated locations even though Geo-Ind has been satisfied. \looseness = -1

We consider a scenario where a target vehicle reports its location multiple times at a sequence of time slots $t_1, ..., t_N$, where the actual locations and the obfuscated locations of the vehicle are denoted by $\mathbf{x}_{{1}: {N}} = \{{x}_{{1}}, ...,  {x}_{{N}}\}$ and $\tilde{\mathbf{y}}_{{1}: {N}} = \{\tilde{y}_{{1}}, ...,  \tilde{y}_{{N}}\}$, respectively ($x_n, \tilde{y}_n \in \mathcal{V}$, for each $t_n = t_1,...,t_N$). Given the observation of the vehicle's obfuscated locations $\tilde{\mathbf{y}}_{{1}: {N}}$, VehiTrack aims to find the vehicle's actual location sequence $\mathbf{x}_{{1}: {N}}$. To achieve this goal, VehiTrack consists of the following two main phases: \looseness = -1

\begin{itemize}
\item [\textbf{Ph1:}] VehiTrack estimates the posterior $p(x_{n}| \tilde{y}_{n})$ of the vehicle's location at each time slot $t_n$, by considering the \textbf{short-term correlation} of the vehicle's locations using a \emph{mobility \newline restriction-aware Bayesian inference model} (Section \ref{subsec:BeyesianInference}). 
\item [\textbf{Ph2:}] VehiTrack improves the accuracy of the posterior sequence $p(x_{1}| \tilde{y}_{1}), ..., p(x_{N}| \tilde{y}_{N})$, by considering the \textbf{long-term correlation} of the vehicle's locations using \emph{Long Short-Term Memory (LSTM)} neural networks (Section \ref{subsec:models}). 
\end{itemize}
\vspace{-0.05in}
Before introducing the details of the above two phases,  in Section \ref{subsec:models}, we first describe the mathematical models used in VehiTrack, including the main notations and assumptions. 
\vspace{-0.10in}
\subsection{Models}
\label{subsec:models}
\vspace{-0.00in}
\subsubsection{Threat model} 
To estimate the target vehicle's true locations $\mathbf{x}_{{1}: {N}} = \{{x}_{{1}}, ...,  {x}_{{N}}\}$, we assume that the attacker has access to the following information at the time slots ${t_{1}, ..., t_{N}}$:  
\newline (1) the vehicle's obfuscated locations $\tilde{\mathbf{y}}_{{1}: {N}} = \{\tilde{y}_{{1}}, ...,  \tilde{y}_{{N}}\}$; 
\newline (2) the obfuscation matrices $\mathbf{Z}_{{1}: {N}} = \{\mathbf{Z}_{{1}}, ..., \mathbf{Z}_{{N}}\}$, where $\mathbf{Z}_{n}$ denotes the obfuscation matrix at time slot $t_n$;
\newline (3) the background information including the vehicle's mobility restrictions in the road networks (e.g., speed limits). 
We assume that the attacker has access to the public vehicle trajectory dataset \cite{roma-taxi-20140717} to obtain historical traffic flow information. 

For simplicity, we use $p(x_n)$ to represent the prior probability that the vehicle is located at $x_{n}$ at time $t_n$ and use $p(\mathbf{x}_{{1}: {n}})$ to represent the prior joint distribution of the vehicle being located at $\mathbf{x}_{{1}: {n}}$ in the time slots $\{t_{1}, ..., t_{n}\}$. 

\vspace{-0.05in}
\subsubsection{Vehicle's mobility model} We describe vehicles' mobility in the road network as a directed graph $\mathcal{G} = \left(\mathcal{V}, \mathcal{E}\right)$, where $\mathcal{V}$ and $\mathcal{E}$ denote the \emph{node (location) set} and the \emph{edge set}, respectively. Each edge $e_{i,j} \in \mathcal{E}$ represents that $v_i$ is adjacent to $v_j$ in the road network, meaning that a vehicle can travel from $v_i$ to $v_j$ without visiting any other location in $\mathcal{V}$. Each edge $e_{i,j} \in \mathcal{E}$ is assigned a weight $w_{i,j}$, representing vehicles' minimum travel time through the edge $e_{i,j}$. The shortest travel time from $v_i$ to $v_j$ (which are unnecessarily adjacent), denoted by $c_{v_i,v_j}$, equals the length of the shortest path from $v_i$ to $v_j$ in $\mathcal{G}$. Here, the \emph{length} of a path is defined as the sum weight of all the edges along the path. \looseness = -1


Note that due to the change of traffic conditions (e.g., peak hours versus off-peak hours on weekdays), the edge weight $w_{i,j}$ can vary over time, rendering the mobility graph $\mathcal{G}$ a \emph{time-varying graph}. Given $\mathcal{G}$, we call a location $v_j$ is \emph{reachable} by $v_i$ during a time interval $[t_{n-1}, t_{n}]$ if the shortest travel time from $v_i$ to $v_j$ during $[t_{n-1}, t_{n}]$ is no larger than $t_{n} - t_{n-1}$, i.e., $c_{v_i,v_j} \leq t_{n} - t_{n-1}$. We use $\mathcal{R}^i_{{n}}$ to denote the set of locations reachable by $v_i$ (or called the \emph{reachable set} of $v_i$) during $[t_{n-1}, t_{n}]$.

\DEL{
\begin{definition} 
(GPS-based Trajectory) A GPS based trajectory $\tau$ is a sequence that consists of $N$ consecutive road segments, denoted as $\tau = v_{{k_1}: {k_N}} = \{v_{k_1},v_{k_2},...,v_{k_N}\}$ \chx{Notations need to be discussed??}
\end{definition}
}
\vspace{-0.10in}
\subsection{Phase 1: Mobility Restriction-Aware Bayesian Inference}
\label{subsec:BeyesianInference}
\vspace{-0.02in}
By leveraging the vehicles' mobility restrictions and the obfuscation matrices, VehiTrack first estimates the posteriors of the target vehicle's locations at the time slots $\left\{{t_{1}, ..., t_{N}}\right\}$ via a \emph{Bayesian inference model}. Note that deriving a posterior over the entire location set $\mathcal{V}$ imposes a substantial computational burden. Indeed, due to the restrictions of the vehicle's mobility and its limited obfuscation range, the possible true location of the vehicle can be confined to a smaller area. As a result, VehiTrack only needs to compute the posteriors of the locations within this reduced area and treat the posteriors of the locations outside this area as negligible.

\begin{figure}[t]
\centering
\begin{minipage}{0.40\textwidth}
\centering
  \subfigure{
\includegraphics[width=1.00\textwidth]{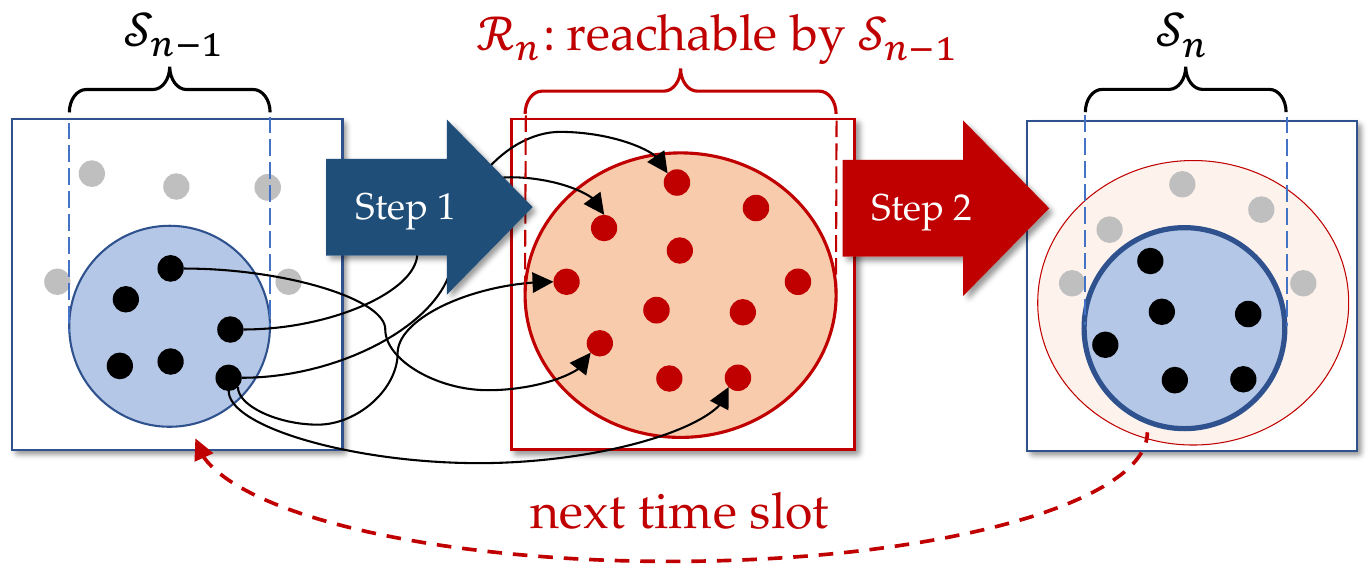}}
\label{}
\end{minipage}
\vspace{-0.20in}
\caption{Framework of deriving  $\mathcal{S}_{n}$ from $\mathcal{S}_{n-1}$. 
\vspace{0.05in}
\newline \small Step 1: Find the location set $\mathcal{R}_{n}$ that are reachable by $\mathcal{S}_{n-1}$; 
\newline \small Step 2: Derive the posterior of each location in $\mathcal{R}_{n}$ and find the ones of which the posteriors are higher than the threshold $\xi$. }
\label{fig:BayesFramework}
\vspace{-0.10in}
\end{figure}

We use $\mathcal{S}_{n}$ to represent the set of the vehicle's possible locations (identified by VehiTrack) at the time slot $t_n$. Fig. \ref{fig:BayesFramework} shows the framework of how VehiTrack iteratively derives $\mathcal{S}_{n}$ from $\mathcal{S}_{n-1}$, which is composed of the following two steps: 

\subsubsection{Step 1 - Identify the location set $\mathcal{R}_{n}$ that is reachable by the locations in $\mathcal{S}_{n-1}$ during the time interval $[t_{n-1}, t_n]$} Here, $\mathcal{R}_{n}$ is the union of the reachable sets of all the locations in $\mathcal{S}_{n-1}$, i.e., $\mathcal{R}_{n} = \cup_{v_i \in \mathcal{S}_{n-1}}\mathcal{R}^i_{{n}}$, i.e., each location in $\mathcal{R}_{n}$ is \emph{reachable} by at least one location in $\mathcal{S}_{n-1}$. To determine $\mathcal{R}^i_{{n}}$ for each $v_i$, VehiTrack builds a \emph{shortest path tree} $SPT_i$ in the graph $\mathcal{G}$ rooted at $v_i$ using the \emph{Dijkstra's algorithm} \cite{Algorithm}, of which the time complexity is $O(|\mathcal{V}'_i|^2)$. Here, $\mathcal{V}'_i \subset \mathcal{V}$ represents the set of nodes included in the $SPT_i$. 
For the sake of computation efficiency, VehiTrack limits $\mathcal{V}'_i$ to the location set of which the Haversine distance is no larger than $(t_n - t_{n-1})s_{\mathrm{limit}}$, i.e., which are reachable by the vehicle with its maximum speed $s_{\mathrm{limit}}$ during $[t_{n-1}, t_n]$ without considering the mobility restriction imposed by the road network, i.e.,  
\vspace{-0.03in}
\begin{equation}
\label{eq:V_i}
\mathcal{V}'_i = \left\{v_j \in \mathcal{V} \left|d_{i,j} \leq (t_n - t_{n-1})s_{\mathrm{limit}}\right. \right\}. 
\vspace{-0.03in}
\end{equation}
VehiTrack first creates an \emph{induced subgraph} $\mathcal{G}'_i$ of $\mathcal{G}$ formed from the node set $\mathcal{V}'_i$, where all of the edges (from $\mathcal{G}$) connect pairs of vertices in $\mathcal{V}'_i$. We then build $SPT_i$ on $\mathcal{G}'_i$ instead of the original graph $\mathcal{G}$.
\vspace{0.00in}
\vspace{-0.05in}
\begin{proposition}
\label{prop:SPT_i}
$SPT_i$ is \emph{sufficient} to identisfy $\mathcal{R}^i_{n}$.
\end{proposition}
\vspace{-0.05in}
\textbf{Proof Sketch}: 
We prove that for $\forall v_k \in \mathcal{R}^i_{n}$, if $c_{v_i, v_k} \leq t_n - t_{n-1}$, then $v_k$ is included in $SPT_i$, and also its distance to $v_i$ is equal to $c_{v_i, v_k}$ in $SPT_i$. We prove it by contradiction, where the detailed proof can be found in Section \ref{sec:proof_SPT_i} in Appendix.

\vspace{0.05in}

\vspace{-0.1in}
\subsubsection{Step 2 - Determine the possible location set $\mathcal{S}_{n}$ using the obfuscation matrices}
Given the observed (obfuscated) location $\tilde{y}_n$ and the obfuscation matrix $\mathbf{Z}_n$ at each time slot $t_n$, VehiTrack derives the posterior probabilities of all the locations $x \in \mathcal{R}_{n}$ using the Bayes' formula:
\vspace{-0.05in}
\begin{equation}
p\left(x|\tilde{y}_{{n}}\right) = \frac{ p(x)z_{x, \tilde{y}_n} }{\sum_{x' \in \mathcal{R}_{n}} p(x') z_{x', \tilde{y}_n}}, ~ \forall x  \in \mathcal{R}_{n}. 
\end{equation}
Here, we consider $x$ as a ``possible location'' of the vehicle in $\mathcal{S}_{{n}}$ only if its posterior value $p\left(x|\tilde{\mathbf{y}}_{{1}: {n}}\right)$ is higher than a pre-determined threshold $\xi > 0$. Therefore, $\mathcal{S}_{{n}}$ is given by
\vspace{-0.00in}
\begin{equation}
\mathcal{S}_{{n}} = \left\{x \in \mathcal{R}_{n} \left|p\left(x|\tilde{y}_{{n}}\right) \geq \xi \right.\right\}. 
\vspace{-0.00in}
\end{equation}


\subsection{Phase 2: Posterior Refinement using LSTM}
\label{subsec:PosteriorRefine}

\begin{figure}[t]
\centering
\begin{minipage}{0.49\textwidth}
\centering
  \subfigure{
\includegraphics[width=1.00\textwidth]{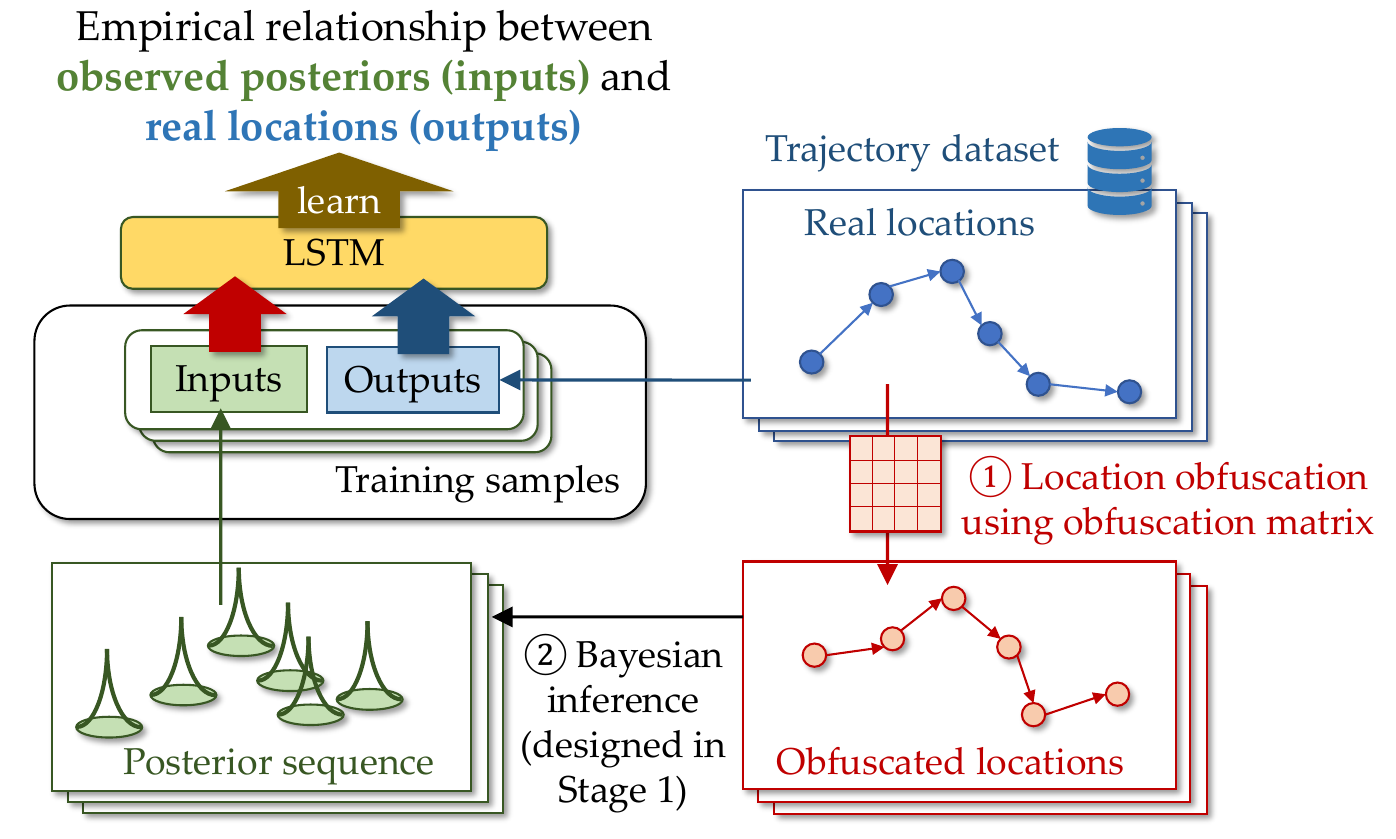}}
\label{}
\end{minipage}
\vspace{-0.15in}
\caption{Training data generation in Phase 2. 
}
\label{fig:Phase2trainingFramework}
\vspace{-0.05in}
\end{figure}

Although VehiTrack considers the mobility restrictions of the target vehicle in adjacent time slots in Phase 1, it falls short of capturing the long-term correlation of the vehicle's locations. In Phase 2, VehiTrack aims to further improve the accuracy of the posterior estimation by incorporating the LSTM networks, due to their strong capabilities of learning both short-term and long-term correlation 
in sequence data \cite{Zhao-ITS2017}. 
Specifically, VehiTrack takes the estimated posterior sequence obtained in Phase 1 as the \textbf{inputs} of the LSTM models and infers the real location sequence as the \textbf{outputs}. Achieving this goal entails training the LSTM model to establish the \emph{empirical relationship between the observed posteriors calculated by Phase 1 and the vehicle's real locations}. \looseness = -1

\subsubsection{Training dataset generation} 
Following the threat model outlined in Section 1, we assume that the attacker has access to the historical vehicle mobility dataset in the region \cite{Yan-Ubicomp2018}. Moreover, we assume that the target vehicle follows similar mobility patterns with other vehicles in the dataset, despite potential individual variations. This allows VehiTrack to infer the target vehicle's locations by LSTM trained by the historical vehicle mobility data. 

VehiTrack generates training samples for an LSTM model by obfuscating real locations and using Bayesian inference to calculate location posteriors.

As Fig.  \ref{fig:Phase2trainingFramework} shows, to obtain the training inputs (location posteriors), 
VehiTrack first obfuscates each real location in the trajectory using the obfuscation matrix (\textbf{Step \textcircled{1}}) and then derives the corresponding posteriors based on the obfuscated locations using the Bayesian inference model in Phase 1 (\textbf{Step \textcircled{2}}). 
The model is trained with one-hot encoded real locations as outputs, and multiple samples are created for each trajectory to reduce variance from the obfuscation process.
To reduce the sample variance stemming from the stochastic obfuscated location selection process, we let VehiTrack generate multiple training samples (e.g., 20 samples in our experiments) for each trajectory. \looseness = -1

\subsubsection{LSTM network architecture} 
\begin{figure}[t]
\centering
\begin{minipage}{0.46\textwidth}
\centering
  \subfigure{
\includegraphics[width=1.00\textwidth]{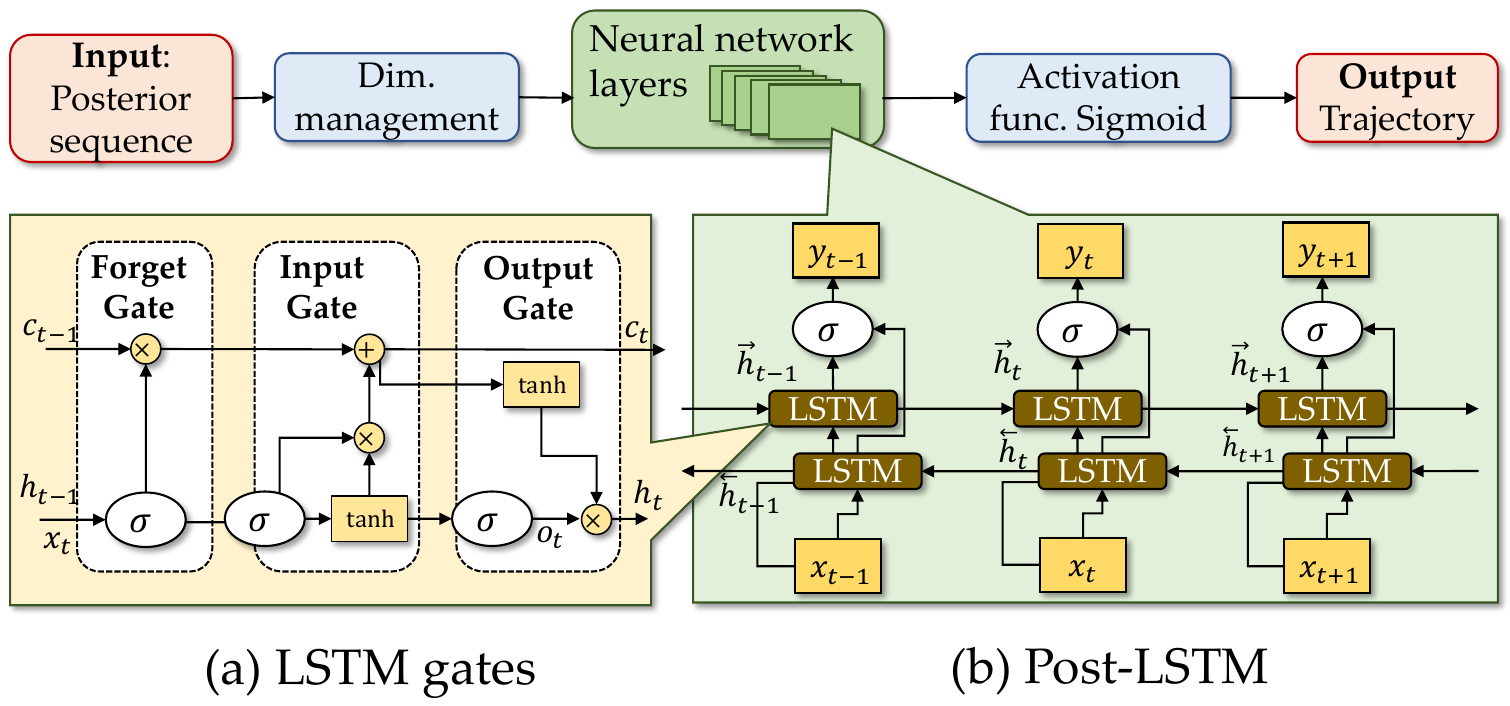}}
\label{}
\end{minipage}
\vspace{-0.1in}
\caption{Framework of Phase 2. 
}
\label{fig:LSTMframework}
\vspace{-0.10in}
\end{figure}

Fig. \ref{fig:LSTMframework} shows the framework of LSTM. The input posterior sequences undergo an initial processing step within the \emph{dimensionality management block}. This block employs a padding method to enforce a standardized input format, specifically by aligning all trajectories with the longest one in the dataset. The padded posterior vectors are then passed to the neural network layer, a combination of 5 LSTM layers. 

\DEL{
\vspace{0.03in}
\noindent \textbf{Posterior Gated Recurrent Unit (Post-GRU)}: Fig. \ref{fig:LSTMframework}(a) shows the architecture of each GRU layer, 
which comprises a candidate hidden state, carrying the information that is needed to be considered for the next hidden state, along with two main gates: \emph{reset} and \emph{update} gates. The reset gate determines how much of the previous hidden state should be reset at the current time slot in a trajectory. The update gate decides how much of the new information should be included in the current hidden state vector. Both the reset gate and the update gate operate upon the current time slot’s posterior vector and previously hidden state vector using the sigmoid activation function. 
}

Fig. \ref{fig:LSTMframework}(a) and \ref{fig:LSTMframework}(b) illustrates our LSTM architecture, which, unlike conventional LSTM, processes entire posterior vectors instead of scalar values. This allows element-wise operations within the LSTM cells. We also use a BiLSTM model with two parallel layers (forward and backward) \cite{huang2015bidirectional} to capture bidirectional patterns. The forget, input, and output gates dynamically manage the flow of information, deciding what to retain, add, or pass as the next hidden state.

\DEL{
\begin{wrapfigure}{r}{0.23\textwidth}
\vspace{-0.0in}
\begin{minipage}{0.23\textwidth}
\centering
    \subfigure{
\includegraphics[width=1.00\textwidth]{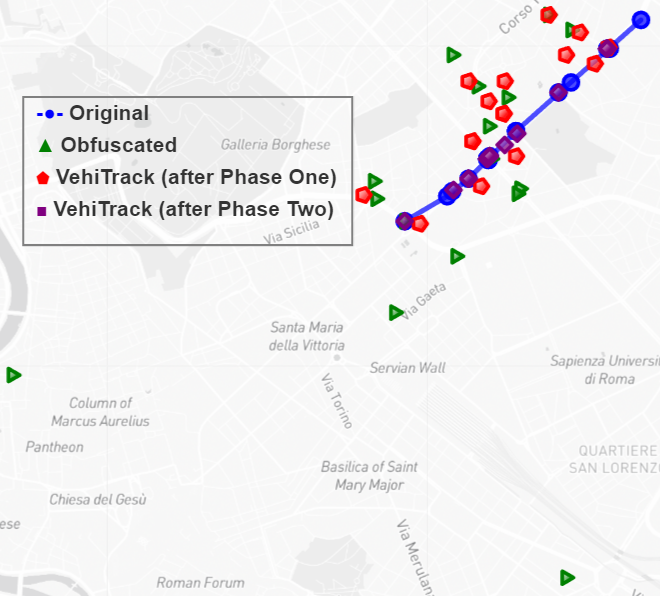}}
\vspace{-0.15in}
\caption{Example of location tracking using VehiTrack.}
\vspace{-0.05in}
\label{fig:VehiTrackExample}
\end{minipage}
\end{wrapfigure}}
To train Post-LSTM, we define the \emph{loss function} as the \emph{cross entropy} between predicted and actual vehicle location. The output of the neural network layer is directed to the sigmoid activation function block to constrain the output within the range $[0,1]$. The result is then passed to the output block where an $\arg\max$ operation is performed upon the output to get the final estimation of the trajectory.
\DEL{
After obtaining the refined posterior sequence of the trajectory $p(x_{1}| \tilde{y}_{1}), ..., p(x_{N}| \tilde{y}_{N})$, VehiTrack picks up the location that maximizes the current posterior as the estimated location of the vehicle, i.e., 
\vspace{-0.05in}
\begin{equation}
\textstyle \hat{x}_n = \arg \max_{x \in \mathcal{S}_{{n}}} p(x| \tilde{y}_{1}), ~ n = 1, ..., N.
\vspace{-0.00in}
\end{equation}
}
\DEL{
\subsection{Performance and Observation}
\begin{figure}[h]
\centering
\begin{minipage}{0.350\textwidth}
\centering
  \subfigure{
\includegraphics[width=1.00\textwidth]{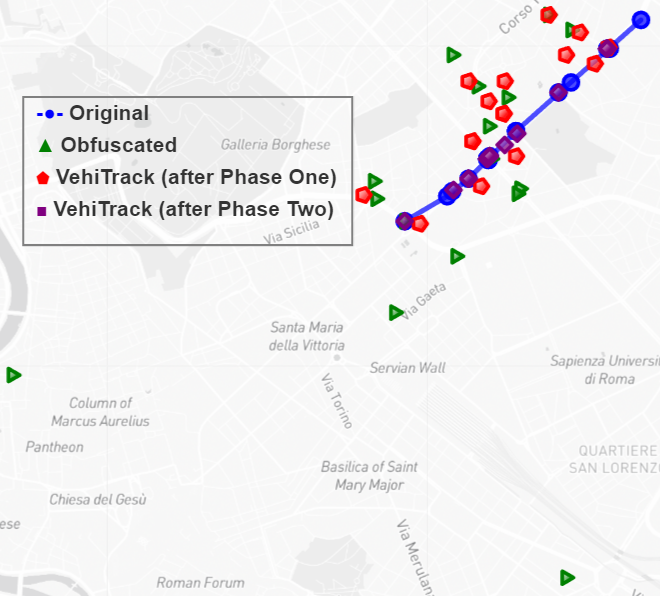}}
\end{minipage}
\vspace{-0.10in}
\caption{Example of location tracking using VehiTrack.}
\label{fig:VehiTrackExample}
\vspace{-0.00in}
\end{figure}}

\vspace{-0.00in}
\subsection{Performance of VehiTrack}

\DEL{Fig. \ref{fig:VehiTrackExample} gives an illustrative example of how accurately VehiTrack recovers a vehicle's actual locations, when the vehicle's location has been obfuscated using Laplacian noise \cite{Andres-CCS2013}. The location data of the vehicle is retrieved from the Rome and San Francisco taxicab dataset \cite{roma-taxi-20140717} \cite{sf-taxi}.} 

As demonstrated in our experiments detailed in Section \ref{subsec:expVehiTrack}, on average, using rome dataset (resp. using the San Francisco dataset), VehiTrack achieves a 65.54\% and 45.93\% (resp. 56.86\% and 48.78\%)
reduction in inference errors corresponding to the Laplacian and Linear Programming methods respectively, compared to the classic Bayesian inference algorithm. Our findings also reveal that, by incorporating contextual information such as the road network and traffic flow, VehiTrack can eliminate a significant percentage of locations within the obfuscation range. For instance, in our experiment in Section \ref{sec:exp}, using rome dataset (resp. using the San Francisco dataset) on average, 81.99\% (resp. 81.39) of locations within the obfuscation range are eliminated by considering vehicles' mobility restrictions. This factor contributes significantly to the high inference accuracy performed by VehiTrack. 



\DEL{
\sourabh{We can drop this below portion as it goes towards the encoder-decoder explanations which we are not following}

We have employed the encoder-decoder-like architecture for handling the pivotal sequence-to-sequence tasks so that sequential dependencies of data can be handled efficiently (\chx{it is better to have a figure here, edited by PPT}). The input sequence (sequence of posterior vectors for each location) is processed by encoders to distill the essential features of the trajectory and the decoder deduces the output sequence (one hot-encoded vector-based trajectory). Gated Recurrent Units (GRUs) are chosen for preparing the fully connected neural network due to their ability to understand the long-range dependencies in trajectory sequence and help in preventing the vanishing gradient problems.

\sourabh{Loss function explanation with proper equations (pending)}
Cross-entropy loss is employed due to its effectiveness in sequence prediction tasks. This loss function measures dissimilarity between predicted and actual distributions. In our case, it quantifies the difference between the model-generated node predictions and the true node sequences. \chx{It is better to put the equation here} When training the parameters of GRU, the goal is to minimize this dissimilarity, ensuring accurate trajectory tracking.
}
\vspace{-0.00in}
\section{TransProtect: A Countermeasure of VehiTrack}
\label{sec:countermeasure}

As analyzed in Section \ref{sec:threat}, Geo-Ind proves susceptible to privacy breaches by VehiTrack when protecting the location privacy of vehicles. This vulnerability stems from the inclusion of ``unrealistic'' locations in its obfuscation range, which are prone to elimination by VehiTrack. Motivated by this insight, in this section, we introduce \emph{TransProtect}, which aims to identify a set of ``candidate obfuscated locations'' that closely adhere to the realistic mobility patterns of vehicles, making them difficult for attackers to distinguish from actual locations. \looseness = -1
\vspace{-0.08in}
\subsection{The Framework of TransProtect}
\vspace{-0.02in}

\begin{figure}[t]
\centering
\includegraphics[width=0.48\textwidth]{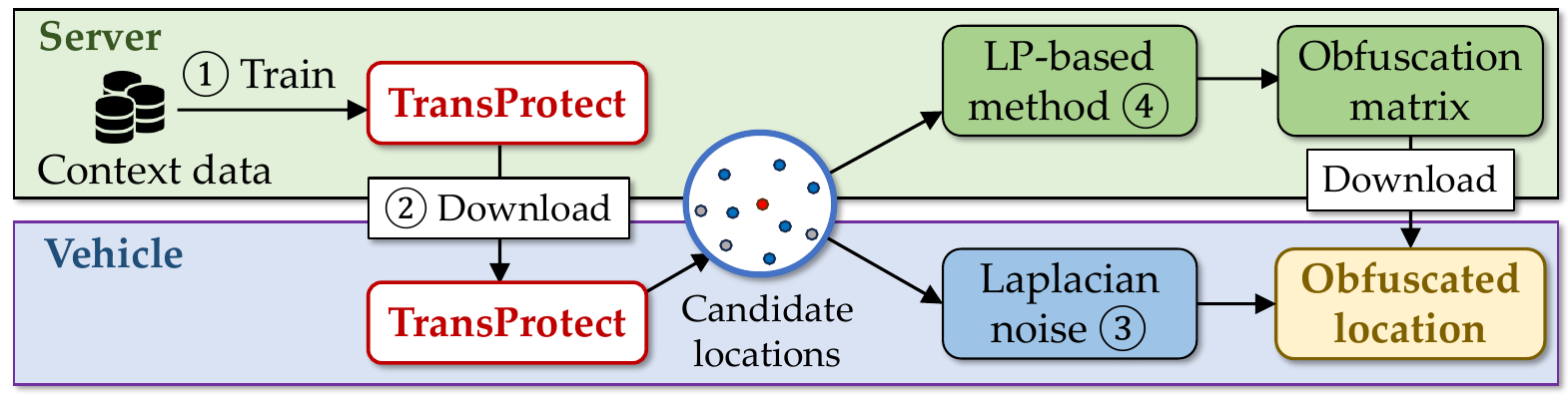}
\vspace{-0.10in}
\caption{Integraing TransProtect in geo-obfuscation.}
\vspace{-0.12in}
\label{fig:transformerinte}
\end{figure}

As illustrated by Fig. \ref{fig:transformerinte}, TransProtect can be integrated into the current geo-obfuscation framework, such as LP-based geo-obfuscation \cite{Qiu-TMC2020} or Laplacian noise \cite{Andres-CCS2013}. Using the context data including local historical traffic flow data and LBS target distributions, the server initially trains a ``TransProtect'' model (\textcircled{1}). This model takes a vehicle's trajectory as input and outputs a set of candidate obfuscated locations for the vehicle's current location within the trajectory. Before reporting the location, each participating vehicle needs to download the trained ``TransProtect'' model to identify the candidate location set for obfuscation (\textcircled{2}). Then, the vehicle can locally obfuscate its location within the candidate location set using Laplacian noise  (\textcircled{3}), which requires a low computational load that doesn't necessitate global LBS service information. Alternatively, the vehicle can report the candidate location set, prompting the server to compute the obfuscation matrix using LP (\textcircled{2}), which incurs a relatively higher computational load and relies on global target information. In both cases, 
The integration of TransProtect into the geo-obfuscation framework allows for the restriction of the obfuscation range to a specific set of locations, aligning with vehicles' realistic mobility features while minimizing utility loss. \looseness = -1

\begin{figure}[t]
\centering
\includegraphics[width=0.46\textwidth]{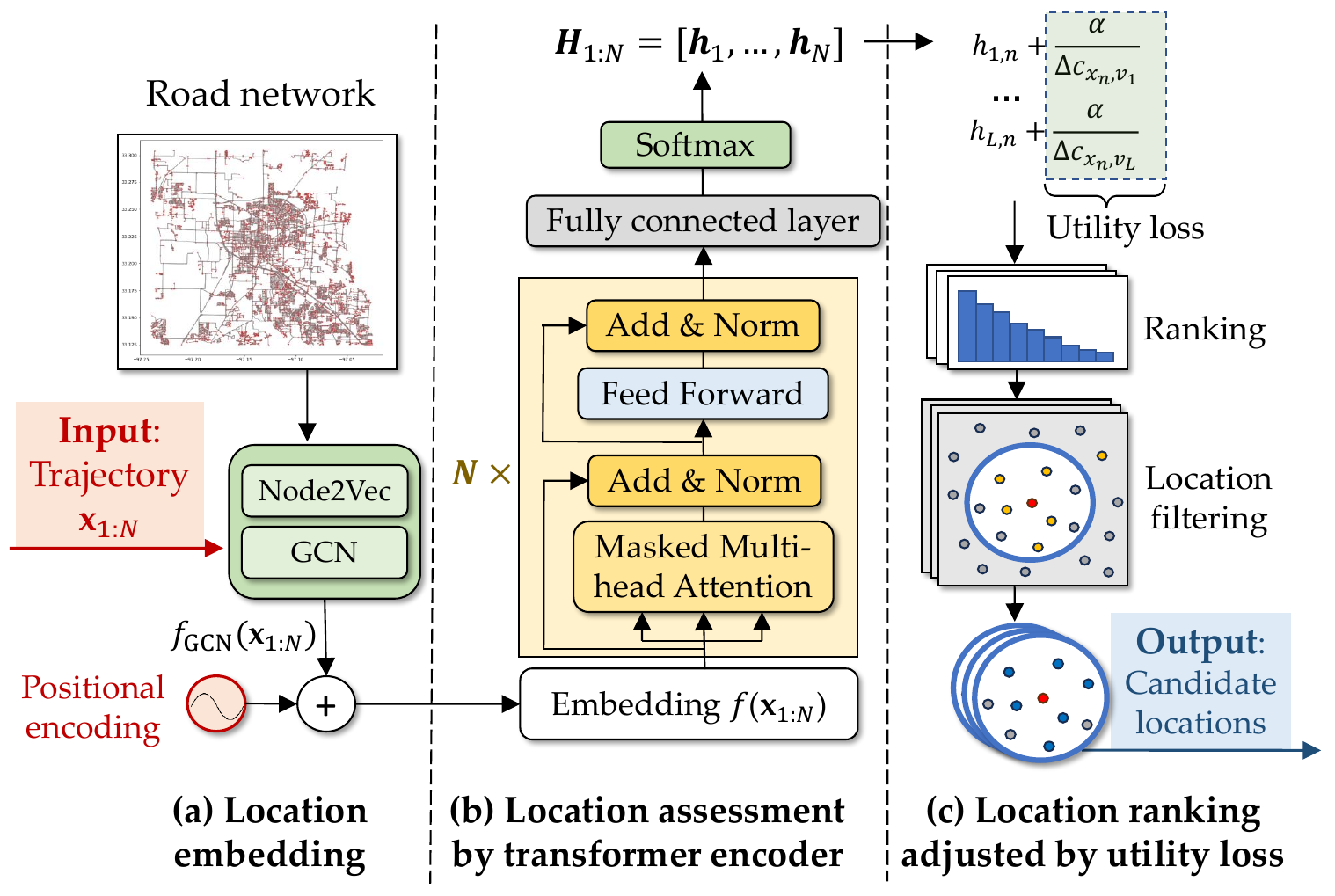}
\vspace{-0.10in}
\caption{TransProtect framework.}
\label{fig:transformer}
\vspace{-0.13in}
\end{figure}

Fig. \ref{fig:transformer} shows the framework of TransProtect. TransProtect first takes the vehicle's real \emph{location sequence} (or the \emph{trajectory}), $\mathbf{x}_{{1}: {N}} = \left\{x_{{1}}, ..., x_{{N}}\right\}$, as the \textbf{input}. During each time slot $t_{n}$, TransProtect assesses both the utility loss and the likelihood of each location $v_i \in \mathcal{V}$ being the actual location, based on the vehicle's historical locations $\mathbf{x}_{1:n-1} = \left\{x_{1}, \ldots, x_{{n-1}}\right\}$. After this assessment, TransProtect \textbf{outputs} a maximum of $K$ locations as the ``candidate locations'' for the obfuscated location, with $K$ representing the maximum allowable number of locations within the obfuscation range.


As shown by Fig. \ref{fig:transformer}(a)(b)(c), TransProtect mainly comprises the following three components: (a) \emph{location embedding}, (b) \emph{location assessment by transformer encoder}, and (c) \emph{location ranking adjusted by utility loss}. Next, we introduce the details of the three components in Section \ref{subsec:locationembed}, Section \ref{subsec:locgenerate}, and Section \ref{subsec:locgenerate}, respectively. 

\vspace{-0.05in}
\subsection{Location Embedding}
\label{subsec:locationembed}

The objective of \emph{location embedding} is to map the nodes (locations) in the road network graph $\mathcal{G}$ to a \emph{low dimensional feature space}, where the neighborhood information of each node in $\mathcal{G}$ can be well-preserved. Here, we let $f: \mathcal{V} \rightarrow \mathbb{R}^g$ be the \emph{map} from the locations to their feature representations, where $g$ denotes the dimension of the resulting embeddings. 

\noindent \textbf{Node2Vec}. To achieve the above objective, 
as Fig. \ref{fig:transformer}(a) shows, we first map locations to vectors using \emph{Node2Vec} \cite{grover2016node2vec},  a semi-supervised algorithm designed for scalable feature learning in network structures. Node2Vec seeks to maximize the log-probability of observing a network neighborhood $\mathcal{N}_{i}$ for each node $v_i$ conditioned on its feature representation $f_{N2V}(v_i)$, i.e., 
\begin{equation}
\label{eq:N2Vobj}
\max_f ~ \sum_{v_i \in \mathcal{V}} \log \mathrm{Pr}\left(\mathcal{N}_{i}|f_{N2V}(v_i)\right). 
\end{equation}
Directly minimizing the objective function in Equ. (\ref{eq:N2Vobj}) results in significant computational overhead when the location set $\mathcal{V}$ is large. Equ. (\ref{eq:N2Vobj}) can be simplified to 
\begin{equation}
\max_f ~ \sum_{v_i \in \mathcal{V}} \left[-\log Z_{v_i} + \sum_{v_j \in \mathcal{N}_{i}} f_{N2V}(v_i)f_{N2V}(v_j)\right]
\end{equation}
by assuming the conditional independence of the likelihood of observing the neighbors in $\mathcal{N}_{i}$ and the symmetry in feature space, where $Z_{v_i} = \sum_{v_j \in \mathcal{V}} \exp(f_{N2V}(v_i)f_{N2V}(v_j))$ can be further approximated using negative sampling \cite{NIPS2013_9aa42b31}.

\DEL{
\begin{definition} 
\label{def:locembedding}
(Location embedding) Given a set of historical trajectories $\mathcal{T}$, we aim at mapping all the nodes in a graph $\mathcal{G} = \left(\mathcal{V}, \mathcal{E}\right)$ into a vector space $X \in \mathbb{R}^{|\mathcal{V}|\times d }$, where $|\mathcal{V}|$ and $d$ denote the size of the node set $\mathcal{V}$ and the dimension of the resulting embeddings, respectively. 
\end{definition}}

\vspace{0.05in}
\noindent \textbf{Graph Convolutional Network}. Following the initial embedding of nodes via Node2Vec, we proceed to improve these embeddings using a \emph{Graph Convolutional Network (GCN)} \cite{kipf2017semisupervised}. The primary goal of GCN is to incorporate both the edge weights and the neighborhood information into the node representations, thus achieving a more contextually comprehensive embedding. 

Specifically, GCN processes the Node2Vec embeddings using a series of convolutional layers. Each layer in GCN updates the node embeddings by aggregating information from their respective neighborhoods, with an emphasis on the connectivity patterns as dictated by the graph structure. This process is formally expressed through the following convolution operation in each layer:
\vspace{-0.02in}
\begin{equation}
\label{eq:Hl}
\mathbf{S}^{(l+1)} = \sigma\left(\hat{\mathbf{D}}^{-\frac{1}{2}} \hat{\mathbf{E}} \hat{\mathbf{D}}^{-\frac{1}{2}} \mathbf{S}^{(l)} \boldsymbol\Theta^{(l)}\right), 
\vspace{-0.05in}
\end{equation}
where $\hat{\mathbf{D}}^{-\frac{1}{2}} \hat{\mathbf{E}} \hat{\mathbf{D}}^{-\frac{1}{2}}$ denotes the symmetric normalized Laplacian matrix. Here,  $\hat{\mathbf{E}} = \mathbf{E} + \mathbf{I}$ includes the addition of the identity matrix $\mathbf{I}$ to incorporate self-connections $\mathbf{E}$, and $\hat{\mathbf{D}}$ is the diagonal node degree matrix of $\hat{\mathbf{E}}$. The matrix $\mathbf{S}^{(l)}$ represents the activations from the $ l $-th layer, $ \boldsymbol\Theta^{(l)} $ is the layer's trainable weight matrix, and $ \sigma(\cdot) $ is a non-linear activation function such as the sigmoid. The initial layer activations are set to the node embeddings $ \mathbf{S}^{(0)} = f_{N2V}(\mathbf{x}_{{1}: {N}}) $.

By applying Node2Vec followed by GCN, we obtain the embedding of each trajectory $\mathbf{x}_{{1}: {N}}$, denoted as $f_{GCN}(\mathbf{x}_{{1}: {N}}) \in \mathbb{R}^{N \times g}$, which captures the spatial 
nuances of the trajectory within the embedding space. \looseness = -1

\vspace{0.02in}
\noindent \textbf{Positional embedding}. As Fig. \ref{fig:transformer}(a) shows, after Node2Vec and GCN, each trajectory $\mathbf{x}_{{1}: {N}}$ is initially transformed into a vector space representation $f_{GCN}(\mathbf{x}_{{1}: {N}})$. 
To incorporate the sequential order of the locations in the trajectory, positional encodings are added to the embedding vectors. These encodings provide a unique position signature that allows the model to consider the order of locations within each trajectory. 
The positional encodings are calculated as follows:
\begin{equation}
PE_{(pos, 2i)} = \sin\left(\frac{pos}{10000^{2i/g}}\right), PE_{(pos, 2i+1)} = \cos\left(\frac{pos}{10000^{2i/g}}\right)
\end{equation}
where $pos$ and $i$ are the location's position in the trajectory and the dimension index, respectively. These encodings are added to the embedding vectors to produce the final location embeddings $
f(\mathbf{x}_{1:N}) = f_{GCN}(\mathbf{x}_{1:N}) + PE$.
\vspace{-0.08in}
\subsection{Location Assessment by Transformer}
\label{subsec:locgenerate}
\DEL{
In the standard multi-head self-attention of the Transformer encoder, given the input trajectory sequence $X$, the $H$ attention heads transform $X$ into $H$ set of $Q$, $K$ and $V$. This process can be formulated as follows:
\begin{equation}
\label{att}
Attention(Q,K,V) = softmax(\frac{QK^T}{\sqrt{d_k}})V
\end{equation}
\begin{equation}
\label{head}
head_i = Attention(XW_i^Q,XW_i^K,XW_i^V)
\end{equation}
\begin{equation}
\label{Xprime}
X' = Multihead(Q,K,V) = ||_{i=1}^H head_i W^O
\end{equation}
where $W_i^Q, W_i^k, W_i^V \in R^{d\times d_k}$ are weight parameters for query, key and value vectors and $d_k = d/H$. $||$ is the concatenation operation that combines multi-head attention outputs. $W^O \in R^{d\times d}$ is the output weight.
}

As Fig. \ref{fig:transformer}(b) shows, taking the location embeddings $f(\mathbf{x}_{1:N})$ as the inputs, the \emph{transformer encoder} in the second component outputs the score matrix $\mathbf{H}_{1:N} = [\mathbf{h}_1, ..., \mathbf{h}_N]$, where each vector  $\mathbf{h}_n = [h_{1,n}, ..., h_{L,n}] \in \mathbb{R}^{L}$ contains the predictive \emph{probability scores} across all the locations in $\mathcal{V}$ at each time slot $t_n$. Here, $L$ is the number of locations in $\mathcal{V}$, and each probability score $h_{j,n} =  \hat{p}(v_j|\mathbf{x}_{1:n-1})$ ($j = 1, ..., L$) 
reflects $v_j$'s likelihood of being the real location at $t_n$ given the observation of the vehicle's historical locations $\mathbf{x}_{1:n-1}$. 

\vspace{0.02in}
\noindent \textbf{Detailed steps of the Transformer encoder}.  As illustrated by Fig. \ref{fig:transformer}(b), $f(\mathbf{x}_{1:N})$ is first passed to a \emph{multi-head attention mechanism}. In each attention head, the input sequence is linearly transformed into \emph{queries} $\mathbf{Q}$, \emph{keys} $\mathbf{K}$, and \emph{values} $\mathbf{V}$ using respective weight matrices. 
The scaled dot-product attention for each head is computed as:
\vspace{-0.05in}
\begin{equation}
Attention(\mathbf{Q},\mathbf{K},\mathbf{V}) = \text{softmax}\left(\frac{\mathbf{Q} \mathbf{K}^T}{\sqrt{g_k}}\right)\mathbf{V}.
\vspace{-0.05in}
\end{equation}
Each attention head $head_i$ processes the sequence independently using the following transformation:
\vspace{-0.00in}
\begin{equation}
\label{att}
head_i = Attention\left(f(\mathbf{x}_{1:N})\mathbf{W}_i^Q, f(\mathbf{x}_{1:N})\mathbf{W}_i^K, f(\mathbf{x}_{1:N})\mathbf{W}_i^V\right). 
\end{equation}
The outputs from each head are then concatenated and linearly transformed to produce the final representation for each position in the sequence: $h_{\mathbf{x}}' = \text{Concatenate}(head_1, \ldots, head_B)\mathbf{W}^O$. 

Here, $B$ is the number of headers, and $\mathbf{W}_i^Q$, $\mathbf{W}_i^K$, $\mathbf{W}_i^V$, and $\mathbf{W}^O$ are the trainable parameters of the model. The dimensionality of each head's output, $g_k$, is set to $g/B$ to maintain a consistent dimensionality across different heads. After processing the sequence through the Transformer's multi-head attention module, the derived representation $ h_{\mathbf{x}}' $ is then delivered to a fully connected layer. This layer applies a learned linear transformation characterized by weight matrix $\mathbf{W}^{FC} $ and bias $\mathbf{b}^{FC}$, producing a set of logits for each location in the sequence: $\mathbf{a}_n = \mathbf{W}^{FC} h_{\mathbf{x},n}' + \mathbf{b}^{FC}$, where $\mathbf{a}_n$ represents the logits at time slot $t_n$, and $h_{\mathbf{x},n}'$ denotes the $n$-th vector in the sequence after the attention mechanism. For each time step $t_n$, the logits $\mathbf{a}_n$ are then passed through a softmax function to yield a probability distribution over the set of all possible locations $\mathcal{V}$, i.e., $\mathbf{h}_n = \text{softmax}(\mathbf{a}_n)$. 
By stacking the vectors $\mathbf{h}_1, ..., \mathbf{h}_N$ for all the $N$ time slots, we construct the probability score matrix $\mathbf{H}_{1:N} = [\mathbf{h}_1, ..., \mathbf{h}_N]$. 
$\mathbf{H}_{1:N}$ serves as the final output from the Transformer encoder, encapsulating the predictive distribution over the location set $\mathcal{V}$ at each time step within the trajectory.

\DEL{
\begin{equation}
\mathbf{H}_{1:N} = [\mathbf{h}_1, \mathbf{h}_2, ..., \mathbf{h}_N]
\end{equation}}

\vspace{0.02in}
\noindent \textbf{Loss and Training}. During the training process, we aim to minimize the \emph{cross entropy loss} $\mathcal{L}_{CE}(\mathbf{H}_{1:N}, \hat{\mathbf{X}}_{1:N})$ between the predicted location probability distribution $\mathbf{H}_{1:N}$ and the location ground truth $\hat{\mathbf{X}}_{1:N}$ (which is obtained from the real vehicle trajectory dataset \cite{roma-taxi-20140717}), i.e., $\min ~\mathcal{L}_{CE}(\mathbf{H}_{1:N}, \hat{\mathbf{X}}_{1:N}) = -\sum_{n=1}^{N} \sum_{v_j \in \mathcal{V}} \hat{x}_{j,n} \log(h_{j,n})$. 

Here, $\hat{\mathbf{X}}_{1:N} = \left\{\hat{x}_{j,n}\right\}_{K\times N}$ is a one-hot encoded matrix of the true locations with $\hat{x}_{j,n}$ indicating the ground truth presence (or absence) of a location $v_j$ at time slot $t_n$ in the trajectory. 
\vspace{-0.10in}
\subsection{Location Ranking Adjusted by Utility Loss}
\label{subsec:locfilter}
\vspace{-0.03in}
As depicted in Fig. \ref{fig:transformer}(c), after assessing the probability scores of the locations using the Transformer encoder, TransProtect adjusts the scores by the data utility associated with each location. Here, we use $\Delta c_{x_n, v_j}$ to denote the utility loss caused by the obfuscated location $v_j$ given the real location $x_n$. Each location $v_j$ is assessed by the weighted sum of the probability score and the inverse of the utility loss: $h_{l,n} + \frac{\alpha}{\Delta c_{x_n, v_j}}$, where the weight $\alpha>0$ is a predefined constant, reflecting the user's emphasis on data utility during location obfuscation. TransProtect then ranks all the locations in $\mathcal{V}$ based on their weighted scores and selects the top $K$ locations as the ``candidate locations'' for obfuscation.

\noindent \textbf{Measurement of utility loss}. The assessment of utility loss is contingent on the specific manner in which location data is used in downstream decision-making. As an example, in this paper, we consider the LBS applications where vehicles need to physically travel to designated locations to receive desired services such as navigation \cite{To-TMC2017}, or to fulfill tasks in spatial crowdsourcing \cite{Qiu-TMC2020}. In those applications, data utility loss can be quantified by the discrepancy between the estimated and actual travel costs to reach the designated locations. Note that our framework is also readily extended to other LBS applications with slight adjustments, provided that a clear relationship between data utility loss and obfuscated data can be established. \looseness = -1

We let $q_l$ denote the prior probability that the target location is located at the location $v_l$ ($l = 1,..., N$). Given a real location $x_n$, the utility loss caused by an obfuscated location $v_j$ is defined as the expected error of the traveling costs to the target location, calculated by
\vspace{-0.10in}
\begin{equation}
\label{eq:utilityloss}
\Delta c_{x_n,v_j} = \sum_{l=1}^N q_l \left|c_{x_n,v_l} - c_{v_j,v_l}\right|. 
\vspace{-0.05in}
\end{equation}

\noindent \textbf{Location filtering}. After calculating the weighted score $h_{l,n}+\frac{\alpha}{ \Delta c_{x_n, v_j}}$ of each location $v_j$, TransProtect identifies the set of candidate locations using a \emph{min heap} \cite{Algorithm}, mainly with the two features: (f1) the top element has the minimum score in the heap; (f2) the min heap has two types of operations: \emph{push} to insert a new element, and \emph{pop} to remove the top element from the heap. 
The min heap is initialized by empty. TransProtect then pushes each location in $\mathcal{V}$ onto the heap. Once the heap reaches its capacity $K$ and determines whether to add a new location $v_j \in \mathcal{V}$, TransProtect first checks whether the top location in the current heap has a higher score than $v_j$. If NO, $v_j$ won't be pushed onto the heap since it has a lower score than the $K$ locations in the current heap; If YES, the top location is popped off and $v_j$ is pushed onto the heap. Note that the popped location cannot have the $K$ highest score, since it has a lower score than the other $K-1$ locations in the current heap and the newly added location $v_j$.

\DEL{
\vspace{-0.10in}
\begin{algorithm}[h]
\SetKwFunction{push}{push}
\SetKwFunction{pop}{pop}
\SetKwFunction{top}{top}
\SetKwInOut{Input}{Input}
\SetKwInOut{Output}{Output}
\small
\Input{The weighted scores of all the locations in $\mathcal{V}$}
\Output{$K$ locations with the highest scores}
    The max heap is initialized by empty\; 
    \For{each candidate location $v_j \in \mathcal{V}$}{
            \If{the heap has not reached capacity $K$}{
            Push $v_j$ onto the heap\; 
            }
            \Else{
                \If{$v_j$ has a lower score than that of the top location in the heap}{
                    Pop the top location off the heap\;
                    Push $v_j$ onto the heap\;
                }
            }
        }
\Return the set of locations in the heap\; 
\normalsize
\caption{Location selection. }
\label{al:elimination}
\end{algorithm}
\vspace{-0.10in}}

\vspace{-0.00in}
\DEL{
Algorithm \ref{al:trajmanager} gives the pseudo-code of the trajectory update algorithm in each round $t$: 

At the beginning of each round $t$, the min heap $\mathbf{q}^{t}$ is initialized by empty (line 1). If it is the first time slot $t_a$ (line 2--7), FTraj randomly picks up $M$ locations around the current true location: $s^{t_a}_{f_1}$, ..., $s^{t_a}_{f_M}$ (line 4), calculates their fitness values $\mathrm{Fit}\left(s^{t_a}_{f_1}\right)$, ..., $\mathrm{Fit}\left(s^{t_a}_{f_M}\right)$ (line 6), and pushes the corresponding nodes $\left(s^{t_a}_{f_i}, \phi, \mathrm{Fit}\left(s^{t_a}_{f_i}\right)\right)$ ($i = 1, ..., M$) sequentially onto $\mathbf{q}^{t_a}$ (line 7). After the first time slot, FTraj needs to carry out the following two steps in each time slot:
\vspace{0.10in}
\begin{itemize}
    \item [S1.] \emph{Production} (line 9--13): \newline For each node $\left(v_{f^{t-1}_i}, s^{t-1}_{f_{i'}}, \mathrm{Fit}\left(\vec{\mathbf{s}}^{[t_a, t-1]}_{f_i}\right)\right)$ in the min heap $\mathbf{q}^{t-1}$, FTraj initializes an empty location set $\mathcal{R}\left(\vec{\mathbf{s}}^{[t_a,t-1]}_{f_i}\right)$. Given the locations of fake trajectories in the last time slot in $\mathbf{q}^{t-1}$ (line 11), FTraj finds a set of locations that are reachable (line 12) and add them to $\mathcal{R}\left(\vec{\mathbf{s}}^{[t_a,t-1]}_{f_i}\right)$ (line 13); 
    \item [S2.] \emph{Selection} (line 14--23): \newline For each location $v_j \in \mathcal{R}\left(\vec{\mathbf{s}}^{[t_a,t-1]}_{f_i}\right)$, FTraj calculates its fitness value $\mathrm{Fit}\left(v_j\right)$ (line 17). The first $M$ nodes are pushed directly onto $\mathbf{q}^{t}$ (line 19). Once $\mathbf{q}^{t}$ reaches its capacity, i.e., $|\mathbf{q}^{t}|=M$, FTraj first checks whether the top node in $\mathbf{q}^{t}$ has higher fitness value than $\left(\mathrm{Fit}\left(\vec{\mathbf{s}}^{[t_a, t-1]}_{f_i}\right) + \mathrm{Fit}\left(v_j\right)\right)$ (line 21): 
    \vspace{0.05in}
    \begin{itemize}
    \item [$\triangleright$] If NO, $q^t_{\mathrm{new}}$ won't be pushed onto $\mathbf{q}^{t}$, since $q^t_{\mathrm{new}}$ has a lower fitness value than any of the $M$ nodes in $\mathbf{q}^{t}$.
    \item [$\triangleright$] If YES, $q^{t}_{\mathrm{top}}$ will be popped off (line 22) and 
    \newline $q^{t}_{\mathrm{new}} = \left(v_{f^{t-1}_{j'}}, v_{j}, \left(\mathrm{Fit}\left(\vec{\mathbf{s}}^{[t_a, t-1]}_{f_i}\right) + \mathrm{Fit}\left(v_j\right)\right)\right)$ will be pushed onto $\mathbf{q}^{t}$  (line 23). $q^{t}_{\mathrm{top}}$ cannot be one of the $M$ highest nodes, since $q^{t}_{\mathrm{top}}$ has a lower fitness value than $q^t_{\mathrm{new}}$'s and the other $M-1$ nodes in $\mathbf{q}^{t}$ (based on feature F1).
\end{itemize}
\end{itemize} 

\vspace{-0.00in}
\begin{algorithm}[h]
\SetKwFunction{push}{push}
\SetKwFunction{pop}{pop}
\SetKwFunction{top}{top}
\SetKwInOut{Input}{Input}
\SetKwInOut{Output}{Output}
\footnotesize
\Input{$t$, $\mathbf{q}^1$, ..., $\mathbf{q}^{t-1}$, $s^{t_a}_f$}
\Output{$\mathbf{q}^{t}$}
$\mathbf{q}^{t}$ is initialized by empty\; 
\If{$t$ equals $t_a$ }{
\tcp{Initialization in the first round}
Randomly pick up $M$ locations around $s^{t_a}_f$: $s^{t_a}_{f_1}$, ..., $s^{t_a}_{f_M}$\;
    \For{each $s^{t_a}_{f_i}$ ($i =1 ,..., M$)}{
    Calculate $s^{t_a}_{f_i}$'s fitness value $\mathrm{Fit}\left(s^{t_a}_{f_i}\right)$ using Equ.  (\ref{eq:fitness})\;
    $\mathbf{q}^{t}$.\push{$\left(s^{t_a}_{f_i}, \phi, \mathrm{Fit}\left(s^{t_a}_{f_i}\right)\right)$}\; 
    }
}
\Else{
    \tcp{S1: Production}
    \For{each $\left(v_{f^{t-1}_i}, s^{t-1}_{f_{i'}}, \mathrm{Fit}\left(\vec{\mathbf{s}}^{[t_a, t-1]}_{f_i}\right)\right)$ in $\mathbf{q}^{t-1}$}{
        Initialize the location set $\mathcal{R}\left(\vec{\mathbf{s}}^{[t_a,t-1]}_{f_i}\right)$ by an empty set\;
        \For{each $v_j \in \mathcal{V}$ with $p^{t-1,t}_{f_i^{t-1},j} > 0$}{
            Add $v_j$ to $\mathcal{R}\left(\vec{\mathbf{s}}^{[t_a,t-1]}_{f_i}\right)$\; 
        }
    }
    \tcp{S2: Selection}
    \For{each $\left(v_{f^{t-1}_i}, s^{t-1}_{f_{i'}}, \mathrm{Fit}\left(\vec{\mathbf{s}}^{[t_a, t-1]}_{f_i}\right)\right)$ in $\mathbf{q}^{t-1}$}{
        \For{$\forall v_j \in \mathcal{R}\left(\vec{\mathbf{s}}^{[t_a,t-1]}_{f_i}\right)$}{
            Calculate $v_j$'s fitness value $\mathrm{Fit}\left(v_j\right)$ using Equ. (\ref{eq:fitness})\;
            \If{$\left|\mathbf{q}^{t}\right| < M$}{
            $\mathbf{q}^{t}$.\push{$\left(s^{t-1}_{f_{j'}}, v_{j}, \mathrm{Fit}\left(\vec{\mathbf{s}}^{[t_a, t-1]}_{f_i}\right) + \mathrm{Fit}\left(v_j\right)\right)$}; 
            }
            \Else{
                \If{$\mathbf{q}^{t}$.\top{} has higher fitness value than $\left(\mathrm{Fit}\left(\vec{\mathbf{s}}^{[t_a, t-1]}_{f_i}\right) + \mathrm{Fit}\left(v_j\right)\right)$}{
                    $\mathbf{q}^{t}$.\pop{}\;
                    $\mathbf{q}^{t}$.\push{$s^{t-1}_{f_{j'}}, v_{j}, \mathrm{Fit}\left(\vec{\mathbf{s}}^{[t_a, t-1]}_{f_i}\right) + \mathrm{Fit}\left(v_j\right)$}\;
                }
            }
        }
    }
}
\Return $\mathbf{q}^{t}$\; 
\normalsize
\caption{Fake trajectory update. }
\label{al:trajmanager}
\end{algorithm}
\vspace{0.03in}}
\noindent \textbf{Time complexity}. 
In min heap, both push and pop operations take $O(\log K)$ time complexity and $O(1)$ space complexity. 
Suppose that there are $L$ candidate locations to check by the transformer encoder. To find the $K$ locations with the highest scores, it takes up to $L$ push/pop operations, amounting to $O(L\log K)$ operations. As both $L$ and $K$ are not large in practice, e.g. they are set by up to 50 and 1739 respectively in our experiment (Section \ref{sec:exp}), such a computation load is acceptable to vehicle-equipped devices like smartphones. 

\DEL{
\vspace{-0.08in}
\subsection{Discussion: Geo-Ind Guarantee}
\label{subsec:discussionGeoInd}
\vspace{-0.03in}
In theory, the obfuscation matrices generated by TransProtect might not fully meet Geo-Ind, which requires that the obfuscation probabilities for each pair of actual locations adhere to ``indistinguishability'' constraints (Equ. (\ref{eq:Geo-Ind-LP})). This partial compliance arises because the obfuscated locations for each location are confined to a predetermined range (identified by the transformer), with Geo-Ind constraints applied solely within this specified area. Consequently, locations outside the designated obfuscation range cannot ensure adherence to the Geo-Ind requirements.

However, it is important to note that the obfuscation probabilities for locations beyond the specified range are generally very low (approaching 0), which minimally impacts the process of selecting obfuscated locations. As a result, while complete compliance with Geo-Ind constraints in TransProtect is not achieved, the occurrence rate of such violations is exceedingly low. Empirical evidence (detailed in Table \ref{Tb:exp:GVR} in Appendix) shows that, on average, the proportion of $z_{i,k}$ and $z_{j,k}$ pairs that do not adhere to the Geo-Ind constraint in Equ. (\ref{eq:Geo-Ind-LP}) in TransProtect is merely up to 0.00028\%.
}

\vspace{-0.00in}
\section{Performance Evaluation}
\label{sec:exp}
\vspace{-0.00in}
We carry out an extensive simulation to assess the performance of our location inference algorithm \textbf{VehiTrack} and our new LPPM \textbf{TransProtect} in Section \ref{subsec:expVehiTrack} (\textbf{Experiment I}) and Section \ref{subsec:expTransProtect} \newline (\textbf{Experiment II}), respectively, with the comparison of a list of state-of-the-art methods. In Section \ref{subsec:settings}, we first introduce the settings of the experiment, including the real-world dataset used in the simulation, the benchmarks, and the performance metrics\footnote{The source code of both VehiTrack and TransProtect is available at: \url{https://github.com/sourabhy1797/VehiTrack}.}. 

\vspace{-0.10in}
\subsection{Experimental Settings}
\label{subsec:settings}
\vspace{-0.02in}
\subsubsection{Vehicle location dataset} We adopt two vehicle trajectory datasets: (1) Rome taxicab  dataset \cite{roma-taxi-20140717}, which includes 367,052 trajectories from approximately 320 taxis, covering 30+ days, and (2) San Francisco dataset \cite{SanFranciscotaxi}, including 34,564 trajectories from 536 taxis, covering 30 days. 
For both datasets, the road network information of the target region is extracted by OpenStreetMap \cite{openstreetmap}, which provides fine-grained location (node) and road (edge) information. To crop the road map data of Rome (resp. San Francisco), we compute the bounding area keeping a coordinate \emph{(latitude = 41.9028, longitude = 12.4964)} (resp. \emph{(latitude = 37.7739, longitude = -122.4312)}) as the center and computed all the nodes and edges within a 20-kilometer (resp. 10-kilometer) radius distance from the center.

\vspace{-0.08in}
\subsubsection{TransProtect model training setting} The experiments are conducted on Ubuntu 22.04 with an NVIDIA GeForce 4090 GPU.
We implement TransProtect using PyTorch 2.1 \cite{PyTorch21}. We set the embedding dimensionalities to 128 and batch size to 50. The initial learning rate is 0.001.
\DEL{
\vspace{-0.00in}
\begin{table*}[t]
\caption{Expected inference error (km)}
\vspace{-0.00in}
\label{Tb:exp:EIE}
\centering
\small  
\begin{tabular}{ c|c|c|c||c|c|c}
\hline
\multicolumn{1}{ c  }{}& \multicolumn{6}{ c }{Location privacy protection algorithms} \\
\cline{2-7}
\multicolumn{1}{ c|  }{Location}
 & \multicolumn{3}{ |c|| }{\textbf{Experiment I}}
 & \multicolumn{3}{ |c }{\textbf{Experiment II}}
 \\ 
\cline{2-7}
\multicolumn{1}{ c|  }{inference}
 & \multicolumn{3}{ |c|| }{Laplace}
 & \multicolumn{3}{ |c }{Laplace+\textbf{TransProtect}}
 \\ 
\cline{2-7}
\multicolumn{1}{ c|  }{algorithm}  & $\epsilon = 50$& $\epsilon = 100$& $\epsilon = 150$& $\epsilon = 50$& $\epsilon = 100$& $\epsilon = 150$\\ 
\hline
\hline
\multicolumn{1}{ c|  }{ \textbf{VehiTrack}} & \textbf{0.2051 $\pm$ 0.0018} & \textbf{0.1828 $\pm$ 0.0061} & \textbf{0.1611 $\pm$ 0.0121} & \textbf{1.2485
$\pm$ 0.9739} & \textbf{1.2168 $\pm$ 0.9534} & \textbf{1.2132 $\pm$ 0.9561} \\ 
\multicolumn{1}{ c|  }{VehiTrack-I} & 0.2985 $\pm$ 0.0076 & 0.2427 $\pm$ 0.0435& 0.2059 $\pm$ 0.0562
& 0.8691 $\pm$ 0.1274 & 0.6473 $\pm$ 0.1349& 0.6309 $\pm$ 0.1202\\
\multicolumn{1}{ c|  }{Bayes} & 0.4243 $\pm$ 0.2059 & 0.3893 $\pm$ 0.2989 & 0.3339 $\pm$ 0.1566& 2.2492 $\pm$ 1.3639 & 2.7109 $\pm$ 1.6767 & 2.3202 $\pm$ 1.4909\\
\multicolumn{1}{ c|  }{HMM} & 0.2207 $\pm$ 0.2417 & 0.2010 $\pm$ 0.2559& 0.2003 $\pm$ 0.2858& 0.4767 $\pm$ 0.5767 & 0.3293 $\pm$ 0.3187& 0.2593 $\pm$ 0.3427 \\
\multicolumn{1}{ c|  }{TCA} & 0.3752 $\pm$ 0.1282 & 0.3516
 $\pm$  0.1547& 0.3212 $\pm$ 0.1291& 0.4980 $\pm$ 0.4798 & 0.3609 $\pm$ 0.2313& 0.3381 $\pm$ 0.3154 \\
\hline
\end{tabular}
\end{table*}}

\DEL{
\vspace{-0.00in}
\begin{table*}[t]
\caption{Expected inference error of Laplace and Laplace + TransProtect (km). }
\vspace{-0.10in}
\label{Tb:exp:EIELaplace}
\centering
\small 
\begin{tabular}{ c|c|c|c||c|c|c }
\hline
\multicolumn{1}{ c  }{}& \multicolumn{6}{ c }{Location privacy protection algorithms} \\
\cline{2-7}
\multicolumn{1}{ c|  }{Location}
 & \multicolumn{3}{ |c|| }{\textbf{Experiment I}}
 & \multicolumn{3}{ |c }{\textbf{Experiment II}}
 \\  
\cline{2-7}
\multicolumn{1}{ c|  }{inference}
 & \multicolumn{3}{ |c|| }{Laplace}
 & \multicolumn{3}{ |c }{Laplace+\textbf{TransProtect}}
 \\ 
\cline{2-7}
\multicolumn{1}{ c|  }{algorithms}  & $\epsilon = 5.0$km$^{-1}$& $\epsilon = 7.5$km$^{-1}$& $\epsilon = 10.0$km$^{-1}$& $\epsilon = 5.0$km$^{-1}$& $\epsilon = 7.5$km$^{-1}$& $\epsilon = 10.0$km$^{-1}$\\ 
\hline
\hline
\multicolumn{1}{ c|  }{ \textbf{VehiTrack}} & \textbf{0.2167} & \textbf{0.2144} & \textbf{0.2116} & \textbf{0.3324$_{(+53.4\%)}$
} & \textbf{0.3199}$_{(+49.2\%)}$ & \textbf{0.3101}$_{(+46.6\%)}$ \\ 
\multicolumn{1}{ c|  }{VehiTrack-I} &0.6299 & 0.6214& 0.6166
& 0.6545$_{(+3.91\%)}$ & 0.6501$_{(+4.62\%)}$& 0.6474$_{(+5.00\%)}$\\
\multicolumn{1}{ c|  }{Bayes} & 0.6478 & 0.6392 & 0.6365& 0.6854$_{(+5.80\%)}$ & 0.6515$_{(+1.92\%)}$ & 0.6396$_{(+0.49\%)}$ \\
\multicolumn{1}{ c|  }{HMM} & 0.2563 & 0.2476& 0.2466& 0.7047$_{(+175.0\%)}$ & 0.6895$_{(+178.5\%)}$& 0.6812$_{(+176.2\%)}$\\
\hline
\end{tabular}
\vspace{-0.10in}
\end{table*}
}
\DEL{
\vspace{-0.00in}
\begin{table*}[t]
\caption{Expected inference error of Laplace and Laplace + TransProtect (km). }
\vspace{-0.10in}
\label{Tb:exp:EIELaplace}
\centering
\small 
\begin{tabular}{ c|c|c|c|c|c|c||c|c|c|c|c|c }
\toprule
\multicolumn{1}{ c  }{}& \multicolumn{12}{ c }{Location privacy protection algorithms} \\
\cline{2-13}
\multicolumn{1}{ c|  }{Location}
 & \multicolumn{6}{ |c|| }{\textbf{Experiment I}}
 & \multicolumn{6}{ |c }{\textbf{Experiment II}}
 \\  
\cline{2-13}
\multicolumn{1}{ c|  }{inference}
 & \multicolumn{6}{ |c|| }{Laplace}
 & \multicolumn{6}{ |c }{Laplace+\textbf{TransProtect}}
 \\ 
\cline{2-13}
\multicolumn{1}{ c|  }{algorithms}  
& \multicolumn{2}{ |c }{$\epsilon = 5.0$km$^{-1}$}
& \multicolumn{2}{ |c }{$\epsilon = 7.5$km$^{-1}$}
& \multicolumn{2}{ |c|| }{$\epsilon = 10.0$km$^{-1}$}
& \multicolumn{2}{ |c }{$\epsilon = 5.0$km$^{-1}$}
& \multicolumn{2}{ |c }{$\epsilon = 7.5$km$^{-1}$}
& \multicolumn{2}{ |c }{$\epsilon = 10.0$km$^{-1}$}\\ 
\cline{2-13}
\multicolumn{1}{ c|  }{}  
& RM& SF
& RM& SF
& RM& SF
& RM& SF
& RM& SF
& RM& SF\\ 
\hline
\hline
\multicolumn{1}{ c|  }{ \textbf{VehiTrack}} & \textbf{0.22} & \textbf{0.22}& \textbf{0.21}& \textbf{0.22}& \textbf{0.21} 
&\textbf{0.21}
&\textbf{0.33$_{(+53.4\%)}$
} &\textbf{0.39}$_{(+74.5\%)}$ & \textbf{0.32}$_{(+49.2\%)}$ &\textbf{0.36}$_{(+64.22\%)}$ & \textbf{0.31}$_{(+46.6\%)}$ &\textbf{0.34}$_{(+58.56\%)}$\\ 
\multicolumn{1}{ c|  }{VehiTrack-I} &0.63 & 0.31 & 0.62& 0.31 & 0.62 & 0.30 
& 0.66$_{(+3.91\%)}$ &0.53$_{(+70.45\%)}$ & 0.65$_{(+4.62\%)}$&0.53$_{(+70.45\%)}$ & 0.65$_{(+5.00\%)}$&0.52$_{(+71.33\%)}$\\
\multicolumn{1}{ c|  }{Bayes} & 0.65 & 0.51 & 0.64 & 0.51 & 0.64& 0.50 & 0.69$_{(+5.80\%)}$ &0.70$_{(+36.79\%)}$ & 0.65$_{(+1.92\%)}$ &0.69$_{(+35.54\%)}$ & 0.64$_{(+0.49\%)}$ &0.65$_{(+29.42\%)}$ \\
\multicolumn{1}{ c|  }{HMM} & 0.26 &0.30 &  0.25&0.30 &  0.25& 0.28 & 0.71$_{(+175\%)}$ &0.68$_{(+124\%)}$ & 0.69$_{(+179\%)}$&0.66$_{(+121\%)}$ & 0.68$_{(+176\%)}$ &0.64$_{(+130\%)}$ \\
\hline
\end{tabular}
\vspace{-0.10in}
\end{table*}}

\vspace{-0.00in}
\begin{table*}[t]
\caption{Expected inference error of Laplace and Laplace + TransProtect (km)}
\vspace{-0.00in}
\label{Tb:exp:EIELaplace}
\centering
\small 
\begin{tabular}{ c|c|c|c|c|c|c||c|c|c|c|c|c }
\toprule
\multicolumn{1}{ c  }{}& \multicolumn{12}{ c }{Location privacy protection algorithms} \\
\cline{2-13}
\multicolumn{1}{ c|  }{Location}
 & \multicolumn{6}{ |c|| }{\textbf{Experiment I}}
 & \multicolumn{6}{ |c }{\textbf{Experiment II}}
 \\  
\cline{2-13}
\multicolumn{1}{ c|  }{inference}
 & \multicolumn{6}{ |c|| }{Laplace}
 & \multicolumn{6}{ |c }{Laplace+\textbf{TransProtect}}
 \\ 
\cline{2-13}
\multicolumn{1}{ c|  }{algorithms}  
& \multicolumn{3}{ c|  }{Rome }
& \multicolumn{3}{ c||  }{San Francisco }
& \multicolumn{3}{ c|  }{Rome }
& \multicolumn{3}{ c  }{San Francisco }\\ 
\cline{1-13}
\multicolumn{1}{ c|  }{$\epsilon$ (km$^{-1}$)}  & 5.0 & 7.5 & 10.0 & 
5.0 & 7.5 & 10.0 & 5.0 & 7.5 & 10.0 & 
5.0 & 7.5 & 10.0 \\
\hline
\hline
\multicolumn{1}{ c|  }{ \textbf{VehiTrack}} & \textbf{0.22} & \textbf{0.21}& \textbf{0.21}& \textbf{0.22}& \textbf{0.22} 
&\textbf{0.21}
&\textbf{0.33$_{(+53.4\%)}$} & \textbf{0.32}$_{(+49.2\%)}$ & \textbf{0.31}$_{(+46.6\%)}$ &\textbf{0.39}$_{(+74.5\%)}$ &\textbf{0.36}$_{(+64.22\%)}$ &\textbf{0.34}$_{(+58.6\%)}$\\ 
\multicolumn{1}{ c|  }{VehiTrack-I} &0.63 & 0.62 & 0.62& 0.31 & 0.31 & 0.30 & 0.66$_{(+3.91\%)}$ & 0.65$_{(+4.62\%)}$ & 0.65$_{(+5.00\%)}$ &0.53$_{(+70.45\%)}$ &0.53$_{(+70.45\%)}$ &0.52$_{(+71.33\%)}$\\
\multicolumn{1}{ c|  }{Bayes} & 0.65 & 0.64 & 0.64 & 0.51 & 0.51& 0.50 & 0.69$_{(+5.80\%)}$ & 0.65$_{(+1.92\%)}$ & 0.64$_{(+0.49\%)}$ &0.70$_{(+36.79\%)}$  &0.69$_{(+35.54\%)}$ &0.65$_{(+29.42\%)}$ \\
\multicolumn{1}{ c|  }{HMM} & 0.26 &0.25 &  0.25&0.30 &  0.30& 0.28 & 0.71$_{(+175\%)}$ & 0.69$_{(+179\%)}$ & 0.68$_{(+176\%)}$ &0.68$_{(+124\%)}$ &0.66$_{(+121\%)}$  &0.64$_{(+130\%)}$ \\
\hline
\end{tabular}
\vspace{-0.10in}
\end{table*}

\DEL{
\vspace{-0.00in}
\begin{table*}[t]
\caption{Expected inference error of LP and LP + TransProtect (km)}
\vspace{-0.10in}
\label{Tb:exp:EIELP}
\centering
\small 
\begin{tabular}{ c|c|c|c||c|c|c}
\hline
\multicolumn{1}{ c  }{}& \multicolumn{6}{ c }{Location privacy protection algorithms} \\
\cline{2-7}
\multicolumn{1}{ c|  }{Location}
 & \multicolumn{3}{ |c|| }{\textbf{Experiment I}}
 & \multicolumn{3}{ |c }{\textbf{Experiment II}}
 \\ 
\cline{2-7}
\multicolumn{1}{ c|  }{inference}
&\multicolumn{3}{ |c|| }{LP}
&\multicolumn{3}{ |c }{LP+\textbf{TransProtect}}
 \\ 
\cline{2-7}
\multicolumn{1}{ c|  }{algorithms}  & $\epsilon = 5.0$km$^{-1}$& $\epsilon = 7.5$km$^{-1}$& $\epsilon = 10.0$km$^{-1}$& $\epsilon = 5.0$km$^{-1}$& $\epsilon = 7.5$km$^{-1}$& $\epsilon = 10.0$km$^{-1}$\\ 
\hline
\hline
\multicolumn{1}{ c|  }{ \textbf{VehiTrack}} & \textbf{0.2146} & \textbf{0.2073}& \textbf{0.1893} & \textbf{0.2894}$_{(+34.9\%)}$ & \textbf{0.2589}$_{(+24.9\%)}$ & \textbf{0.2512}$_{(+32.7\%)}$ \\ 
\multicolumn{1}{ c|  }{VehiTrack-I} 
&0.2983& 0.2604&0.2489& 0.3901$_{(+30.8\%)}$& 0.3513$_{(+34.9\%)}$&0.3276$_{(+31.6\%)}$ \\
\multicolumn{1}{ c|  }{Bayes} & 0.3969 & 0.3674 & 0.3145 & 0.4534$_{(+14.2\%)}$& 0.4389$_{(+19.5\%)}$&0.4173$_{(+32.7\%)}$ \\
\multicolumn{1}{ c|  }{HMM} & 0.2987& 0.2598&0.2318& 0.4489$_{(+50.3\%)}$& 0.4303$_{(+65.6\%)}$& 0.4129$_{(+78.1\%)}$ \\
\hline
\end{tabular}
\vspace{-0.1in}
\end{table*}}

\DEL{

\vspace{-0.00in}
\begin{table*}[t]
\caption{Expected inference error of LP and LP + TransProtect (km)}
\vspace{-0.10in}
\label{Tb:exp:EIELP}
\centering
\small 
\begin{tabular}{ c|c|c|c|c|c|c||c|c|c|c|c|c}
\toprule
\multicolumn{1}{ c  }{}& \multicolumn{12}{ c }{Location privacy protection algorithms} \\
\cline{2-13}
\multicolumn{1}{ c|  }{Location}
 & \multicolumn{6}{ |c|| }{\textbf{Experiment I}}
 & \multicolumn{6}{ |c }{\textbf{Experiment II}}
 \\ 
\cline{2-13}
\multicolumn{1}{ c|  }{inference}
&\multicolumn{6}{ |c|| }{LP}
&\multicolumn{6}{ |c }{LP+\textbf{TransProtect}}\\
\cline{2-13}
\multicolumn{1}{ c|  }{algorithms}  
& \multicolumn{2}{ |c }{$\epsilon = 5.0$km$^{-1}$}
& \multicolumn{2}{ |c }{$\epsilon = 7.5$km$^{-1}$}
& \multicolumn{2}{ |c|| }{$\epsilon = 10.0$km$^{-1}$}
& \multicolumn{2}{ |c }{$\epsilon = 5.0$km$^{-1}$}
& \multicolumn{2}{ |c }{$\epsilon = 7.5$km$^{-1}$}
& \multicolumn{2}{ |c }{$\epsilon = 10.0$km$^{-1}$}\\ 
\cline{2-13}
\multicolumn{1}{ c|  }{}  
& RM& SF
& RM& SF
& RM& SF
& RM& SF
& RM& SF
& RM& SF\\ 
\hline
\hline
\multicolumn{1}{ c|  }{ \textbf{VehiTrack}} & \textbf{0.21} & \textbf{0.21} & \textbf{0.21}& \textbf{0.20} & \textbf{0.19}& \textbf{0.20} & \textbf{0.29}$_{(+34.9\%)}$ &\textbf{0.26}$_{(+22.43\%)}$ 
 & \textbf{0.26}$_{(+24.9\%)}$ &\textbf{0.25}$_{(+23.73\%)}$  & \textbf{0.25}$_{(+32.7\%)}$ &\textbf{0.25}$_{(+24.58\%)}$  \\ 
\multicolumn{1}{ c|  }{VehiTrack-I} 
&0.30& 0.32 & 0.26 & 0.31 &0.25 & 0.32 & 0.39$_{(+30.8\%)}$&0.44$_{(+37.56\%)}$  & 0.35$_{(+34.9\%)}$&0.42$_{(+34.27\%)}$  &0.33$_{(+31.6\%)}$&0.40$_{(+25.53\%)}$  \\
\multicolumn{1}{ c|  }{Bayes} & 0.40 & 0.41 & 0.37 & 0.40  & 0.31 & 0.40 & 0.45$_{(+14.2\%)}$&0.64$_{(+53.96\%)}$ 
 & 0.44$_{(+19.5\%)}$ &0.63$_{(+54.70\%)}$  &0.42$_{(+32.7\%)}$ &0.62$_{(+56.64\%)}$  \\
\multicolumn{1}{ c|  }{HMM} & 0.30 & 0.34 & 0.26 &0.34 &0.23 & 0.33 & 0.45$_{(+50.3\%)}$ &0.57$_{(+67.51\%)}$ 
 & 0.43$_{(+65.6\%)}$ &0.56$_{(+64.26\%)}$  & 0.41$_{(+78.1\%)}$ &0.55$_{(+66.01\%)}$  \\
\hline
\end{tabular}
\vspace{-0.1in}
\end{table*}}

\vspace{-0.00in}
\begin{table*}[t]
\caption{Expected inference error of LP and LP + TransProtect (km)}
\vspace{-0.00in}
\label{Tb:exp:EIELP}
\centering
\small 
\begin{tabular}{ c|c|c|c|c|c|c||c|c|c|c|c|c}
\toprule
\multicolumn{1}{ c  }{}& \multicolumn{12}{ c }{Location privacy protection algorithms} \\
\cline{2-13}
\multicolumn{1}{ c|  }{Location}
 & \multicolumn{6}{ |c|| }{\textbf{Experiment I}}
 & \multicolumn{6}{ |c }{\textbf{Experiment II}}
 \\ 
\cline{2-13}
\multicolumn{1}{ c|  }{inference}
&\multicolumn{6}{ |c|| }{LP}
&\multicolumn{6}{ |c }{LP+\textbf{TransProtect}}\\
\cline{2-13}
\multicolumn{1}{ c|  }{algorithms}  
& \multicolumn{3}{ c|  }{Rome }
& \multicolumn{3}{ c||  }{San Francisco }
& \multicolumn{3}{ c|  }{Rome }
& \multicolumn{3}{ c  }{San Francisco }\\ 
\cline{1-13}
\multicolumn{1}{ c|  }{$\epsilon$ (km$^{-1}$)}  & 5.0 & 7.5 & 10.0 & 
5.0 & 7.5 & 10.0 & 5.0 & 7.5 & 10.0 & 
5.0 & 7.5 & 10.0 \\
\hline
\hline
\multicolumn{1}{ c|  }{ \textbf{VehiTrack}} & \textbf{0.21} & \textbf{0.21} & \textbf{0.19}& \textbf{0.21} & \textbf{0.20}& \textbf{0.20} & \textbf{0.29}$_{(+34.9\%)}$ & \textbf{0.26}$_{(+24.9\%)}$ & \textbf{0.25}$_{(+32.7\%)}$  &\textbf{0.26}$_{(+22.43\%)}$  & \textbf{0.25}$_{(+23.73\%)}$  & \textbf{0.25}$_{(+24.58\%)}$  \\ 
\multicolumn{1}{ c|  }{VehiTrack-I} & 0.30 & 0.26 & 0.25 & 0.32  & 0.31 & 0.32 & 0.39$_{(+30.8\%)}$ & 0.35$_{(+34.9\%)}$ & 0.33$_{(+31.6\%)}$ & 0.44$_{(+37.56\%)}$ & 0.42$_{(+34.27\%)}$   & 0.40$_{(+25.53\%)}$  \\
\multicolumn{1}{ c|  }{Bayes} & 0.40 & 0.37 & 0.31 & 0.41 & 0.40 & 0.40 & 0.45$_{(+14.2\%)}$ & 0.44$_{(+19.5\%)}$ & 0.42$_{(+32.7\%)}$ & 0.64$_{(+53.96\%)}$ & 0.63$_{(+54.70\%)}$  & 0.62$_{(+56.64\%)}$  \\
\multicolumn{1}{ c|  }{HMM} & 0.30 & 0.26 & 0.23 & 0.34  &0.34  & 0.33 & 0.45$_{(+50.3\%)}$ & 0.43$_{(+65.6\%)}$ & 0.41$_{(+78.1\%)}$ & 0.57$_{(+67.51\%)}$  &0.56$_{(+64.26\%)}$  &0.55$_{(+66.01\%)}$  \\
\hline
\end{tabular}
\vspace{-0.1in}
\end{table*}

\DEL{
\vspace{-0.00in}
\begin{table*}[t]
\caption{Expected inference error of Laplace and Laplace + TransProtect (km). ?? To update by Sourabh}
\vspace{-0.10in}
\label{Tb:exp:EIELaplaceSan}
\centering
\small 
\begin{tabular}{ c|c|c|c||c|c|c }
\hline
\multicolumn{1}{ c  }{}& \multicolumn{6}{ c }{Location privacy protection algorithms} \\
\cline{2-7}
\multicolumn{1}{ c|  }{}
 & \multicolumn{3}{ |c|| }{\textbf{Experiment I}}
 & \multicolumn{3}{ |c }{\textbf{Experiment II}}
 \\  
\cline{2-7}
\multicolumn{1}{ c|  }{}
 & \multicolumn{3}{ |c|| }{Laplace}
 & \multicolumn{3}{ |c }{Laplace+\textbf{TransProtect}}
 \\ 
\cline{2-7}
\multicolumn{1}{ c|  }{Location inference algorithms}  & $\epsilon = 5.0$km$^{-1}$& $\epsilon = 7.5$km$^{-1}$& $\epsilon = 10.0$km$^{-1}$& $\epsilon = 5.0$km$^{-1}$& $\epsilon = 7.5$km$^{-1}$& $\epsilon = 10.0$km$^{-1}$\\ 
\hline
\hline
\multicolumn{1}{ c|  }{ \textbf{VehiTrack}} & {\rd \textbf{0.2234}} & \textbf{0.2205} & \textbf{0.2143} & {\rd \textbf{0.3898$_{(+74.5\%)}$}
} & \textbf{0.3621}$_{(+64.22\%)}$ & \textbf{0.3398}$_{(+58.56\%)}$ \\ 
\multicolumn{1}{ c|  }{VehiTrack-I} &0.3134 & 0.3102& 0.3021
& 0.5342$_{(+70.45\%)}$ & 0.5253$_{(+69.34\%)}$& 0.5176$_{(+71.33\%)}$\\
\multicolumn{1}{ c|  }{Bayes} & 0.5128 & 0.5087 & 0.5012& 0.7015$_{(+36.79\%)}$ & 0.6895$_{(+35.54\%)}$ & 0.6487$_{(+29.42\%)}$ \\
\multicolumn{1}{ c|  }{HMM} & 0.3019 & 0.3001& 0.2784& 0.6754$_{(+123.71\%)}$ & 0.6634$_{(+121.05\%)}$ & 0.6412$_{(+130.31\%)}$\\
\hline
\end{tabular}
\vspace{-0.10in}
\end{table*}

\vspace{-0.00in}
\begin{table*}[t]
\caption{Expected inference error of LP and LP + TransProtect (km) ?? To update by Sourabh}
\vspace{-0.10in}
\label{Tb:exp:EIELPSan}
\centering
\small 
\begin{tabular}{ c|c|c|c||c|c|c}
\hline
\multicolumn{1}{ c  }{}& \multicolumn{6}{ c }{Location privacy protection algorithms} \\
\cline{2-7}
\multicolumn{1}{ c|  }{}
 & \multicolumn{3}{ |c|| }{\textbf{Experiment I}}
 & \multicolumn{3}{ |c }{\textbf{Experiment II}}
 \\ 
\cline{2-7}
\multicolumn{1}{ c|  }{}
&\multicolumn{3}{ |c|| }{LP}
&\multicolumn{3}{ |c }{LP+\textbf{TransProtect}}
 \\ 
\cline{2-7}
\multicolumn{1}{ c|  }{Location inference algorithms}  & $\epsilon = 5.0$km$^{-1}$& $\epsilon = 7.5$km$^{-1}$& $\epsilon = 10.0$km$^{-1}$& $\epsilon = 5.0$km$^{-1}$& $\epsilon = 7.5$km$^{-1}$& $\epsilon = 10.0$km$^{-1}$\\ 
\hline
\hline
\multicolumn{1}{ c|  }{ \textbf{VehiTrack}} & \textbf{0.2113} & \textbf{0.2031} & \textbf{0.1989} & \textbf{ 0.2587$_{(+22.43\%)}$
} & \textbf{0.2513}$_{(+23.73\%)}$ & \textbf{0.2478}$_{(+24.58\%)}$ \\ 
\multicolumn{1}{ c|  }{VehiTrack-I} &0.3189 & 0.3110& 0.3176
& 0.4387$_{+37.56\%)}$ & 0.4176$_{(+34.27\%)}$ & 0.3987$_{(+25.53\%)}$ \\
\multicolumn{1}{ c|  }{Bayes} & 0.4123 & 0.4042 & 0.3951& 0.6348$_{(+53.96\%)}$ & 0.6253$_{(+54.70\%)}$ & 0.6189$_{(+56.64\%)}$ \\
\multicolumn{1}{ c|  }{HMM} & 0.3423 & 0.3389& 0.3312& 0.5734$_{(+67.51\%)}$ & 0.5567$_{(+64.26\%)}$ & 0.5498$_{(+66.01\%)}$ \\
\hline
\end{tabular}
\vspace{-0.1in}
\end{table*}}

\DEL{
\vspace{-0.00in}
\begin{table*}[t]
\caption{Expected inference error (km)}
\vspace{-0.10in}
\label{Tb:exp:EIE}
\centering
\begin{tabular}{ c|c|c|c|c|c|c||c|c|c|c|c|c }
\hline
\multicolumn{1}{ c  }{}& \multicolumn{12}{ c }{Location privacy protection algorithms} \\
\cline{2-13}
\multicolumn{1}{ c|  }{Location}
 & \multicolumn{6}{ |c|| }{\textbf{Exp. I}}
 & \multicolumn{6}{ |c }{\textbf{Exp. II}}
 \\ 
\cline{2-13}
\multicolumn{1}{ c|  }{inference}
 & \multicolumn{3}{ |c| }{Laplace}&\multicolumn{3}{ |c|| }{LP}
 & \multicolumn{3}{ |c| }{Laplace+\textbf{TransProtect}}&\multicolumn{3}{ |c }{LP+\textbf{TransProtect}}
 \\ 
\cline{2-13}
\multicolumn{1}{ c|  }{algorithm}  & $\epsilon = 5$& $\epsilon = 7.5$& $\epsilon = 10$& $\epsilon = 5$& $\epsilon = 7.5$& $\epsilon = 10$& $\epsilon = 5$& $\epsilon = 7.5$& $\epsilon = 10$& $\epsilon = 5$& $\epsilon = 7.5$& $\epsilon = 10$\\ 
\hline
\hline
\multicolumn{1}{ c|  }{ \textbf{VehiTrack}} & \textbf{0.2167} & \textbf{0.2144} & \textbf{0.2116} & \textbf{0.2146} & \textbf{0.2073}& \textbf{0.1893} & \textbf{0.3324
} & \textbf{0.3199} & \textbf{0.3101} & \textbf{0.2894} & \textbf{0.2589} & \textbf{0.2512} \\ 
\multicolumn{1}{ c|  }{VehiTrack-I} &0.6299 & 0.6214& 0.6166
&0.2983& 0.2604&0.2489& 0.6545 & 0.6501& 0.6474&0.3901& 0.3513&0.3276 \\
\multicolumn{1}{ c|  }{Bayes} & 0.6478 & 0.6392 & 0.6365& 0.3969 & 0.3674 & 0.3145 & 0.6854 & 0.6515 & 0.6396&0.4534& 0.4389&0.4173 \\
\multicolumn{1}{ c|  }{HMM} & 0.2563 & 0.2476& 0.2466&0.2987& 0.2598&0.2318& 0.7047 & 0.6895& 0.6812
&0.4489& 0.4303&0.4129 \\
\hline
\end{tabular}
\end{table*}}

\vspace{-0.1in}
\subsubsection{Benchmarks} In {\bf Experiment I}, we test the performance of VehiTrack against the two conventional geo-obfuscation methods
\newline \textbf{(i)  Planar Laplacian noise} (labeled as ``{\bf Laplace}'') \cite{Wang-WWW2017}, which uses $\epsilon$-Geo-Ind as the privacy criterion. Laplace assumes the obfuscation probabilities $z_{i,k}  \propto e^{-\epsilon\frac{d_{i, k}}{\Lambda_{\max}}}$, where $\epsilon$ is the \emph{privacy budget}, and $\Lambda_{\max}$ is the maximum distance between any two locations in the target region. 
\newline \textbf{(ii) LP-based geo-obfuscation} (labeled as ``{\bf LP}'') \cite{Qiu-TMC2020}: LP (defined in Equ. (\ref{eq:OMGobj})) aims to minimize the data utility loss of a single vehicle with the $\epsilon$-GeoInd constraints being satisfied.
\vspace{0.02in}

We compare the performance of VehiTrack with the following four classic location inference algorithms. 
\newline \textbf{(i) Bayesian Inference attack} \cite{Yu-NDSS2017}, labeled as ``{\bf Bayes}'': Given the vehicle's obfuscated location, Bayes derives the posterior of the vehicle's real location by the Bayes' formula and estimates the vehicle's real location as the location that maximizes the posterior. 
\newline \textbf{(ii) Hidden Markov Model-based location inference} \cite{Qiu-SIGSPATIAL2022}, labeled as ``{\bf HMM}'': HMM assumes the vehicle's mobility follows a Markov process, of which the transition matrix can be learned explicitly using publicly accessible traffic flow data \cite{Yan-IoTJ2022}. In HMM, the vehicle's real locations and obfuscated locations are considered \emph{hidden states} and \emph{observable states}, respectively. Under these assumptions, the vehicle's real trajectory can be recovered from the obfuscated locations by the Viterbi algorithm \cite{Viterbi-TIT1967}. 
\newline \textbf{(iii) VehiTrack in Phase 1}, labeled as ``{\bf VehiTrack-I}'': To conduct an {\bf ablation study}, we test VehiTrack-I, wherein the vehicles' locations are inferred directly using the posterior sequence generated by Phase 1 of VehiTrack. A comparative analysis between VehiTrack and VehiTrack-I allows us to assess the extent of improvement attributed to the incorporation of LSTM in Phase 2. 

\vspace{0.03in}
In {\bf Experiment II}, we integrate TransProtect into Laplace and LP, labeled as ``{\bf Laplace+TransProtect}'' and ``{\bf LP+TransProtect}'', respectively. Specifically, we limit the obfuscation range of Laplace and LP to the candidate location set output by TransProtect. We test the four inference algorithms when vehicles' locations are protected by ``Laplace+TransProtect'' and ``LP+TransProtect''.

\vspace{-0.08in}
\subsubsection{Metrics} In both Experiments I\&II, we measure two metrics: (1) \emph{expected inference errors (EIE)}, which is defined as the expected error between the estimated locations (by attackers) and the vehicles' actual locations, and (2) \emph{data utility loss}, which is defined as the expected distortion of estimated traveling cost (in Equ. (\ref{eq:utilityloss})). 
\vspace{-0.00in}

The main experimental results regarding inference error and data utility loss in Experiments I and II are listed in Table \ref{Tb:exp:EIELaplace}\&\ref{Tb:exp:EIELP} and Table \ref{Tb:exp:QL}, respectively. 

\vspace{-0.00in}
\subsection{Experiment I: Evaluation of VehiTrack}
\label{subsec:expVehiTrack}
\vspace{-0.02in}
We randomly select 100 trajectories from the Rome and San Francisco Taxicab datasets to simulate the vehicles' mobility. We use Laplace and LP to obfuscate all the locations within each trajectory, with locations recorded approximately every 20 seconds for both datasets. We then apply the four location inference algorithms, VehiTrack, VehiTrack-I, HMM, and Bayes to infer the vehicles' real locations from the obfuscated locations, of which the expected inference errors are compared in ``Experiment I'' in Table \ref{Tb:exp:EIELaplace} (Laplace) and Table \ref{Tb:exp:EIELP} (LP). Based on the two tables, we have the following observations: \looseness = -1

\vspace{0.02in}
\noindent \textbf{(1) Context-free location inference method Bayes has the highest inference error}. On average,
if we apply Laplacian noise as obfuscation methods, the inference error of Bayes is respectively 199.19\%, 2.97\%, and 156.33\% (resp. 131.81\%, 64.51\%, and 70.00\%) higher than that of VehiTrack, VehiTrack-I, and HMM using the Rome dataset (resp. using the San Francisco dataset). 
If we apply LP as obfuscation methods, the inference error of Bayes is 76.10\%, 36.60\%, and 36.65\% (resp. 95.23\%, 28.12\%, and 36.66\%) higher than that of VehiTrack, VehiTrack-I, and HMM using the Rome dataset (resp. using the San Francisco dataset).  
Unlike Bayes mainly focusing on single-location inference, the four context-aware inference methods achieve lower inference errors since they all account for the vehicles' location correlation using either the road network mobility model (VehiTrack and VehiTrack-I) or Markov Model (HMM).  \looseness = -1
In addition, Fig. \ref{fig:percentagedrop} in Appendix shows that, on average, VehiTrack-I (and also VehiTrack) eliminates 81.69\% (resp. 89.39\%)of locations across 100 trajectories of Rome dataset (resp. San Francisco dataset) by considering vehicles' mobility restrictions due to the road network. This substantial reduction aids attackers in narrowing down the search range for the vehicles' actual locations. \looseness = -1

\DEL{
\begin{figure*}[t]
\centering
\begin{minipage}{0.25\textwidth}
\centering
  \subfigure{
\includegraphics[width=1.00\textwidth, height = 0.13\textheight]{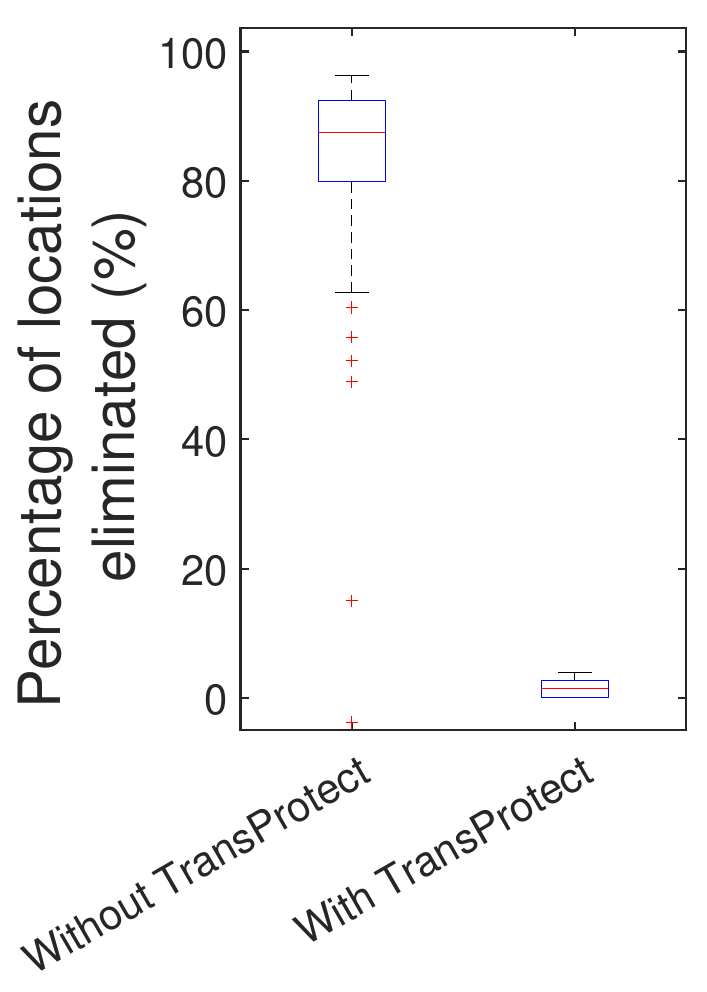}}
\vspace{-0.08in}
\caption{Percentage of locations eliminated in VehiTrack-I. }
\label{fig:percentagedrop}
\end{minipage}
\vspace{-0.12in}
\hspace{0.03in}
\begin{minipage}{0.70\textwidth}
\centering
  \subfigure[When $\epsilon = 5.0$km$^{-1}$.]{
\includegraphics[width=0.30\textwidth, height = 0.12\textheight]{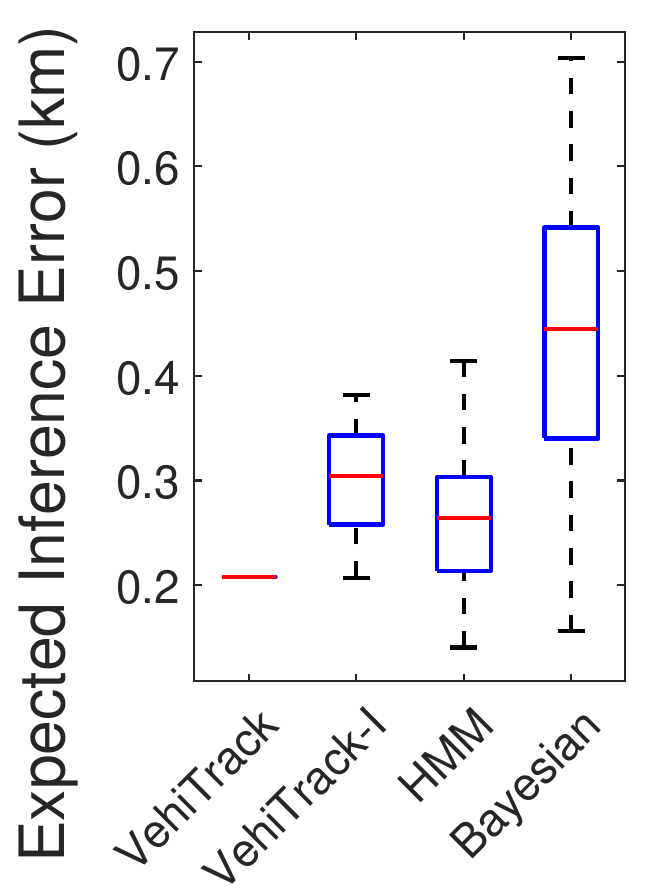}}
  \subfigure[When $\epsilon = 7.5$km$^{-1}$.]{
\includegraphics[width=0.30\textwidth, height = 0.12\textheight]{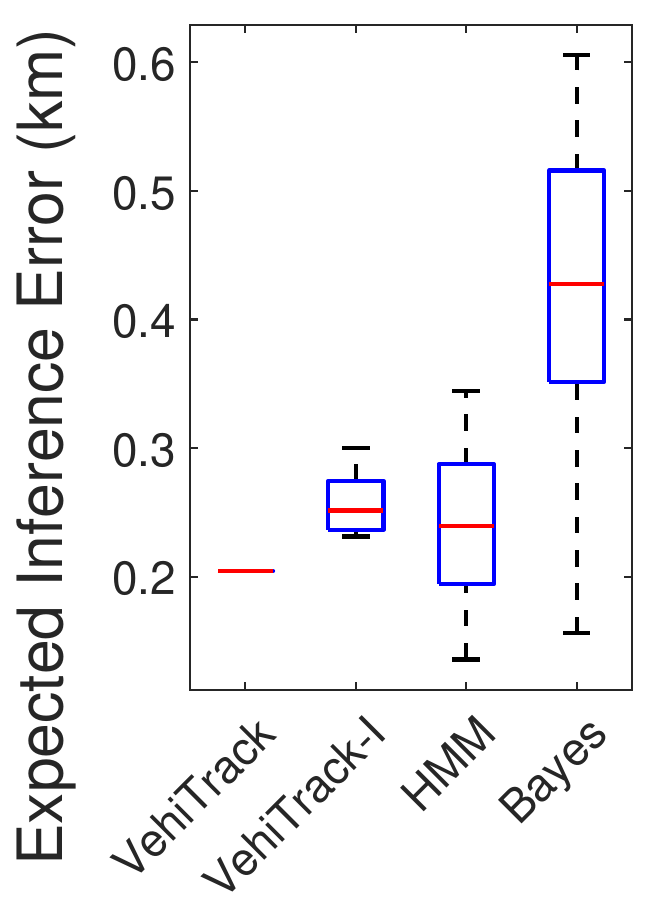}}
  \subfigure[When $\epsilon = 10$km$^{-1}$.]{
\includegraphics[width=0.30\textwidth, height = 0.12\textheight]{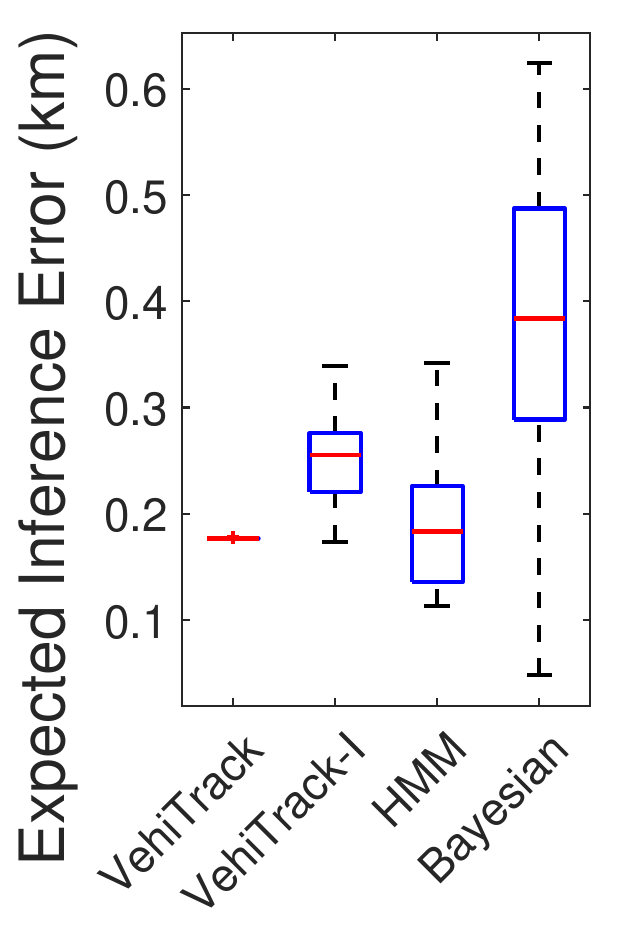}}
\vspace{-0.14in}
\caption{Comparison of expected inference error of different location inference algorithms when the length of trajectories $\geq 40$.}
\label{fig:EIEVehiTrackvsBayes}
\end{minipage}
\vspace{-0.00in}
\end{figure*}}

\vspace{0.01in}
\noindent \textbf{(2) VehiTrack achieves an even lower expected inference error compared to the Markov-based method HMM}. On average, using the Rome dataset (resp. the San Francisco dataset), the EIE of VehiTrack is 14.34\% and 22.22\% (resp. 26.66\% and 38.23\%) lower than that of HMM when Laplacian and LP are applied, respectively. 
HMM has higher inference error because assuming Markov property in HMM can only capture the correlation of vehicles' locations in adjacent time slots (short-term), while LSTM in Phase 2 of VehiTrack can additionally capture the long-term correlations of vehicles' locations, further improving the VehiTrack's inference accuracy. \looseness = -1

\vspace{0.01in}
\noindent \textbf{(3) VehiTrack outperforms VehiTrack-I in terms of inference accuracy (Ablation study)}. By comparing VehiTrack and VehiTrack-I, we find that LSTM in Phase 2 further reduces the average inference error by 65.58\% and 24.12\% for the Rome dataset and by 29.03\% and 34.37\% for the San Francisco dataset, considering both obfuscation methods (Laplacian noise and LP). Like HMM, VehiTrack-I achieves higher inference error, since it only captures short-term correlations within the location sequence by considering vehicles' mobility restrictions due to the road network conditions, but without considering long-term correlation between locations. \looseness = -1

To demonstrate that VehiTrack can better capture the long-term correlation of vehicles' locations compared to the benchmarks, among the 100 trajectories, we pick up trajectories that have more than 40 locations. In Fig. \ref{fig:EIEVehiTrackvsBayesRM}(a)(b)(c) and Fig. \ref{fig:EIEVehiTrackvsBayesSF}(a)(b)(c) in Appendix, we exclusively evaluate the inference errors of the four algorithms for these selected ``long'' trajectories. The depicted results in the figure highlight that the accuracy advantage of VehiTrack is even more significant compared to the findings in Table \ref{Tb:exp:EIELaplace} and Table \ref{Tb:exp:EIELP}, e.g., using the Rome dataset (resp. San Francisco dataset), VehiTrack's inference error is {\textbf{51.36\%, 33.33\%, and 48.56\%}} (resp. {\textbf{41.93\%, 25\%, and 29.16\%}}) lower than that of Bayes, VehiTrack-I, and HMM, respectively (consider that for all the 100 trajectories of rome dataset, VehiTrack's inference error is 49.77\%, 41.41\%, and 41.52\% lower than that of Bayes, VehiTrack-I, and HMM, respectively). 


\DEL{
\vspace{-0.00in}
\begin{table}[t]
\caption{Average data utility loss (km)}
\vspace{-0.00in}
\label{Tb:exp:QL}
\centering
\small
\begin{tabular}{ c|c||c}
\hline
\multicolumn{1}{ c  }{}& \multicolumn{2}{ c }{Location privacy protection algorithms} \\
\cline{2-3}
\multicolumn{1}{ c|  }{}
 & \multicolumn{1}{ |c|| }{{\bf Experiment I}}
 & \multicolumn{1}{ |c }{{\bf Experiment II}}
 \\ 
\cline{2-3}
\multicolumn{1}{ c|  }{Privacy budget}
 & \multicolumn{1}{ |c|| }{Laplace}
 & \multicolumn{1}{ |c }{Laplace+TransProtect}
 \\ 
\hline
\hline
\multicolumn{1}{ c|  }{ $\epsilon = 50$} & 0.1819 $\pm$ 0.5313 & \textbf{0.2142}  $\pm$ 0.1946\\ 
\multicolumn{1}{ c|  }{ $\epsilon = 100$} & 0.1811 $\pm$ 0.4973 & \textbf{0.1648}  $\pm$ 0.1247\\ 
\multicolumn{1}{ c|  }{ $\epsilon = 150$} & 0.1654 $\pm$ 0.4667 & \textbf{0.1377}  $\pm$ 0.8545\\ 
\hline
\end{tabular}
\vspace{-0.00in}
\end{table}}


\DEL{
\vspace{-0.00in}
\begin{table}[t]
\caption{Expected data utility loss of different methods (km)}
\vspace{-0.00in}
\label{Tb:exp:QL}
\centering
\begin{tabular}{ c|c|c||c|c}
\hline
\multicolumn{1}{ c  }{}& \multicolumn{4}{ c }{Location privacy protection algorithms} \\
\cline{2-5}
\multicolumn{1}{ c|  }{}
 & \multicolumn{2}{ |c|| }{{\bf Exp. I}}
 & \multicolumn{2}{ |c }{{\bf Exp. II}}
 \\ 
\cline{2-5}
\multicolumn{1}{ c|  }{$\epsilon$}
 & \multicolumn{1}{ |c| }{Laplace}&\multicolumn{1}{ |c|| }{LP}
 & \multicolumn{1}{ |c| }{Lap.+\textbf{TransProt.}}&\multicolumn{1}{ |c }{LP+\textbf{TransProt.}}
 \\ 

\hline
\multicolumn{1}{ c|  }{ 5.0} & 0.2423 & 0.5313 & \textbf{0.2530}$_{(+4.41\%)}$ & \textbf{0.5835}$_{(+9.82\%)}$ \\ 
\multicolumn{1}{ c|  }{ 7.5} & 0.2418 & 0.4659 & \textbf{0.2480}$_{(+2.56\%)}$ & \textbf{0.4921}$_{(+5.62\%)}$ \\ 
\multicolumn{1}{ c|  }{ 10.0} & 0.2412 & 0.2944 & \textbf{0.2431}$_{(+0.79\%)}$ & \textbf{0.3837}$_{(+30.33\%)}$ \\ 
\hline
\end{tabular}
\vspace{-0.20in}
\end{table}}

\DEL{\vspace{-0.00in}
\begin{table}[t]
\caption{Expected data utility loss of different methods (km)}
\vspace{-0.10in}
\label{Tb:exp:QL}
\centering

\begin{tabular}{ c|c|c||c|c}
\hline
\multicolumn{1}{ c  }{}& \multicolumn{4}{ c }{Location privacy protection algorithms} \\
\cline{2-5}
\multicolumn{1}{ c|  }{$\epsilon$}
& \multicolumn{2}{ |c|| }{{\bf Exp. I}}
& \multicolumn{2}{ |c }{{\bf Exp. II}}
\\ 
\cline{2-5}
\multicolumn{1}{ c|  }{ (km$^{-1}$)}
 & \multicolumn{1}{ |c| }{Laplace}&\multicolumn{1}{ |c|| }{LP}
 & \multicolumn{1}{ |c| }{Lap.+\textbf{TransProt.}}&\multicolumn{1}{ |c }{LP+\textbf{TransProt.}}
 \\ 

\hline
\multicolumn{1}{ c|  }{ 5.0} & 0.2423 & 0.5313 & \textbf{0.2530}$_{(+4.41\%)}$ & \textbf{0.5835}$_{(+9.82\%)}$ \\ 
\multicolumn{1}{ c|  }{ 7.5} & 0.2418 & 0.4659 & \textbf{0.2480}$_{(+2.56\%)}$ & \textbf{0.4921}$_{(+5.62\%)}$ \\ 
\multicolumn{1}{ c|  }{ 10.0} & 0.2412 & 0.2944 & \textbf{0.2431}$_{(+0.79\%)}$ & \textbf{0.3837}$_{(+30.33\%)}$ \\ 
\hline
\end{tabular}
\vspace{-0.20in}
\end{table}}

\DEL{
\vspace{-0.00in}
\begin{table}[t]
\caption{Expected data utility loss of different methods (km)}
\vspace{-0.10in}
\label{Tb:exp:QL}
\centering
\small 
\begin{tabular}{ c|c|c|c|c||c|c|c|c}
\hline
\multicolumn{1}{ c  }{}& \multicolumn{8}{ c }{Location privacy protection algorithms} \\
\cline{2-9}
\multicolumn{1}{ c|  }{$\epsilon$}
& \multicolumn{4}{ |c|| }{{\bf Exp. I}}
& \multicolumn{4}{ |c }{{\bf Exp. II}}
\\ 

\cline{2-9}
\multicolumn{1}{ c|  }{ (km$^{-1}$)}
 & \multicolumn{2}{ |c| }{Laplace}&\multicolumn{2}{ |c|| }{LP}
 & \multicolumn{2}{ |c| }{Lap.+\textbf{TransP.}}&\multicolumn{2}{ |c }{LP+\textbf{TransP.}}
 \\ 
\cline{1-9}
\multicolumn{1}{ c|  }{}
& RM & SF & RM & SF & RM & SF & RM & SF
\\ 
\hline
\hline
\multicolumn{1}{ c|  }{ 5.0} & 0.242 & 0.285 & 0.531 & 0.498 & \textbf{0.253} & \textbf{0.307} & \textbf{0.584} & \textbf{0.433} \\ 
\multicolumn{1}{ c|  }{ 7.5} & 0.242 & 0.283 & 0.466 & 0.430 & \textbf{0.248} & \textbf{0.304} & \textbf{0.492} & \textbf{0.417} \\ 
\multicolumn{1}{ c|  }{ 10.0} & 0.241 & 0.281 & 0.294 & 0.398 & \textbf{0.243} & \textbf{0.301} & \textbf{0.384} & \textbf{0.399} \\ 
\hline
\end{tabular}
\vspace{-0.20in}
\end{table}}

\vspace{-0.00in}
\begin{table}[t]
\caption{Expected data utility loss of different methods (km)}
\vspace{-0.00in}
\label{Tb:exp:QL}
\centering
\small 
\begin{tabular}{ c|c|c|c|c||c|c|c|c}
\hline
\multicolumn{1}{ c  }{}& \multicolumn{8}{ c }{Location privacy protection algorithms} \\
\cline{2-9}
\multicolumn{1}{ c|  }{$\epsilon$}
& \multicolumn{4}{ |c|| }{{\bf Exp. I}}
& \multicolumn{4}{ |c }{{\bf Exp. II}}
\\ 

\cline{2-9}
\multicolumn{1}{ c|  }{ (km$^{-1}$)}
 & \multicolumn{2}{ |c| }{Laplace}&\multicolumn{2}{ |c|| }{LP}
 & \multicolumn{2}{ |c| }{Lap.+\textbf{TransP.}}&\multicolumn{2}{ |c }{LP+\textbf{TransP.}}
 \\ 
\cline{1-9}
\multicolumn{1}{ c|  }{}
& RM & SF & RM & SF & RM & SF & RM & SF
\\ 
\hline
\hline
\multicolumn{1}{ c|  }{ 5.0} & 0.24 & 0.29 & 0.53 & 0.50 & \textbf{0.25} & \textbf{0.31} & \textbf{0.58} & \textbf{0.43} \\ 
\multicolumn{1}{ c|  }{ 7.5} & 0.24 & 0.28 & 0.47 & 0.43 & \textbf{0.25} & \textbf{0.30} & \textbf{0.49} & \textbf{0.42} \\ 
\multicolumn{1}{ c|  }{ 10.0} & 0.24 & 0.28 & 0.29 & 0.40 & \textbf{0.24} & \textbf{0.30} & \textbf{0.38} & \textbf{0.40} \\ 
\hline
\end{tabular}
\vspace{-0.15in}
\end{table}

\DEL{
\vspace{-0.00in}
\begin{table}[t]
\rd{\caption{Expected data utility loss of different methods (km) ?? To update by Sourabh}}
\vspace{-0.10in}
\label{Tb:exp:QLSan}
\centering
\small 
\begin{tabular}{ c|c|c||c|c}
\hline
\multicolumn{1}{ c  }{}& \multicolumn{4}{ c }{Location privacy protection algorithms} \\
\cline{2-5}
\multicolumn{1}{ c|  }{$\epsilon$}
& \multicolumn{2}{ |c|| }{{\bf Exp. I}}
& \multicolumn{2}{ |c }{{\bf Exp. II}}
\\ 
\cline{2-5}
\multicolumn{1}{ c|  }{ (km$^{-1}$)}
 & \multicolumn{1}{ |c| }{Laplace}&\multicolumn{1}{ |c|| }{LP}
 & \multicolumn{1}{ |c| }{Lap.+\textbf{TransProt.}}&\multicolumn{1}{ |c }{LP+\textbf{TransProt.}}
 \\ 

\hline
\multicolumn{1}{ c|  }{ 5.0} & 0.2847 & 0.4978 & \textbf{0.3065} & \textbf{0.4328} \\ 
\multicolumn{1}{ c|  }{ 7.5} & 0.2825 & 0.4298 & \textbf{0.3042} & \textbf{0.4168} \\ 
\multicolumn{1}{ c|  }{ 10.0} & 0.2810 & 0.3983 & \textbf{0.3013} & \textbf{0.3992} \\ 
\hline
\end{tabular}
\vspace{-0.20in}
\end{table}}



\vspace{0.01in}
\noindent \textbf{(4) 
As $\epsilon$ increases, the inference errors of all four inference algorithms increase}. This is attributed to higher values of $\epsilon$ allowing for smaller deviations from obfuscated locations to actual locations. Consequently, this leads to a reduction in overall inference errors and also a lesser loss of data utility due to Laplacian noise and Linear Programming, as demonstrated in Table \ref{Tb:exp:QL}. \looseness = -1



\DEL{
{\bl 
Measure the ``expected inference error'' of Bayes, BA-Bayes, HMM, Traffic-Adapter, GRU, LSTM given different vehicle trajectories (different lengths) 
\newline Figure 2(a): location data is obfuscated by Laplace); 
\newline Figure 2(b): location data is obfuscated by LP.  
\newline Figure 3(a)(b): the correlation between the ``expected inference error'' and report frequency, length of trajectories; 
}}
\vspace{-0.00in}
\subsection{Experiment II: Evaluation of TransProtect}
\label{subsec:expTransProtect}
\vspace{-0.00in}
We apply TransProtect to refine the location set of geo-obfuscation and then assess the inference error of the four location inference algorithms, of which the results are shown in ``Experiment II'' in Table \ref{Tb:exp:EIELaplace} (Laplace) and Table \ref{Tb:exp:EIELP} (LP). By comparing the experimental results in Experiments I and II, we can check how much privacy improvement is contributed by TransProtect. In the tables, the subscript $_{(+a\%)}$ means the inference error is increased by $a\%$ after integrating TransProtect. We have the following observations:

\vspace{0.01in}
\noindent \textbf{(1) Integrating TransProtect in Laplace and LP increases the expected inference error of the context-aware inference algorithms}. On average, employing TransProtect increases the inference error of VehiTrack, VehiTrack-I, and HMM by 40.28\%, 18.39\%, and 119.41\% (resp. 65.77\%, 70.74\%, and 125\%) 
using Rome dataset (resp. San Francisco dataset). 
This is because the obfuscation range 
is restricted to the candidate locations (determined by TransProtect) that are difficult to distinguish from real locations using context-aware inference models. Particularly, Fig. \ref{fig:percentagedrop}(a)(b) in Appendix shows that with TransProtect integrated, on average using Rome dataset (resp. San Francisco), only 1.45\% (resp. 1.59\%) locations are eliminated in the obfuscation range by VehiTrack-I when vehicles' mobility restrictions are considered, making it difficult for attackers to narrow down the search range for the target locations. \looseness = -1

\vspace{0.01in}
\noindent \textbf{(2) The integration of TransProtect maintains the utility loss at an acceptable level}. This is attributed to TransProtect's inclination to choose locations with lower data utility loss, considering road network conditions. In contrast, the original obfuscation methods (i.e. Laplacian and LP) don't consider measuring data utility loss in the road network when selecting obfuscated locations. Consequently, TransProtect allows the selection of locations further away from the real location, 
with the data utility loss guaranteed at an acceptable level (as shown in Table \ref{Tb:exp:QL}, i.e., on average, using Rome dataset (resp. San Francisco dataset), TransProtect increases the data utility loss by 1.04, 1.02, 1.30 times (resp. all 1.07) when $\epsilon = $ 5.0km$^{-1}$, 7.5km$^{-1}$, and 10.0km$^{-1}$. 

\DEL{
Subsequently, we randomly select a location and depict the sets of locations associated with different set limits $K$ and various utility loss weights $\alpha$ in Fig. \ref{fig:HeatmapK} (for $K = 5, 10, 15$) and Fig. \ref{fig:HeatmapUtilityweight} (for $\alpha = 1, 10, 100$), respectively. In both figures, the heat colors of the locations represent the weighted scores, encompassing both probability scores and data utility losses, as determined by TransProtect. As expected, upon comparing Fig. \ref{fig:HeatmapK}(a)(b)(c), it becomes apparent that a higher location set limit results in the inclusion of more locations in the candidate location set. Additionally, by comparing Fig. \ref{fig:HeatmapUtilityweight}(a)(b)(c), it is evident that an increase in the utility loss weight leads to lower scores being assigned to locations with higher data utility loss.}

\DEL{
{\bl 
Measure the ``expected inference error'' of VehiTrack given the locations are obfuscated by Laplacian noise, LP, and TransProtect 
\newline Figure 4a: different privacy budget $\epsilon$; 
\newline Figure 4b: different weight (when ranking the locations based on their transformer's scores and utility loss);
\newline Figure 4c: different pool capacity $K$; 
\newline Figure 4d: different report frequency (we can determine the time period between each report, e.g., 1 minutes, 90 seconds, etc.); 
\newline Figure 4f: different trajectory lengths. 
}}

\DEL{
{\bl 
Measure the ``data utility loss'' of VehiTrack given the locations are obfuscated by Laplacian noise, LP, and the transformer; 
\newline Figure 5a: different privacy budget $\epsilon$; 
\newline Figure 5b: different weight (when ranking the locations based on their transformer's scores and utility loss);
\newline Figure 5c: different pool capacity $K$; 
\newline Figure 5d: different trajectory lengths.} }

\DEL{
\begin{figure}[t]
\centering
\begin{minipage}{0.155\textwidth}
\centering
  \subfigure[$K=5$]{
\includegraphics[width=1.00\textwidth, height = 0.11\textheight]{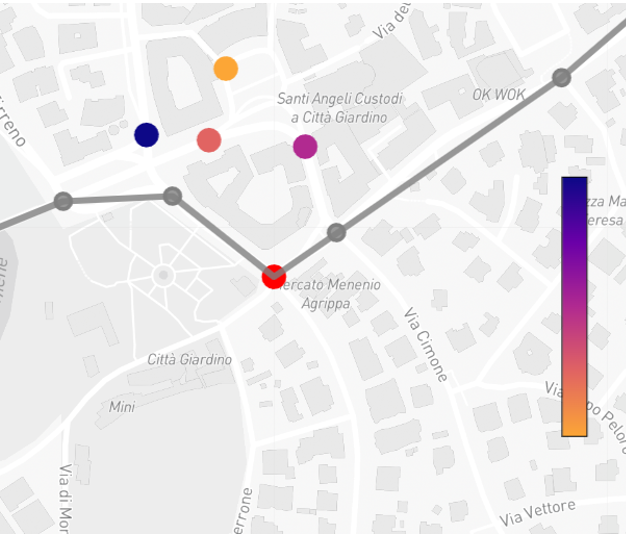}}
\end{minipage}
\begin{minipage}{0.155\textwidth}
\centering
  \subfigure[$K=10$]{
\includegraphics[width=1.00\textwidth, height = 0.11\textheight]{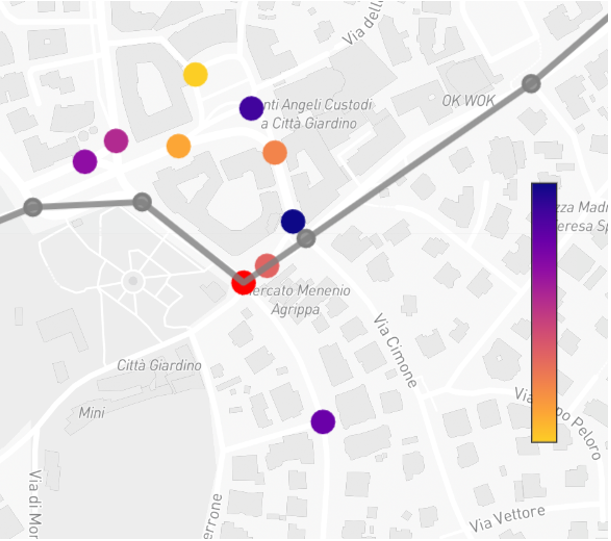}}
\end{minipage}
\hspace{0.00in}
\begin{minipage}{0.155\textwidth}
\centering
  \subfigure[$K=15$]{
\includegraphics[width=1.00\textwidth, height = 0.11\textheight]{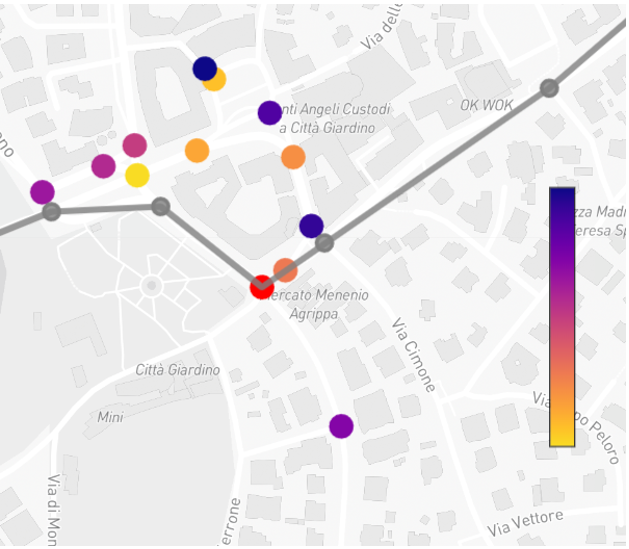}}
\end{minipage}
\caption{Heat map of location scores given different location set size $K$. }
\label{fig:HeatmapK}
\vspace{-0.10in}
\end{figure}

\begin{figure}[t]
\centering
\begin{minipage}{0.155\textwidth}
\centering
  \subfigure[$\alpha=1$]{
\includegraphics[width=1.00\textwidth, height = 0.11\textheight]{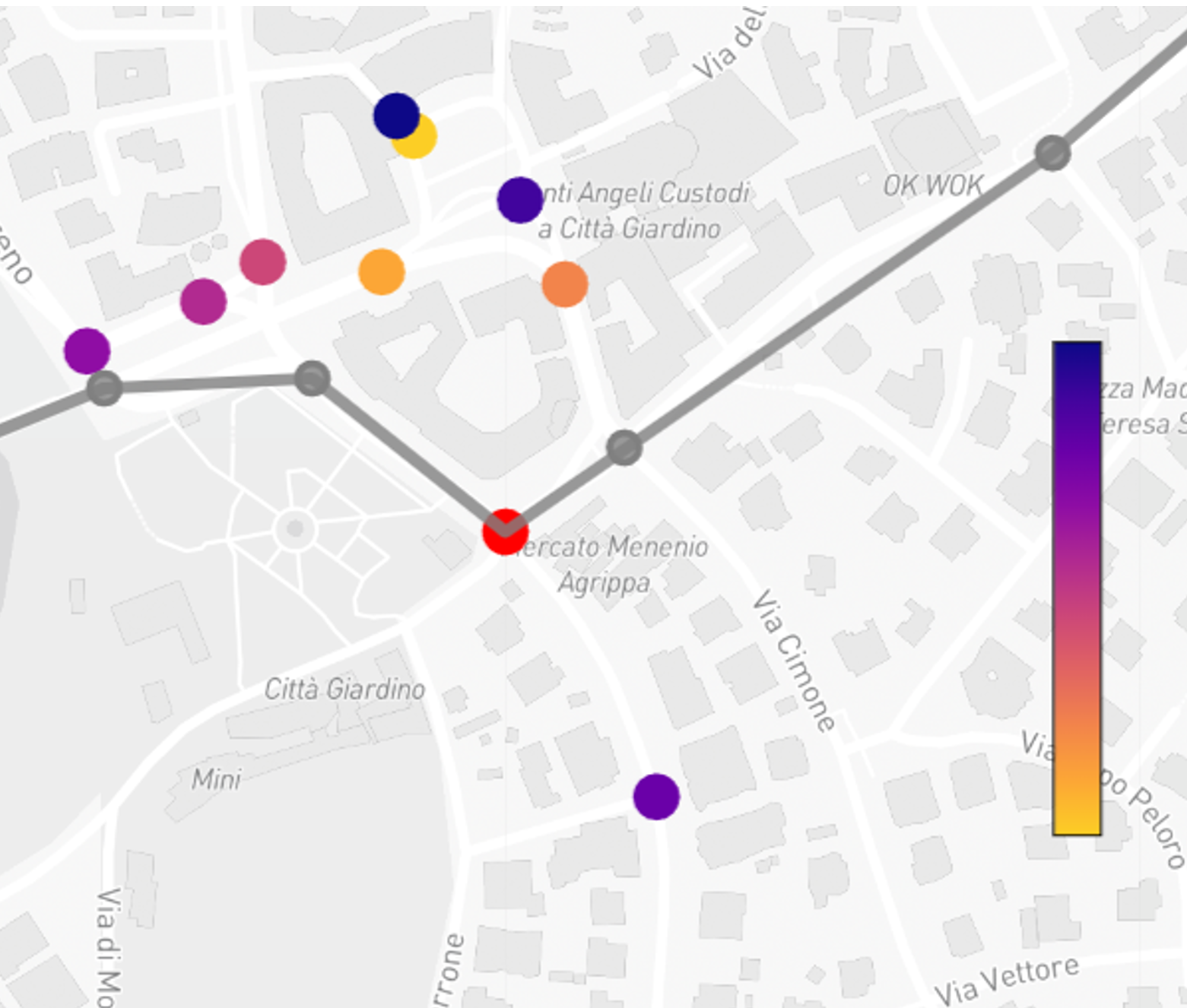}}
\end{minipage}
\begin{minipage}{0.155\textwidth}
\centering
  \subfigure[$\alpha=10$]{
\includegraphics[width=1.00\textwidth, height = 0.11\textheight]{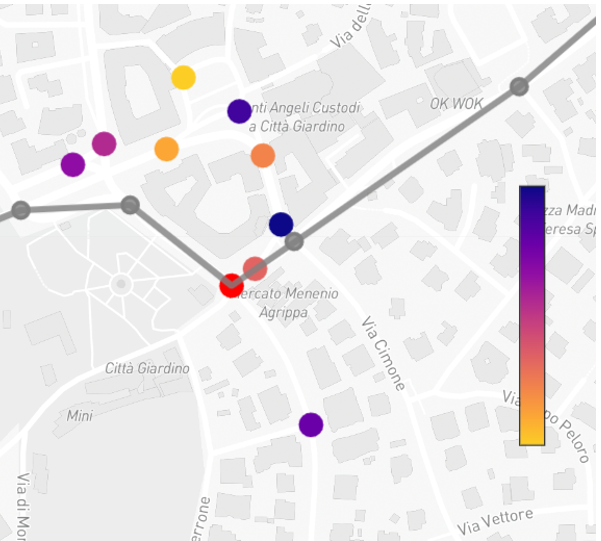}}
\end{minipage}
\hspace{0.00in}
\begin{minipage}{0.155\textwidth}
\centering
  \subfigure[$\alpha=100$]{
\includegraphics[width=1.00\textwidth, height = 0.11\textheight]{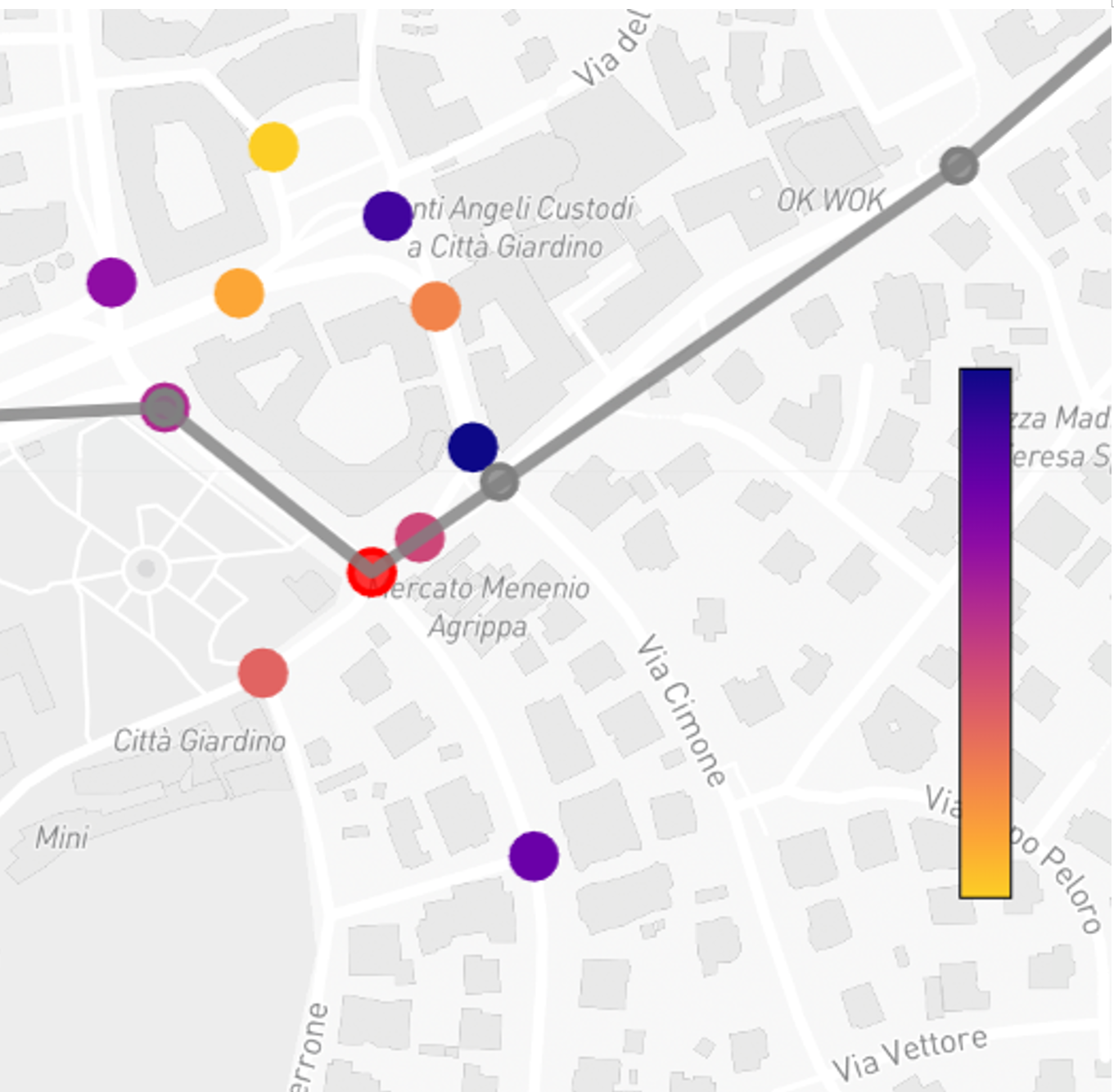}}
\end{minipage}
\caption{Heat map of location scores given different utility loss weight $\alpha$. }
\label{fig:HeatmapUtilityweight}
\vspace{-0.10in}
\end{figure}}

\vspace{-0.12in}
\section{Related Works}
\label{sec:related}
\vspace{-0.02in}
\noindent \textbf{Geo-Ind}. 
The discussion of location privacy criteria dates back nearly two decades to when Gruteser and Grunwald \cite{Gruteser-MobiSys2003} introduced location $k$-anonymity, based on Sweeney's $k$-anonymity \cite{Sweeney2002}. Recently, Andr{'e}s \emph{et al.} \cite{Andres-CCS2013} extended Differential Privacy (DP) to "Geo-Ind" for location privacy protection, spurring the development of new geo-obfuscation methods \cite{Shokri-CCS2012, Andres-CCS2013, Wang-WWW2017, Yu-NDSS2017}. For instance, Andr{'e}s \emph{et al.} \cite{Andres-CCS2013} developed a geo-obfuscation method adding noise from a polar Laplacian distribution to actual locations to achieve Geo-Ind. Considering the diverse sensitivity of data utility loss to obfuscation in LBS, other works discretize the location domain and optimize the obfuscation distribution using LP \cite{Bordenabe-CCS2014, Wang-WWW2017, Qiu-TMC2020, Qiu-CIKM2020, Qiu-ICDCS2019}.

\vspace{0.00in}
\noindent \textbf{Context-aware threat models}. Although effective in protecting sporadic locations, geo-obfuscation based on Geo-Ind is still vulnerable to context-aware inference attacks. Recent efforts have focused on attacking Geo-Ind using the spatiotemporal correlation of users' reported locations, either from a single user over time (e.g., trajectory) \cite{LIAO2007311,Xu-WWW2017,Cao-ICDE2017,Emrich-ICDE2012,Li-Sigspatial2008,Arain-MTA2018} or from multiple users \cite{Cao-ICDE2019,Li-SigSpatial2017}. Some works assume users' mobility follows a Markov process \cite{LIAO2007311, Emrich-ICDE2012}, where current locations depend on previous ones, e.g., 
\DEL{
Some inference algorithms focus on other statistical features of users' mobility (e.g., visiting frequency at certain interests of points \cite{Korkmaz-MIKE2018} or users' location spatial similarity \cite{Li-SigSpatial2017}). For example, Li et al. \cite{Li-VLDB2010} used a frequent mining approach to find moving objects that move within arbitrary shape of clusters for certain timestamps that are possibly nonconsecutive. 
Moreover, in the setting of aggregated data, some recent works have designed location or trajectory inference attacks from aggregated location data \cite{Xu-WWW2017} or proximity query results from location data \cite{Argyros-TPS2017}. \looseness = -1}
our prior work \cite{Qiu-SIGSPATIAL2022} tracks vehicles' locations using an HMM, where 
we learn 
the transition matrix of the Markov chain via publicly accessible traffic flow data. \looseness = -1

\vspace{0.00in}
\noindent \textbf{Context-aware LPPM}. Another approach to context-aware location privacy focuses on new privacy criteria and solutions to protect users' location data \cite{Cao-ICDE2017, Cao-ICDE2019, Ghinita-Sigspatial2009, Xiao-SIGSAC2015}. For instance, assuming attackers use Markov models for users' mobility, Cao \emph{et al.} \cite{Cao-ICDE2017} defined a criterion to quantify privacy levels of existing methods. Cao \emph{et al.} \cite{Cao-ICDE2019} extended DP to new criteria for spatiotemporal event privacy and created a framework to calculate privacy loss in location protection mechanisms. Considering temporal correlations, Xiao \emph{et al.} \cite{Xiao-SIGSAC2015} introduced $\delta$-location set-based DP and a planar isotropic mechanism for geo-obfuscation. In \cite{Qiu-SIGSPATIAL2022}, we proposed generating synthetic trajectories using Markov chain, making it harder for attackers to distinguish real from obfuscated locations using traffic flow information.  
While elegant, context-aware threat models and LPPMs primarily rely on explicit stochastic models like Markov chains, overlooking long-term correlations between locations hidden in context data. In contrast, VehiTrack and TransProtect use DNNs to uncover implicit relationships in sequence data, outperforming existing methods in location inference and privacy protection. \looseness = -1

\vspace{0.01in}
\noindent \textbf{Synthetic data-based privacy protection}. It is worth mentioning that several recent works have used synthetic data to protect users' privacy, especially in high-dimensional, sparse datasets prone to breaches \cite{arnold2021really, choi2018generating, Drechsler-JASA2010, opendatainstitute, UKGovernment, xu2019modeling, Yale2019PrivacyPS}. These studies focus on anonymization, preserving the statistical properties of original data while hiding users' identities \cite{choi2018generating} to protect personally identifiable information. In contrast, TransProtect uses synthetic data to increase indistinguishable pairs in the protected dataset, leading to different research challenges due to the divergent goals.

\vspace{-0.00in}
\section{Discussions and Conclusions}
\label{sec:conclude}
\vspace{-0.02in}
In this work, we studied the context-aware location privacy protection for vehicles in LBS. We introduced a new threat model \textbf{VehiTrack} to show the vulnerability of Geo-Ind. As a countermeasure, we then developed \textbf{TransProtect} to create candidate locations for obfuscation that are hard to distinguish from real locations (by VehiTrack). The simulation results have demonstrated the vulnerability of Geo-Ind to VehiTrack and the effectiveness of TransProtect in protecting vehicles' location privacy against VehiTrack. 

We envision new promising research directions to explore further. 
In addition to LSTM, transformer models provide an alternative method for VehiTrack to track the locations of vehicles. Transformer has demonstrated its strong capability not only in synthetic data generation but also in a variety of inference models \cite{CHITTYVENKATA2023102990}. However, incorporating Transformer into VehiTrack introduces some additional challenges to address. First, we will study how to use Multimodal Transformers \cite{xu2023multimodal} instead of Vanilla transformers, considering the different modalities in VehiTrack's input (location posterior sequence) and output (location sequences). Before directing location posteriors into the Transformer model, we will apply approximation or discretization techniques to map posterior sequences to lower dimensional feature space considering their high dimensions. \looseness = -1

\vspace{-0.00in}
\section{Acknowledgements}
\vspace{-0.02in}
The research was partially sponsored by the Army Research Laboratory and was accomplished under Cooperative Agreement Number W911NF-23-2-0014. The views and conclusions contained in this document are those of the authors and should not be interpreted as representing the official policies, either expressed or implied, of the Army Research Laboratory or the U.S. Government. The U.S. Government is authorized to reproduce and distribute reprints for Government purposes, not withstanding any copyright notation herein.
This research was partially supported by U.S. NSF grants CNS-2136948 and CNS-2313866.

\DEL{
\vspace{-0.02in}
\section{Acknowledgements}
\vspace{-0.02in}
The research was partially sponsored by the Army Research Laboratory and was accomplished under Cooperative Agreement Number W911NF-23-2-0014. The views and conclusions contained in this document are those of the authors and should not be interpreted as representing the official policies, either expressed or implied, of the Army Research Laboratory or the U.S. Government. The U.S. Government is authorized to reproduce and distribute reprints for Government purposes, not withstanding any copyright notation herein.

This work was partially supported by U.S. NSF grants CNS-2136948 and CNS-2313866.}

\clearpage
\bibliographystyle{unsrt}
\bibliography{mybib}

\newpage  
\appendix
\section{Appendix}
\vspace{-0.00in}

\DEL{
\subsection{Math Notations}

\begin{table}[h]
\caption{Main notations and their descriptions}
\vspace{-0.00in}
\label{Tb:Notationmodel}
\centering
\small 
\begin{tabular}{l l}
\hline
\hline
Symbol            & Description \\
\hline
$\mathcal{V}$       & Location set $\mathcal{V} = \{v_1, ..., v_L\}$ \\
$t_n$               & Time slot of the $n^{\mathrm{th}}$ location report \\
$\mathbf{Z}_{n}$    & Obfuscation matrix at time slot $t_n$\\
$z^{n}_{i,k}$       & Probability of taking $v_k$ as the obfuscated location \\ 
                      & given the true location $v_i$ at $t_n$\\ 
$x_{n}$             & Vehicle's true location at $t_n$\\ 
$\tilde{y}_{n}$             & Vehicle's obfuscated (reported)  location at $t_n$ \\
$\mathrm{UL}\left(\mathbf{Z}\right)$ & Utility loss due to the obfuscation matrix $\mathbf{Z}$ \\
$\mathcal{G} = (\mathcal{V}, \mathcal{E})$ & Road network, where $\mathcal{V}$ and $\mathcal{E}$ denote the node set \\
& and the edge set, respectively\\ 
$\mathcal{R}^i_{{n}}$ & Set of locations reachable by $v_i$ during $[t_{n-1}, t_{n}]$.\\ 
$\mathcal{S}_n$ & Set of vehicle's possible locations (identified by \\
& VehiTrack) at time slot $t_n$ \\
$\mathcal{R}_{{n}}$ & Set of locations reachable by any location in $\mathcal{S}_n$ \\
& during $[t_{n-1}, t_{n}]$.\\ 
$h_t$ & Hidden state vector of LSTM at time slot $t_n$ \\
\hline
\end{tabular}
\vspace{-0.15in}
\normalsize
\end{table}

\vspace{-0.00in}
\begin{table}[h]
\caption{Main notations in TransProtect}
\vspace{-0.00in}
\label{Tb:notationtransprotect}
\centering
\small 
\begin{tabular}{l l}
\hline
\hline
Symbol            & Description \\
\hline
$f_{N2V}(v_i)$       & Location embedding of location $v_i$ output by \\
& Node2Vec \\
$f_{N2V}(\mathbf{x})$       & Location embedding of trajectory $\mathbf{x}$ output by \\
& Node2Vec  \\
$f_{GCN}(\mathbf{x})$       & Lcation embedding of trajectory $\mathbf{x}$ output by GCN  \\
$f(\mathbf{x})$ & Final location embedding of trajectory $\mathbf{x}$ \\ 
& (after positional embedding) \\
$\mathcal{N}_{i}$       & Neighbor set of location $v_i$ \\
$g$       & Dimsion of embeddings \\
$h_{j,n}$ & $v_j$'s likelihood of being the real location at $t_n$ \\ 
$\hat{x}_{j,n}$ & The ground truth presence (or absence) of $v_j$ at \\
&  time slot $t_n$ in the trajectory. \\ 
\hline
\end{tabular}
\vspace{-0.10in}
\normalsize
\end{table}
}

\subsection{Proof of Proposition \ref{prop:SPT_i}}
\label{sec:proof_SPT_i}
\vspace{-0.00in}
\begin{proof}
For the sake of contradiction, we assume that there exists a location $v_k \in \mathcal{R}^i_{n}$, i.e., $c_{v_i,v_k} \leq t_n - t_{n-1}$, but not identified by $SPT_i$. There are two cases: 
\newline \textbf{Case 1}: $v_k \notin \mathcal{V}'_i$ and $c_{v_i,v_k} \leq t_n - t_{n-1}$. Due to the restriction of the road network, the travel cost from $v_i$ to $v_k$, denoted by $d'_{i,k}$, should be no smaller than $d_{i,k}$ (Haversine distance). Let $s_{i,k}$ denote a vehicle's average speed from $v_i$ to $v_k$ (note $s_{i,k} \leq s_{\mathrm{limit}}$), then we can obtain that $\frac{d_{i,k}}{s_{\mathrm{limit}}} \leq  \frac{d'_{i,k}}{s_{i,k}} = c_{v_i,v_k} \leq t_n - t_{n-1}$, indicating that $d_{i,k} \leq (t_n - t_{n-1})s_{\mathrm{limit}}$ and $v_k \in \mathcal{V}'_i$ by Equ. (\ref{eq:V_i}), which is a contradiction. 
\DEL{
\begin{eqnarray}
&&\frac{d_{i,k}}{s_{\mathrm{limit}}} \leq  \frac{d'_{i,k}}{s_{i,k}} = c_{v_i,v_k} \leq t_n - t_{n-1} \\ \nonumber 
&\Rightarrow& d_{i,k} \leq (t_n - t_{n-1})s_{\mathrm{limit}} \Rightarrow v_k \in \mathcal{V}'_i~ \mbox{(by Equ. (\ref{eq:V_i}))}
\end{eqnarray}
which is a contradiction.}  
\newline \textbf{Case 2}: $v_k \in \mathcal{V}'_i$ and $c_{v_i,v_k} \leq t_n - t_{n-1}$, but the travel cost from $v_i$ to $v_k$ in $SPT_i$ is larger than $t_n - t_{n-1}$. In this case, there must exist at least one location $v_l \in \mathcal{V}\backslash \mathcal{V}'_i$ that is in the shortest path from $v_i$ to $v_k$. Then, the travel cost from $v_i$ to $v_l$ in $\mathcal{G}$ is no larger than $t_n - t_{n-1}$ since $c_{v_i,v_l} = c_{v_i,v_k} - c_{v_l,v_k} \leq t_n - t_{n-1} - c_{v_l,v_k} \leq  t_n - t_{n-1}$, which is a contradiction that has been proved in \textbf{Case 1} (by considering $v_l$ as $v_k$). 
\end{proof}

\subsection{Additional Experimental Results}
\vspace{-0.10in}

\begin{table}[h]
\caption{Expected data utility loss (km) of Laplace and LP given different $K$ values}
\vspace{-0.10in}
\label{Tb:exp:QL_K}
\centering
\small 
\begin{tabular}{ c|c|c|c|c|c}
\toprule
\multicolumn{1}{ c  }{}& \multicolumn{5}{ c }{Expected data utility loss (km)} \\
\hline
\hline
\multicolumn{1}{ c|  }{}& \multicolumn{5}{ c }{Rome dataset} \\ 
\cline{2-6} 
\multicolumn{1}{ c|  }{}
 & \multicolumn{5}{ |c }{{\bf Laplace+TransProtect}}\\ 
\cline{2-6} 
\multicolumn{1}{ c|  }{ $\epsilon$ } &
$K = 5$ & $K = 10$ & $K = 15$ & $K = 20$ & $K = 25$ \\ 
\hline 
\multicolumn{1}{ c|  }{ 5.0km$^{-1}$} & 0.2530 & 0.3120 & 0.3923  & 0.5207 & 0.6532 \\ 
\multicolumn{1}{ c|  }{ 7.5km$^{-1}$} & 0.2480 & 0.3055 & 0.3892 & 0.5132 & 0.6498 \\ 
\multicolumn{1}{ c|  }{ 10.0km$^{-1}$} & 0.2431 & 0.2991 & 0.3822 & 0.5089 & 0.6412 \\ 
\hline
\cline{2-6} 
\multicolumn{1}{ c|  }{}
 & \multicolumn{5}{ |c }{{\bf LP+TransProtect}}\\ 
\cline{2-6} 
\multicolumn{1}{ c|  }{ $\epsilon$ } &
$K = 5$ & $K = 10$ & $K = 15$ & $K = 20$ & $K = 25$ \\ 
\hline 
\multicolumn{1}{ c|  }{ 5.0km$^{-1}$} & 0.4073 & 0.5835 & 	0.7983	 & 1.019 & 	1.378\\ 
\multicolumn{1}{ c|  }{ 7.5km$^{-1}$} & 0.3591	 & 0.4921	 & 0.6784	 & 0.916 & 	1.342\\ 
\multicolumn{1}{ c|  }{ 10.0km$^{-1}$} & 0.2963	 & 0.3837	 & 0.5429 & 	0.878 & 	1.336\\ 
\hline
\hline
\multicolumn{1}{ c|  }{}& \multicolumn{5}{ c }{San Francisco dataset} \\ 
\cline{2-6} 
\multicolumn{1}{ c|  }{}
 & \multicolumn{5}{ |c }{{\bf Laplace+TransProtect}}\\ 
\cline{2-6} 
\multicolumn{1}{ c|  }{ $\epsilon$ } &
$K = 5$ & $K = 10$ & $K = 15$ & $K = 20$ & $K = 25$ \\ 
\hline 
\multicolumn{1}{ c|  }{ 5.0km$^{-1}$} & 0.2750 & 0.3065 & 0.3516  & 0.5002 & 0.6614 \\ 
\multicolumn{1}{ c|  }{ 7.5km$^{-1}$} & 0.2430 & 0.3042 & 0.3441 & 0.4962 & 0.6535 \\ 
\multicolumn{1}{ c|  }{ 10.0km$^{-1}$} & 0.2391 & 0.3013 & 0.3413 & 0.4909 & 0.6489 \\ 
\hline
\cline{2-6} 
\multicolumn{1}{ c|  }{}
 & \multicolumn{5}{ |c }{{\bf LP+TransProtect}}\\ 
\cline{2-6} 
\multicolumn{1}{ c|  }{ $\epsilon$ } &
$K = 5$ & $K = 10$ & $K = 15$ & $K = 20$ & $K = 25$ \\ 
\hline 
\multicolumn{1}{ c|  }{ 5.0km$^{-1}$} & 0.4266 & 0.5028 & 	0.5373	 & 0.6743 & 	0.7136\\ 
\multicolumn{1}{ c|  }{ 7.5km$^{-1}$} & 0.4213	 & 0.5008	 & 0.5322	 & 0.6721 & 	0.7094\\ 
\multicolumn{1}{ c|  }{ 10.0km$^{-1}$} & 0.4083	 & 0.4992	 & 0.5257 & 	0.6648 & 	0.7025\\ 
\hline
\end{tabular}
\vspace{-0.00in}
\end{table}

\vspace{-0.00in}
\begin{table}[h]
\caption{Expected inference error (km) of Laplace and LP given different $K$ values for Rome dataset}
\vspace{-0.10in}
\label{Tb:exp:EIE_K}
\centering
\small 
\begin{tabular}{ c|c|c|c|c|c}
\toprule
\multicolumn{1}{ c  }{}& \multicolumn{5}{ c }{Expected inference error (km)} \\
\hline
\hline
\multicolumn{1}{ c|  }{}& \multicolumn{5}{ c }{Rome dataset} \\ 
\cline{2-6}
\multicolumn{1}{ c|  }{}
& \multicolumn{5}{ |c }{{\bf Laplace+TransProtect}}
\\ 
\cline{2-6} 
\multicolumn{1}{ c|  }{ $\epsilon$ } &
$K = 5$ & $K = 10$ & $K = 15$ & $K = 20$ & $K = 25$ \\ 
\hline
\multicolumn{1}{ c|  }{ 5.0km$^{-1}$} & 0.3324 & 	0.3515	 & 0.3846	 & 0.4123	 & 0.4532 \\ 
\multicolumn{1}{ c|  }{ 7.5km$^{-1}$} & 0.3148	 & 0.3389	 & 0.3698	 & 0.4087	 & 0.4435\\ 
\multicolumn{1}{ c|  }{ 10.0km$^{-1}$} & 0.3043	 & 0.3243	 & 0.3602	 & 0.3892	 & 0.4369 \\ 
\hline
\multicolumn{1}{ c|  }{}
 & \multicolumn{5}{ |c }{{\bf LP+TransProtect}}\\ 
\cline{2-6} 
\multicolumn{1}{ c|  }{ $\epsilon$ } &
$K = 5$ & $K = 10$ & $K = 15$ & $K = 20$ & $K = 25$\\ 
\hline
\multicolumn{1}{ c|  }{ 5.0km$^{-1}$} & 0.2829 & 0.2894 & 0.3129 & 0.3198 & 0.3301  \\ 
\multicolumn{1}{ c|  }{ 7.5km$^{-1}$} & 0.3301 & 0.2589 & 0.2983 & 0.3047 & 0.3193\\ 
\multicolumn{1}{ c|  }{ 10.0km$^{-1}$} & 0.2339 & 0.2512 & 0.2743 & 0.2983 & 0.3101\\ 
\hline
\hline
\multicolumn{1}{ c|  }{}& \multicolumn{5}{ c }{San Francisco dataset} \\ 
\cline{2-6}
\multicolumn{1}{ c|  }{}
& \multicolumn{5}{ |c }{{\bf Laplace+TransProtect}}
\\ 
\cline{2-6} 
\multicolumn{1}{ c|  }{ $\epsilon$ } &
$K = 5$ & $K = 10$ & $K = 15$ & $K = 20$ & $K = 25$ \\ 
\hline
\multicolumn{1}{ c|  }{ 5.0km$^{-1}$} & 0.3543 & 	0.3898	 & 0.4164	 & 0.4506	 & 0.4914 \\ 
\multicolumn{1}{ c|  }{ 7.5km$^{-1}$} & 0.3212	 & 0.3621	 & 0.4013	 & 0.4474	 & 0.4885\\ 
\multicolumn{1}{ c|  }{ 10.0km$^{-1}$} & 0.3053	 & 0.3398	 & 0.3972	 & 0.4403	 & 0.4834 \\ 
\hline
\multicolumn{1}{ c|  }{}
 & \multicolumn{5}{ |c }{{\bf LP+TransProtect}}\\ 
\cline{2-6} 
\multicolumn{1}{ c|  }{ $\epsilon$ } &
$K = 5$ & $K = 10$ & $K = 15$ & $K = 20$ & $K = 25$\\ 
\hline
\multicolumn{1}{ c|  }{ 5.0km$^{-1}$} & 0.2436 & 0.2587 & 0.2853 & 0.3081 & 0.3284  \\ 
\multicolumn{1}{ c|  }{ 7.5km$^{-1}$} & 0.2418 & 0.2513 & 0.2811 & 0.3013 & 0.3197\\ 
\multicolumn{1}{ c|  }{ 10.0km$^{-1}$} & 0.2394 & 0.2478 & 0.2785 & 0.2965 & 0.3137\\ 
\hline
\end{tabular}
\end{table}
\vspace{-0.10in}

\begin{figure*}[t]
\centering
\begin{minipage}{0.25\textwidth}
\centering
    \subfigure[Rome]{
\includegraphics[width=0.42\textwidth, height = 0.13\textheight]{./fig/exp/percentagedrop.pdf}
}
    \subfigure[San \newline Francisco]{
\includegraphics[width=0.42\textwidth, height = 0.13\textheight]{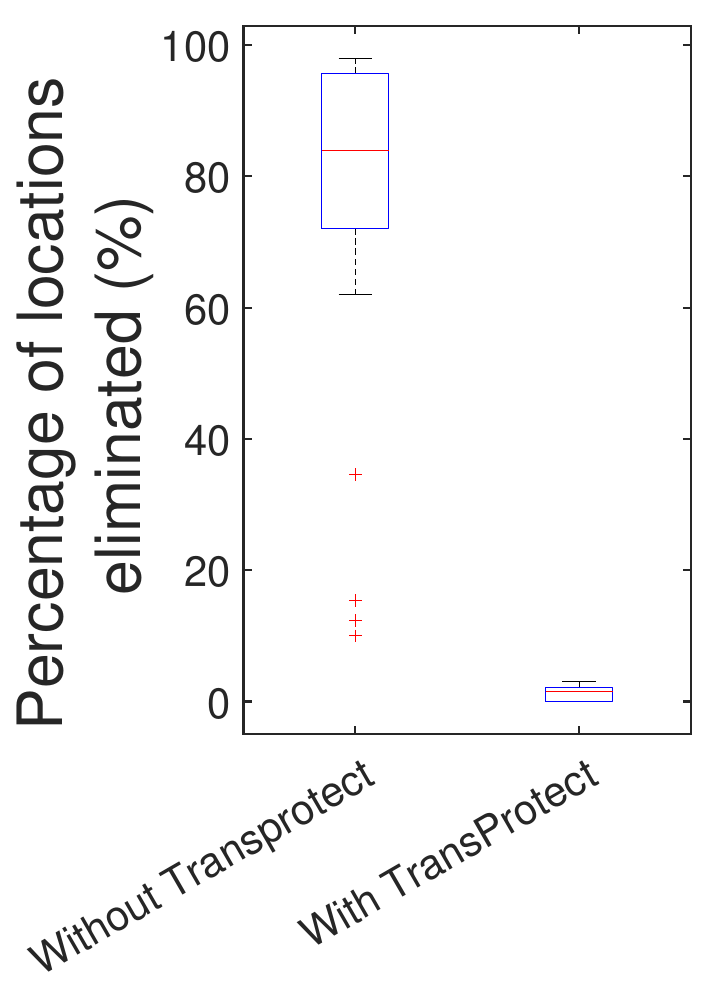}
}
\vspace{-0.18in}
\caption{Percentage of locations eliminated in VehiTrack-I. }
\label{fig:percentagedrop}
\end{minipage}
\vspace{-0.15in}
\hspace{0.03in}
\begin{minipage}{0.34\textwidth}
\centering
  \subfigure[\small $\epsilon = 5.0$km$^{-1}$]{
\includegraphics[width=0.30\textwidth, height = 0.12\textheight]{./fig/exp/EIE_e5_fig10}}
  \subfigure[\small $\epsilon = 7.5$km$^{-1}$]{
\includegraphics[width=0.30\textwidth, height = 0.12\textheight]{./fig/exp/EIE_e7_5_fig10}}
  \subfigure[\small $\epsilon = 10$km$^{-1}$]{
\includegraphics[width=0.30\textwidth, height = 0.12\textheight]{./fig/exp/EIE_e10_fig10}}
\vspace{-0.14in}
\caption{Comparison of EIE of different location inference algorithms when the length of trajectories $\geq 40$ (Rome).}
\label{fig:EIEVehiTrackvsBayesRM}
\end{minipage}
\hspace{0.05in}
\begin{minipage}{0.34\textwidth}
\centering
  \subfigure[\small $\epsilon = 5.0$km$^{-1}$]{
\includegraphics[width=0.30\textwidth, height = 0.12\textheight]{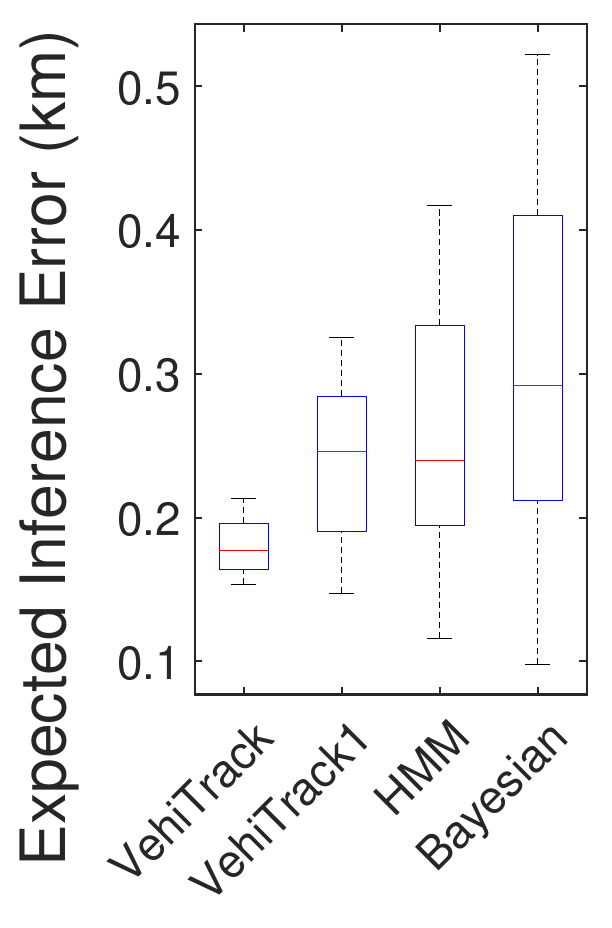}}
  \subfigure[\small $\epsilon = 7.5$km$^{-1}$]{
\includegraphics[width=0.30\textwidth, height = 0.12\textheight]{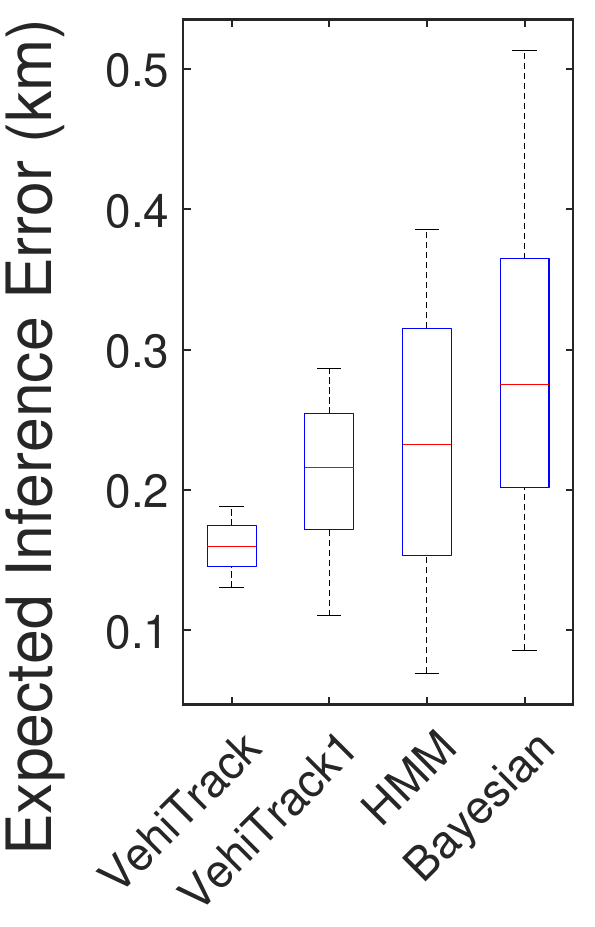}}
  \subfigure[\small $\epsilon = 10$km$^{-1}$]{
\includegraphics[width=0.30\textwidth, height = 0.12\textheight]{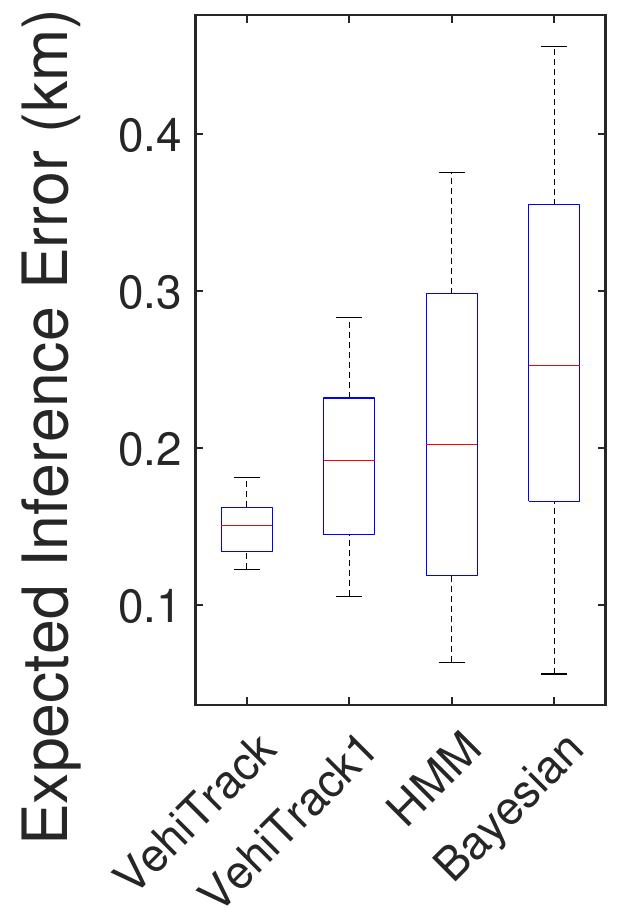}}
\vspace{-0.14in}
\caption{Comparison of EIE of different location inference algorithms when the length of trajectories $\geq 40$ (San Francisco).}
\label{fig:EIEVehiTrackvsBayesSF}
\end{minipage}
\vspace{-0.00in}
\end{figure*}

\begin{figure}[t]
\centering
\begin{minipage}{0.490\textwidth}
\centering
  \subfigure[Laplace + TransProtect]{
\includegraphics[width=0.48\textwidth, height = 0.135\textheight]
{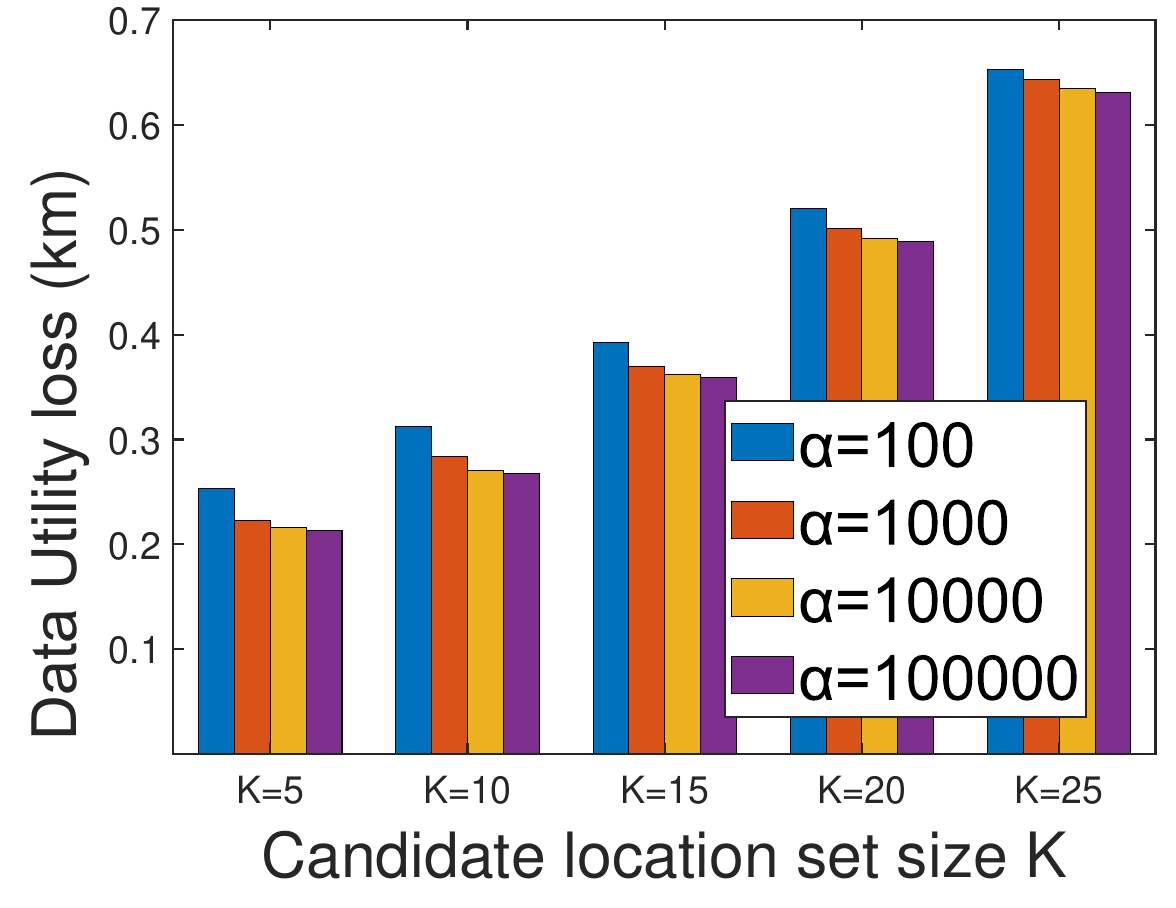}}
\centering
  \subfigure[LP + TransProtect]{
\includegraphics[width=0.48\textwidth, height = 0.135\textheight]{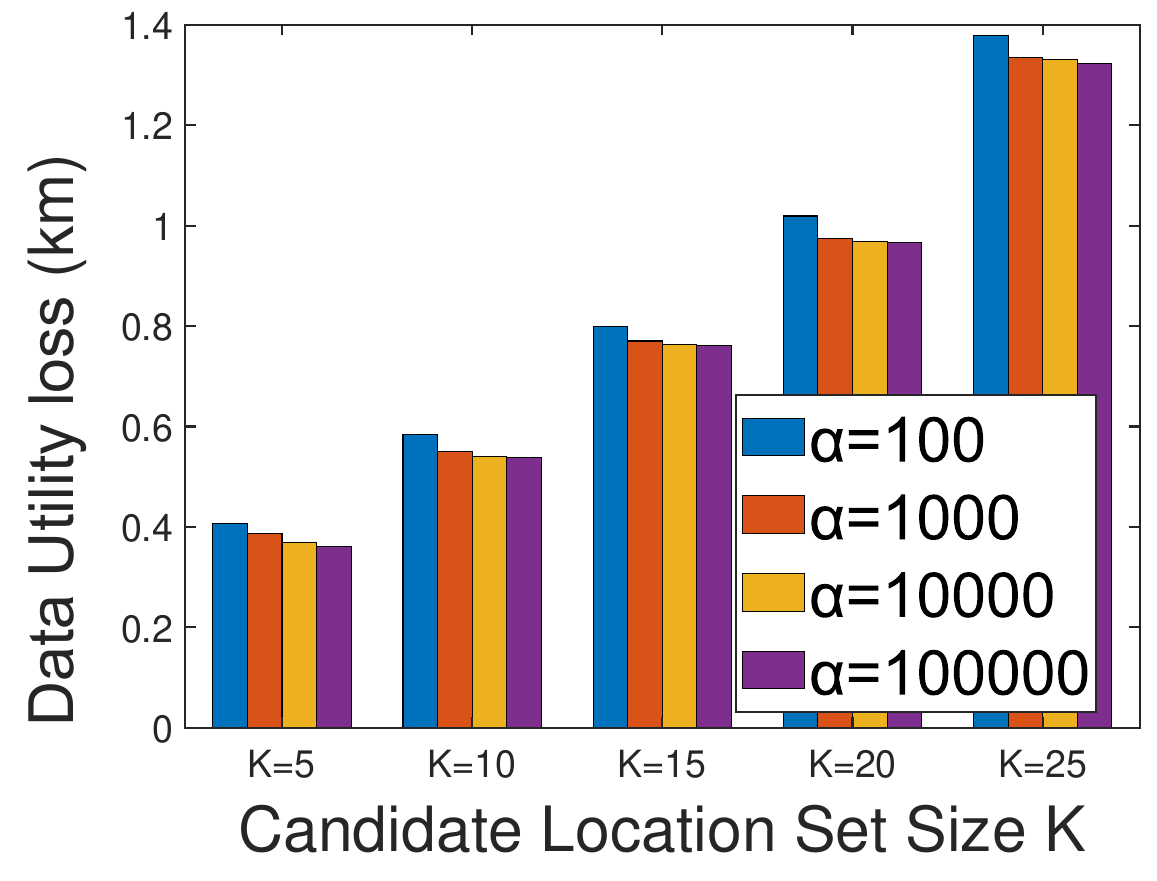}}
\vspace{-0.10in}
\caption{Impacet of $K$ (candidate location set size) and $\alpha$ (utility loss weight) on the data utility loss of TransProtect for Rome Dataset}
\label{fig:UL_K_weight_rome}
\end{minipage}
\hspace{0.00in}
\begin{minipage}{0.490\textwidth}
\centering
  \subfigure[Laplace + TransProtect]{
\includegraphics[width=0.48\textwidth, height = 0.135\textheight]{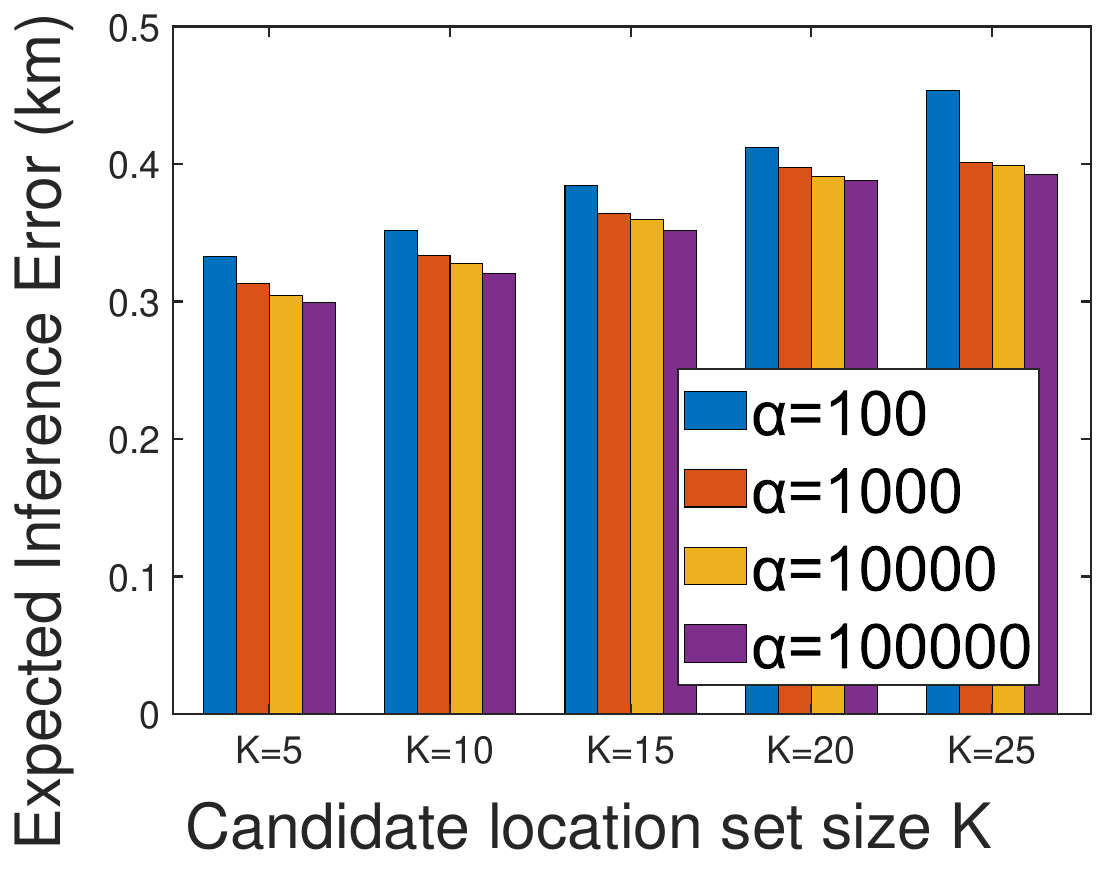}}
\centering
  \subfigure[LP + TransProtect]{
\includegraphics[width=0.48\textwidth, height = 0.135\textheight]{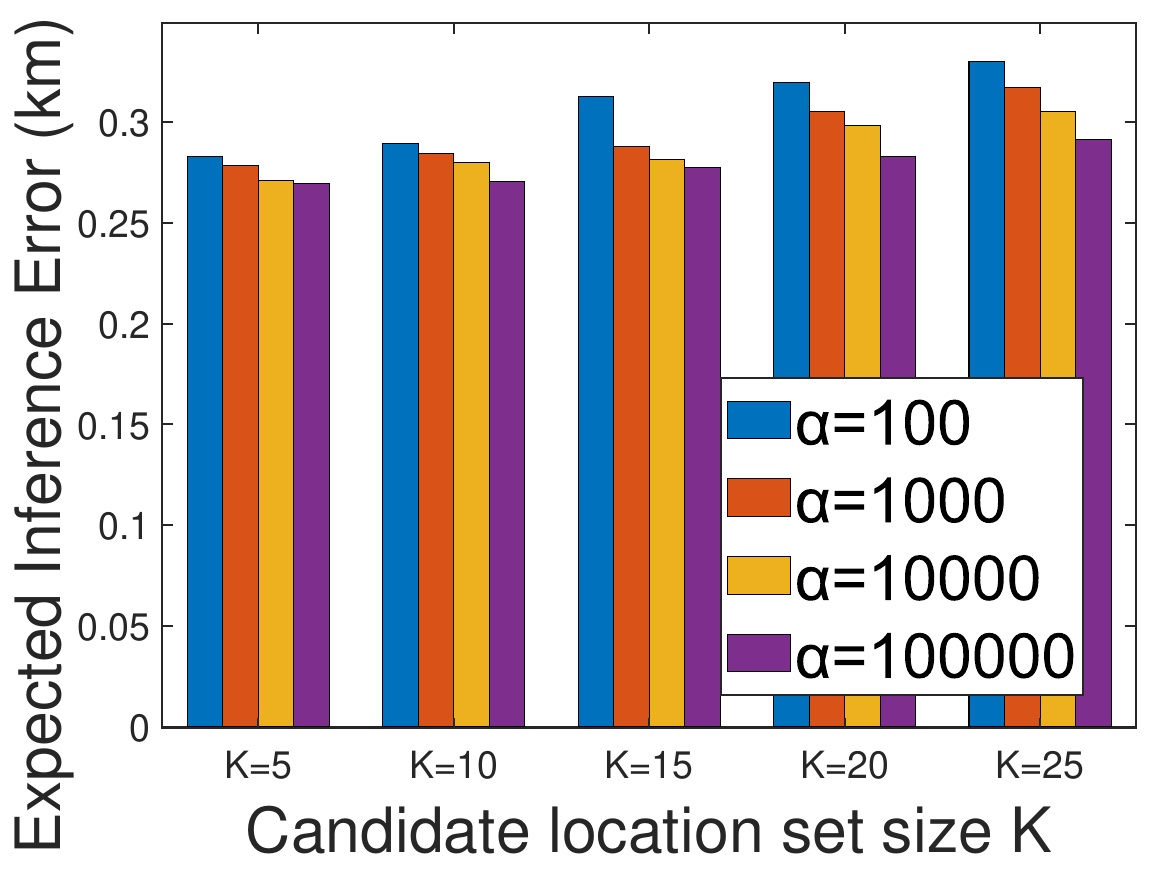}}
\vspace{-0.10in}
\caption{Impacet of $K$ (candidate location set size) and $\alpha$ (utility loss weight) on the expected inference error of TransProtect for the Rome dataset.}
\label{fig:EIE_K_weight_rome}
\end{minipage}
\vspace{-0.10in}
\end{figure}

\begin{figure}[t]
\centering
\begin{minipage}{0.490\textwidth}
\centering
  \subfigure[Laplace + TransProtect]{
\includegraphics[width=0.48\textwidth, height = 0.135\textheight]
{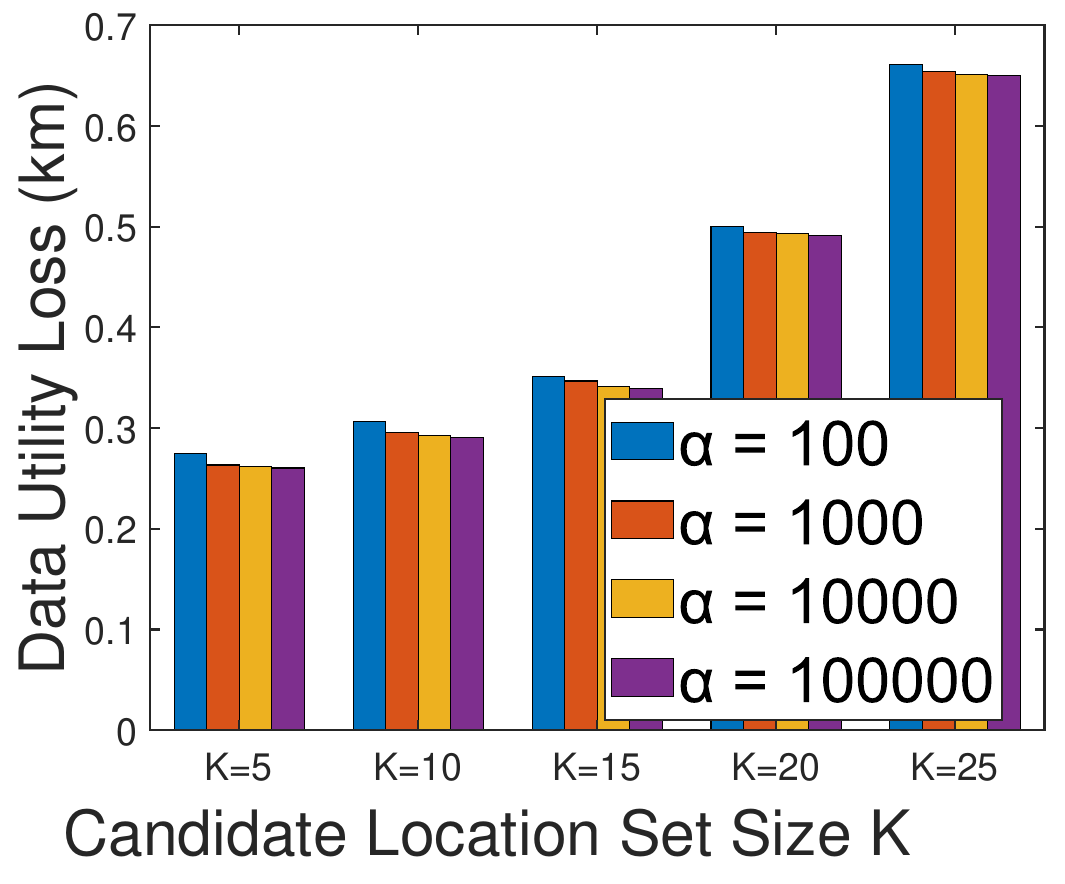}}
\centering
  \subfigure[LP + TransProtect]{
\includegraphics[width=0.48\textwidth, height = 0.135\textheight]{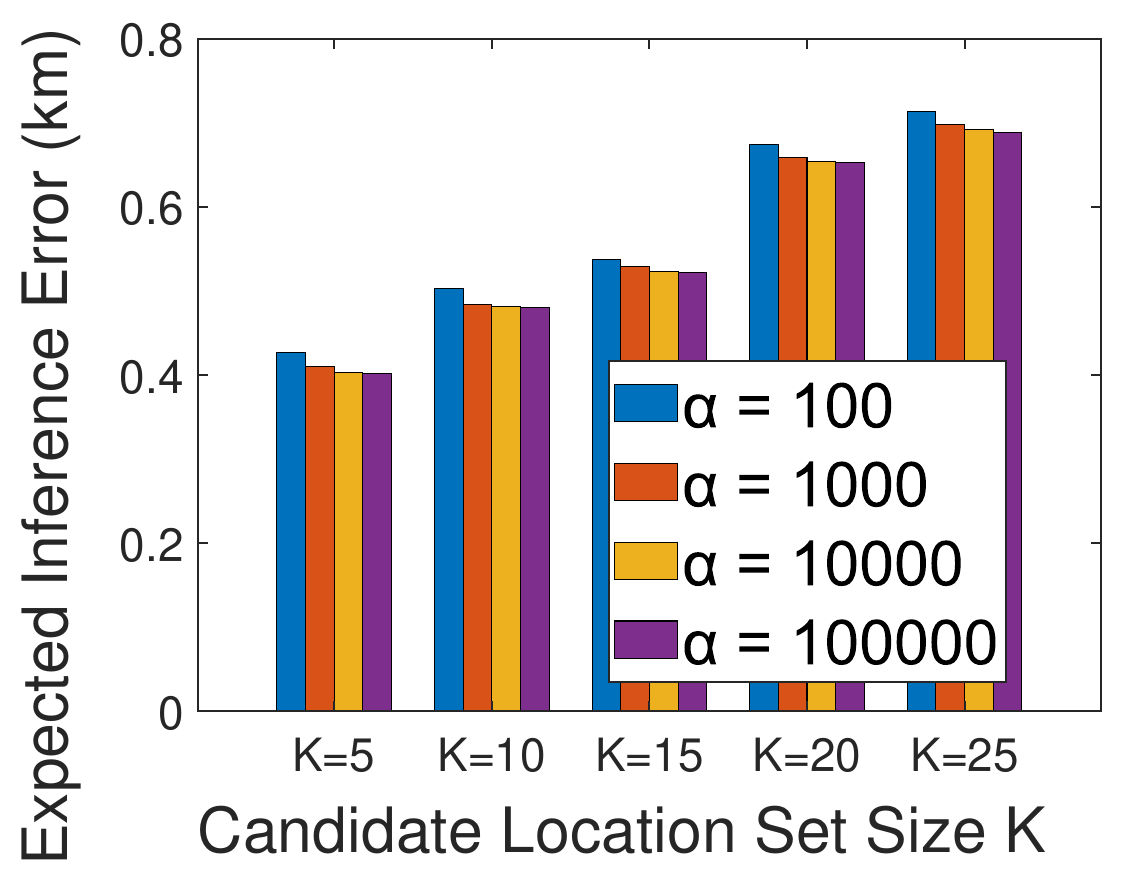}}
\vspace{-0.10in}
\caption{Impacet of $K$ (candidate location set size) and $\alpha$ (utility loss weight) on the data utility loss of TransProtect for SF Dataset}
\label{fig:UL_K_weight_SF}
\end{minipage}
\hspace{0.00in}
\begin{minipage}{0.490\textwidth}
\centering
  \subfigure[Laplace + TransProtect]{
\includegraphics[width=0.48\textwidth, height = 0.135\textheight]{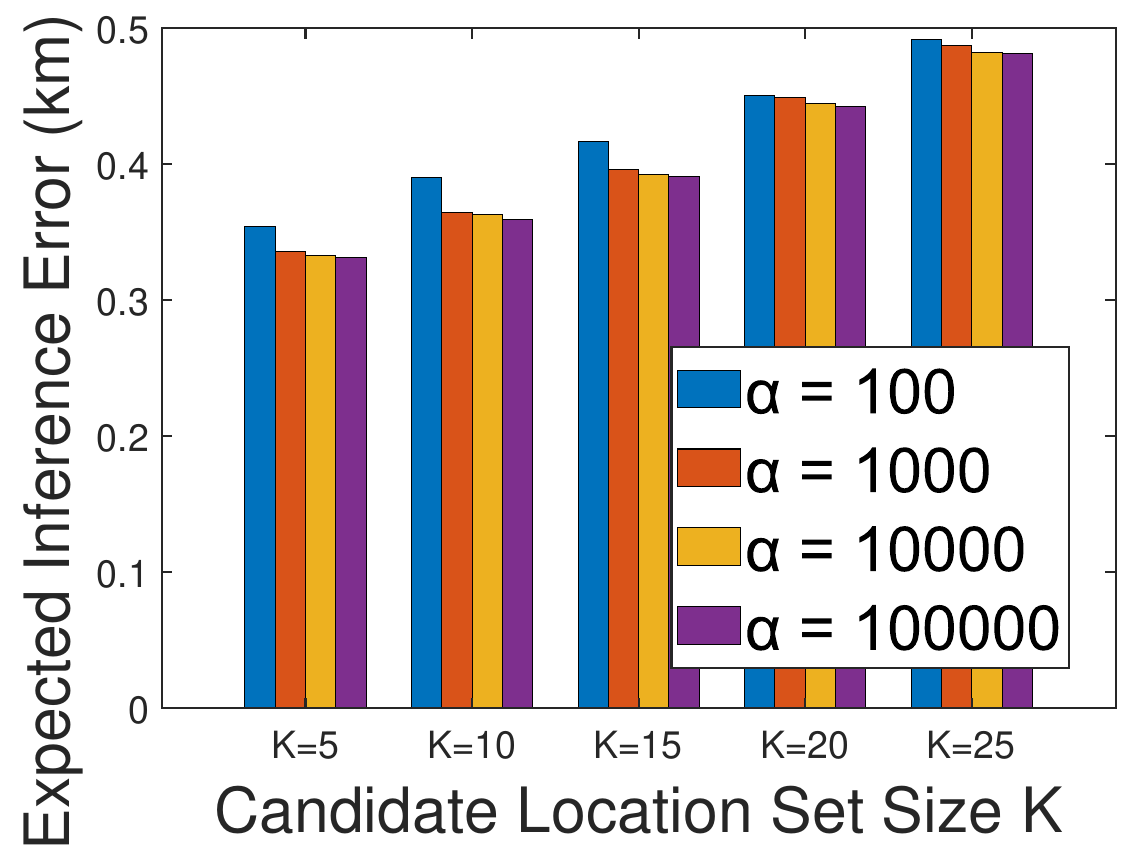}}
\centering
  \subfigure[LP + TransProtect]{
\includegraphics[width=0.48\textwidth, height = 0.135\textheight]{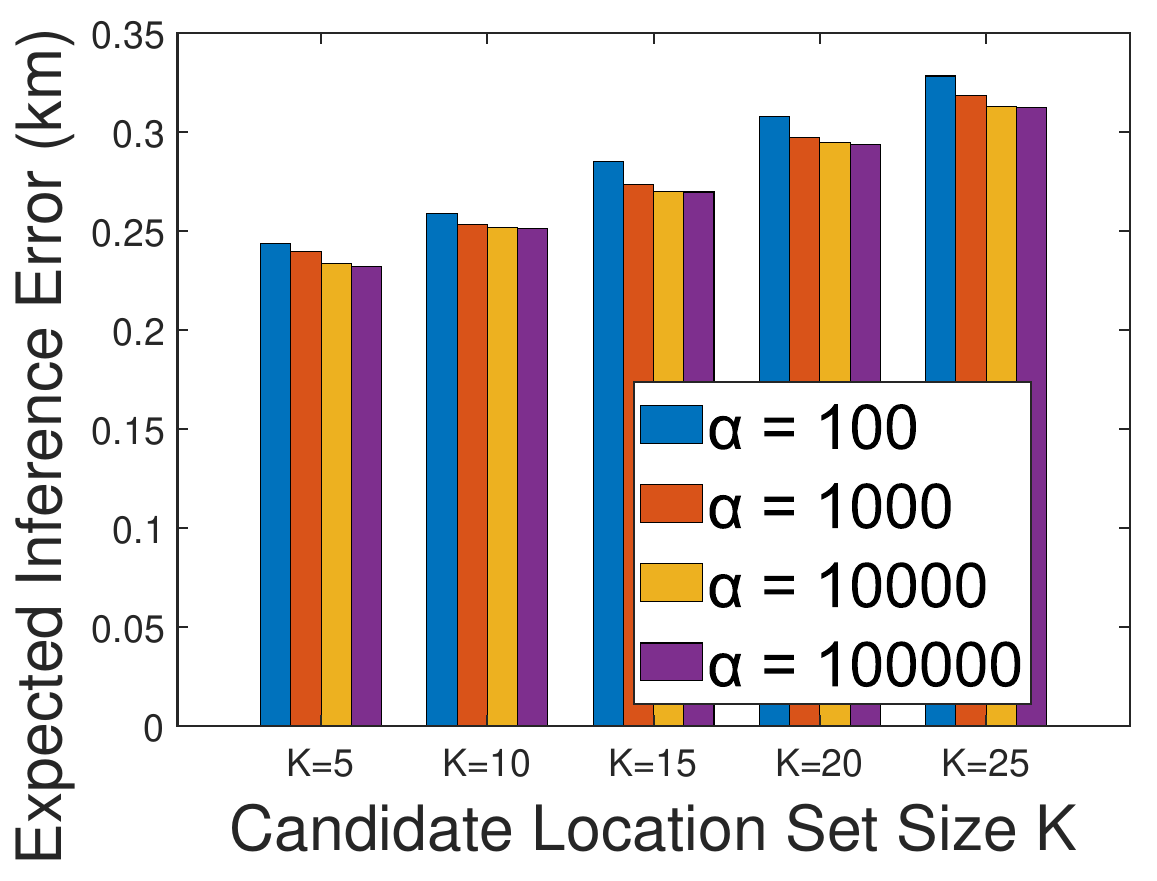}}
\vspace{-0.10in}
\caption{Impacet of $K$ (candidate location set size) and $\alpha$ (utility loss weight) on the expected inference error of TransProtect for the San Francisco dataset.}
\label{fig:EIE_K_weight_SF}
\end{minipage}
\vspace{-0.00in}
\end{figure}

\vspace{0.01in}
\noindent \textbf{(1) The TransProtect parameters $K$ (candidate location set size) and $\alpha$ (utility loss weight) impact the data utiltiy loss}. We present the average data utility loss and the expected inference error of ``Laplace+TransProtect'' and ``LP+TransProtect'' given different $K$ and $\alpha$ for the Rome and San Francisco datasets in Fig. \ref{fig:UL_K_weight_rome}(a)(b) and Fig. \ref{fig:EIE_K_weight_rome}(a)(b), and Fig. \ref{fig:UL_K_weight_SF}(a)(b) and Fig. \ref{fig:EIE_K_weight_SF}(a)(b), respectively. In addition, the expected utility loss and the expected inference error of the two methods with different values of $K$ and $\epsilon$ are shown in Table \ref{Tb:exp:QL_K} and Table \ref{Tb:exp:EIE_K}. The figures and tables show that the average data utility loss and expected inference error of ``Laplace+TransProtect'' and ``LP+TransProtect'' increases with an increase in $K$. This is because a higher value of $K$ expands the candidate location set, providing a chance for locations with higher data utility loss to be selected. 

In Fig. \ref{fig:UL_K_weight_rome}(a)(b), Fig. \ref{fig:EIE_K_weight_rome}(a)(b), Fig. \ref{fig:UL_K_weight_SF}(a)(b) and Fig. \ref{fig:EIE_K_weight_SF}(a)(b), we find that when $\alpha$ increases, both data utility loss and expected inference error of ``Laplace+ TransProtect'' and ``LP+TransProtect" decreases. 
This is because a higher $\alpha$ value results in locations with lower data utility loss having a comparatively higher score than locations with higher probability scores (output by the Transformer encoder), making them more likely to be selected as candidate locations. Fig. \ref{fig:UL_K_weight_rome}(a)(b) and Fig. \ref{fig:UL_K_weight_SF}(a)(b) provides a visual example, illustrating that when $\alpha = 100$, certain locations with higher data utility loss are included in the candidate location set. Conversely, when $\alpha = 10,000$, almost all candidate locations can achieve low data utility loss. Fig. \ref{fig:UL_K_weight_rome}(a)(b) and Fig. \ref{fig:UL_K_weight_SF}(a)(b) also indicates that once $\alpha \geq 10,000$, data utility loss plays a predominant role in candidate location selection in TransProtect, and further increases in $\alpha$ do not significantly impact data utility loss (as observed when comparing data utility loss at $\alpha = 10,000$ and $\alpha = 100,000$).

\begin{figure}[h]
\centering
\hspace{0.00in}
\begin{minipage}{0.145\textwidth}
\centering
  \subfigure[$K = 10, \alpha = 100$]{
\includegraphics[width=1.00\textwidth, height = 0.115\textheight]{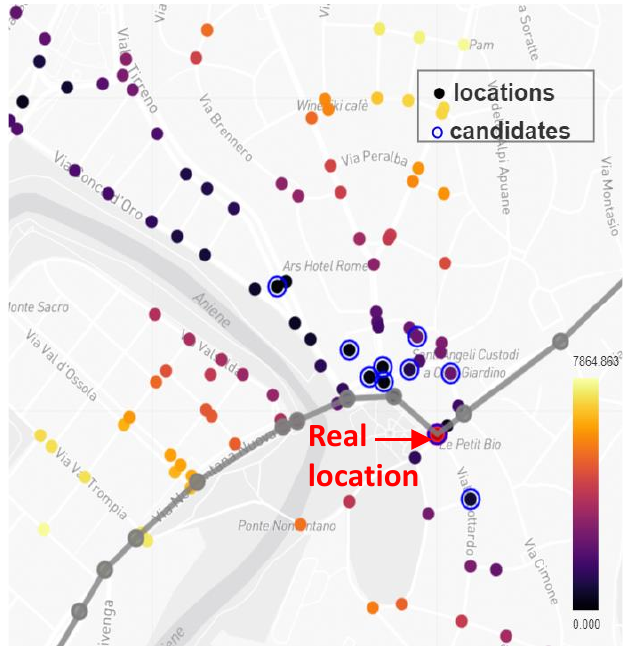}}
\end{minipage}
\hspace{0.00in}
\begin{minipage}{0.145\textwidth}
\centering
  \subfigure[$K = 15, \alpha = 100$]{
\includegraphics[width=1.00\textwidth, height = 0.115\textheight]{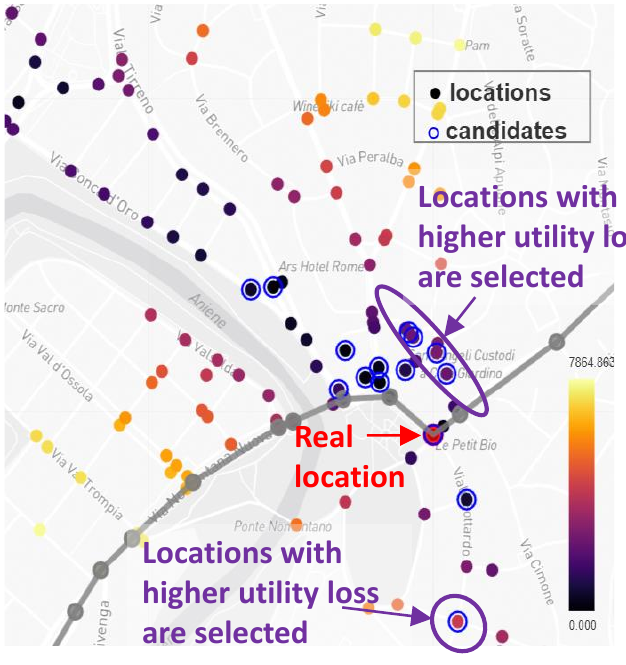}}
\end{minipage}
\hspace{0.00in}
\begin{minipage}{0.145\textwidth}
\centering
  \subfigure[$K = 10, \alpha = 10^5$]{
\includegraphics[width=1.00\textwidth, height = 0.115\textheight]{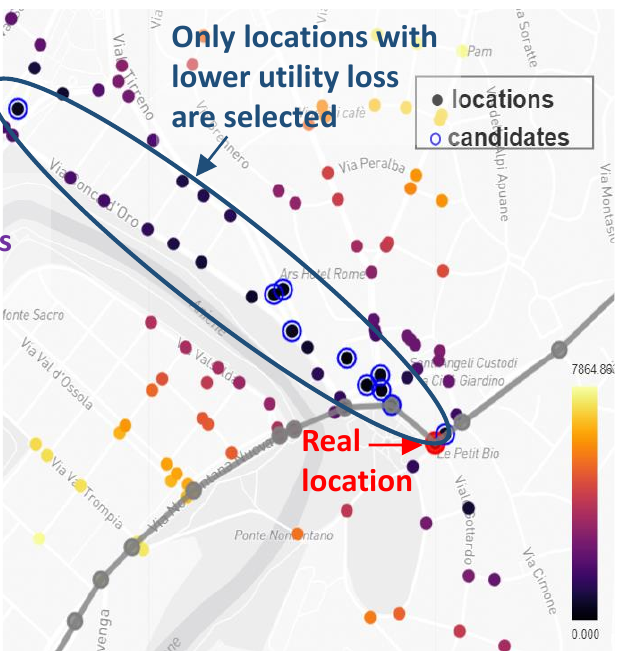}}
\end{minipage}
\vspace{-0.00in}
\caption{Impacet of $K$ (candidate location set size) and $\alpha$ (utility loss weight) on the data utility loss of TransProtect. \newline *(a)(b)(c) shows the heatmap of the data utility loss of the locations around a real location (marked by red).}
\label{fig:UL_K_weight_example}
\vspace{-0.1in}
\end{figure}

In addition, Fig. \ref{fig:UL_K_weight_example}(a)(b)(c) give illustrative examples to show how $K$ and $\alpha$ impact the data utility loss. Fig. \ref{fig:UL_K_weight_example}(a)(b) shows that when $K$ is increased from 10 to 15, more locations with higher data utility loss become part of the candidate location set. Fig. \ref{fig:UL_K_weight_example}(a)(c) shows that when $\alpha = 100$, some locations with higher utility loss are included in the candidate location set, while when $\alpha = 10,000$, almost all the candidate locations can achieve low utility loss. The figure also indicates that when $\alpha \geq 10,000$, utility loss already achieves the major role in candidate location selection in TransProtect, and further increasing $\alpha$ won't impact the utility loss significantly (by comparing the utility loss when $\alpha = 10,000, 100,000$).

\DEL{
\vspace{-0.00in}
\begin{table}[h]
\caption{Geo-Ind violation ratio when TransProtect is applied}
\vspace{-0.10in}
\label{Tb:exp:GVR}
\centering
\begin{tabular}{ c|c|c }
\hline
\hline
\multicolumn{1}{ c  }{}& \multicolumn{2}{ c }{Average Geo-Ind violation ratio (\%)} \\
\cline{2-3} 
\multicolumn{1}{ c|  }{ $\epsilon$ } & Laplace + TransProtect & LP + TransProtect  \\ 
\hline
\multicolumn{1}{ c|  }{ 5.0km$^{-1}$} & 0.0002819 & 0.0000072 \\ 
\multicolumn{1}{ c|  }{ 7.5km$^{-1}$} & 0.0000421 & 0.0000028 \\ 
\multicolumn{1}{ c|  }{ 10.0km$^{-1}$} & 0.0000192 & 0.0000012 \\ 
\hline
\end{tabular}
\vspace{-0.10in}
\end{table}

\vspace{0.03in}
\noindent \textbf{(2) The Geo-Ind violation ratio of integerating TransProtect is low}. As outlined in Section \ref{subsec:discussionGeoInd}, it is acknowledged that TransProtect does not theoretically ensure Geo-Ind. In Experiment II, we evaluate the \emph{Geo-Ind Violation Ratio}, defined as the proportion of $z_{i,k}$ and $z_{j,k}$ pairs that do not adhere to the Geo-Ind constraint in Equ. (\ref{eq:Geo-Ind-LP}), when $\epsilon = $ 5.0km$^{-1}$, 7.5km$^{-1}$, and 10.0km$^{-1}$. The outcomes, presented in Table \ref{Tb:exp:GVR}, reveal that the Geo-Ind violation ratio peaks at a mere 0.00028\%, demonstrating that Geo-Ind can be practically achieved with an exceedingly high likelihood.}

\newpage

\DEL{

Fig. xx represents the GRU architecture employed for posterior refinement. 
The input posterior vectors undergo an initial processing step within the \emph{dimensionality management block}. This block employs a padding method to enforce a standardized input format, specifically by aligning all trajectories with the longest one in the dataset.

The padded posterior vectors are then passed to the \emph{post-GRU neural network layer},  a combination of 5 GRU layers. Each GRU layer performs a sequence of calculations and correspondingly updates the states of update and reset gates dynamically. 
For a particular GRU layer, there are two main gates, i.e. reset and update gate. Moreover, there is a candidate hidden state, which carries the information that is needed to be considered for the next hidden state. The reset gate is responsible for determining how much of the previous hidden state must be forgotten or reset at the current instance of a time slot in a trajectory. The reset gate operates upon the current time slot’s posterior vector and previously hidden state vector using the sigmoid activation function. 

\begin{equation}
    r_t= \sigma(W_{r}.[h_{t-1},X_{t} ]+b_{r} )
\end{equation}
where $W_r$ is the weighted matrix and $b_r$ bias for the reset gate output.

The update gate is mainly responsible for deciding how much of the new information should be included in the current hidden state vector. It also operates with a previous hidden state and current posterior vector utilizing a sigmoid activation function. 

\begin{equation}
u_t= \sigma(W_{u}.[h_{t-1},X_{t} ]+b_{u} )
\end{equation}
where $W_u$ is the weighted matrix and $b_u$ bias for the update gate output.

The candidate hidden state carries the new information that needs to be incorporated into hidden state and is computed by operating upon reset gate output, the current time slot posterior vector and the previous hidden state vector using the hyperbolic tangent (tanh) activation function. At each vector multiplication or addition place, element-wise operation is performed (i.e., either multiplication or addition). This candidate’s hidden state and previous hidden state are blended by the update gate for generating the final hidden state for that time slot in trajectory.

\begin{equation}
    {h_{t}^{'}} = \tanh{(W_h.[r_t*h_{t-1},X_{t} ]+b_{h} )}
\end{equation}
where $W_h$ is the weighted matrix and $b_h$ bias for the candidate state.

\begin{equation}
    h_t=(1- z_t )*h_{t-1}+z_t*h_i^{'}
\end{equation}

After the post-GRU neural network layer, the output is passed to the sigmoid activation function block for constraining the output within the range [0,1]. Then the output of the activation block is passed to the output block where argmax operation is performed upon the output to get the final prediction. For model optimization and creating the comparison between predicted and actual values, Cross entropy is employed as a loss function, for the model’s ultimate output refinement.

\textbf{Posterior Long short-term memory (Post-LSTM)}: Fig xx. gives a high-level overview of the LSTM model architecture.
The architecture first takes the sequence of posterior vectors as an input. As mentioned in the above explanations, these vectors carry information about spatial and statistical information regarding trajectory data. Once the posterior sequences are onboarded, the dimensionality management block manages the inconsistency in the sizes of the trajectory. In this block trajectory padding is performed to ensure the consistency and uniformity of the among all the input sequences. Over here, all the posterior sequences are padded taking reference to the longest trajectory in the dataset. Then comes is Post-BiLSTM layer, which is the BiLSTM-inspired architecture. Typical BiLSTM more often takes the scalar value for a given instance of time and position in the sequence, whereas in this architecture, each posterior vector is treated as a holistic entity. Since the posteriors are in vectorized form, all the other outputs of the gates and hidden states of LSTM cells are also in vectorized form, where element-wise multiplication is performed. Moreover, the BiLSTM architecture is employed where two parallel layers of LSTM blocks (forward and backward) work in parallel to capture the bidirectional patterns of the sequences. 
The data flow of the posterior vectors in intrinsic gates of the LSTM cell i.e., forget, input, and output gates happens dynamically. The forget gate mainly decides to either retain the previous cell state or not. Both, the previous hidden state vector and current posterior vector act as the input for the forget gate, which is activated through the sigmoid function. The weighted matrix and bias are also maintained in parallel. 

\begin{equation}
    f_t= \sigma(W_f.[h_{t-1},X_t ]+b_f )
\end{equation}
where $W_f$ is the weighted matrix and $b_f$ bias for the forget gate respectively.
Similarly, the input gate also operates using the previous hidden state vector and posterior vector, whose main task is to determine which information should be added to the cell state from the current posterior input. The input gate is also activated using the sigmoid function. 

\begin{equation}
    i_t= \sigma(W_I.[h_{t-1},X_t ]+b_I )
\end{equation}
where $W_I$ is the weighted matrix and $b_I$ bias for the input gate respectively.

There is a cell state in every LSTM block, which is responsible for storing and carrying the information across the different time slots in a sequence. The candidate cell state is the representation of the new information that can be added to the cell state of the LSTM. The linear combination of the current posterior input and the previous hidden state passed across the hyperbolic tangent (tanh) activation function, and the output is represented as the candidate cell state. The updated cell state is computed by element-wise multiplication and addition of forget and input gates outputs. 

\begin{equation}
    c_t^{'}= \tanh{(W_c.[h_{t-1},X_t ]+b_c )}
\end{equation}
where $W_c$ is the weighted matrix and $b_c$ bias for the candidate state.
\begin{equation}
    c_t=f_t*c_{t-1}+i_t*c_t^{'}
\end{equation}

The output gate gives the output for the LSTM cell as well as determines the next hidden state vector, where the same sigmoid function is used for activation with the previously hidden state vector and posterior vector. The output of this sigmoid function is considered as output and element-wise multiplication of this output with updated cell state gives the final hidden state.

\begin{equation}
    o_t= \sigma(W_o.[h_{t-},X_t ]+b_o )
\end{equation}
where $W_f$ is the weighted matrix and $b_f$ bias for the output gate respectively.

\begin{equation}
    h_t= o_t*\tanh⁡{c_t}
\end{equation}

These interactions of intrinsic gates give the utility to LSTM to structure and memorize the long-range dependencies of posterior vector data.
After the Post-BiLSTM layer, the model takes the sigmoid activation function for constraining the output, so that the final output converges within the range [0,1]. The final layer in the architecture involves an output block where the argmax operation is performed on activation output. This way the predicted node is yielded from the model.
For optimization of the model at each iteration of training, the Cross entropy Loss function is employed due to its efficacy in performing well for classification tasks.

\begin{equation}
L_t = \sum_{j} Y_t[j].\log{O_t[j]}
\end{equation}

where $Y_t [j]$ is the one hot encoded target indexed at $j$ and $(O_t [j])$ is the prediction probability for the same index.

By updating the model's parameters through backpropagation, training seeks to reduce this loss. 

The training process is iterated over multiple batches of posterior vectors. For the same trajectories, 20 different sets of posterior vectors are prepared for capturing the randomness of the obfuscation mechanism. This iterative training process introduces the unique perspective of generalization across the various trajectories. It enhances the robustness of the model, making it capable of handling the wide range of input variations posed by the multiple posteriors.}

\end{document}